\documentclass[onecolumn,showpacs]{revtex4}
\usepackage{psfig,epsfig,graphicx,float,pst-all}
\usepackage{amsfonts,amssymb}
\usepackage{setspace}
\usepackage{makeidx}
\usepackage{multind}
\renewcommand{\=}{~=~}
\newcommand{\WD}{{\cal D}}
\newcommand{\be}{\begin{equation}}
\newcommand{\ee}{\end{equation}}
\newcommand{\bea}{\begin{eqnarray}}
\newcommand{\eea}{\end{eqnarray}}
\newcommand{\bc}{\begin{center}}
\newcommand{\ec}{\end{center}}
\begin{document}

\title{Solutions of the Bohr hamiltonian, a compendium}
\author{Lorenzo Fortunato}
\affiliation{Vakgroep subatomaire en stralingfysica, Proeftuinstraat 86,
B-9000 Ghent, Belgium}

\begin{abstract}
The Bohr hamiltonian, also called collective hamiltonian, is one of the
cornerstone of nuclear physics and a wealth of solutions (analytic or 
approximated) of the associated eigenvalue equation have been proposed 
over more than half a century (confining ourselves to the 
quadrupole degree of freedom).
Each particular solution is associated
with a peculiar form for the $V(\beta,\gamma)$ potential. 
The large number and the different details of the mathematical
derivation of these solutions, as well as their increased and renewed 
importance for nuclear structure and spectroscopy,  demand a  
thorough discussion. It is the aim of the present monograph to present
in detail all the known solutions in $\gamma-$unstable and $\gamma-$stable 
cases, in a taxonomic and didactical way. In pursuing this task
we especially stressed the mathematical side leaving the discussion of the
physics to already published comprehensive material. 

The paper contains also a new approximate solution for the linear potential,
and a new solution for prolate and oblate soft axial rotors,
as well as some new formulae and comments, and an appendix on the analysis 
of a few interesting numerical sequences appearing in this context.
The quasi-dynamical SO(2) symmetry is proposed in connection with the 
labeling of bands in triaxial nuclei.

\end{abstract}     
\maketitle 

\fbox{PREPRINT - PRELIMINARY -\today }

\tableofcontents

\section{Introduction} 
More than half a century has elapsed from the publication of one of the  
milestones of nuclear physics { \it ``The coupling of nuclear surface  
oscillations to the motion of individual nucleons''} \cite{Bohr1}   
on the danish journal 
{\it ``Matematisk-fysiske Meddelelser''} by Aage Bohr in 1952. 
The cited work contains the foundations of the collective model,  
developed fully in a second important work in collaboration with Ben R.  
Mottelson \cite{Bohr2},  
and especially contains the first derivation of the subject of this  
monograph: the Bohr hamiltonian, also called collective hamiltonian 
or , sometimes, Bohr-Mottelson hamiltonian($H_B$).  
This operator contains  kinetic as well as restoring potential terms in 
some set of  
variables describing the extent of deformation of the nuclear surface. 
The present paper, far from being a comprehensive review (that are numerous  
in the specialized literature \cite{Rev,EG}) of the many physical topics,
successes and applications of the collective model, is rather meant to 
provide the reader with an enumeration and mathematical discussion of  
the analytic and approximate solutions of the stationary eigenvalue equation  
$H_B \Psi=E\Psi$ with various form of the potentials and referring to  
different physical situations. Other developments originated from the 
collective model (as the Frankfurt model, the Baranger-Kumar model or 
the symplectic model) will not be discussed here.

The need for such a taxonomic {\it compendium} is, to our view, 
justified by the recent proliferation of new solutions that makes this 
topics exciting and, at the same time, significantly extended. 
The new solutions have received a considerable attention, both from
theoreticians and from experimentalists, and application of these
ideas or survey of spectroscopic data are constituting a very important
research line.
It is also our belief that these studies have reached 
a considerable degree of completeness and that a classification of the  
subject is therefore timely. To corroborate our views upon the renewed
role of the collective model as a driving force for new research, 
we mention that an interesting paper, that contains some sections 
dedicated to the Bohr hamiltonian, has recently been published \cite{Jolos}
(See section \ref{Jol-sec}).

The problem of the solution of the Bohr hamiltonian has been mainly 
tackled in two ways: direct solution of the  
second order differential equation and use of algebraic techniques to 
exploit dynamical symmetries. The latter approach has also been used as a  
source of 'labels' for various solutions of distinctive importance.  
A novel interest has been raised by the possibility to give analytical 
solution at the critical point of a shape phase transitions between 
various types of quadrupole deformed nuclear surfaces \cite{Iac1,Iac2}. 
Another method that is worth mentioning by virtue of the insight that 
one may get is the numerical diagonalization of the problem for 
complicated potentials that cannot be treated analytically. We will also 
remark on some of these cases.

After a brief discussion on the historical and scientific foundations  
of the collective model (Section II), we will  
be concerned with an analysis of the known solutions of the  
eigenvalue equation for the Bohr hamiltonian for $\gamma-$unstable  
(Section III), axial $\gamma-$stable (Section IV) and triaxial
$\gamma-$stable (Section V) cases, with direct solution 
of the differential equation, which we will lay  
before the reader following, as far as possible, the course of time. 
Some comments on group theoretical techniques and solutions are also  
given (Section VI). Section VII contains a brief description of a few 
important recent advances that are connected with the main theme of 
this compendium.

Writing in a single paper a summary of works and efforts that have been 
devoted to the study of collective models and paying the tribute to every 
researcher involved in the matter would be a formidable task,  
and we have decided to confine ourselves to the work plan cited above,  
apologizing for any unintentional omission.  We decided to label the 
solutions on the basis of their proposers, mainly to avoid confusion
with the corresponding problems in the solution of the Schr\"odinger 
equation in the configuration space. Where the authors were many we used
alternative titles.

This paper is not only a review, but contains some novelties as 
the treatment of the linear potential in Section III.H, that to our knowledge 
has never been treated in this context, or, for instance, formula  
(\ref{mianorma}). Moreover we propose a new solution for prolate and
oblate soft axial rotors (Section \ref{new-sol}) obtaining 
interesting patterns for $\gamma$ and $\beta$ excitations. \\
A critical discussion of the various cases is supplemented
by few comments that may share light on specific points (as for instance 
in section \ref{corrispondenza}).
 In the appendix we discuss from a purely number theoretical perspective, 
the problem of repetitions  
of quantum numbers, that, as far as we know, has never been discussed  
elsewhere: we give two new sequences and we evidence some curious  
connections with other fields.

\section{Foundations of the Collective Model} 
Niels Bohr, father of Aage Bohr and of the atomic theory,  
is among the proposers of the theory of nuclear surface oscillations  
\cite{Niels}. The idea was to treat the nucleus as a liquid droplet, and that 
the fundamental collective modes of motion of the surface were linked to  
nuclear excitations.
Let us quote Aage Bohr's words \cite{Bohr1} about the liquid drop theory: 
\begin{quotation} 
According to the liquid drop model, the fundamental modes of nuclear  
excitation correspond to collective types of motion, such as surface  
oscillations and elastic vibrations. Even if it has not been possible,  
with certainty, to associate observed nuclear levels with particular  
modes of oscillation, the model gives an immediate explanation of the  
rapid increase of level density with increasing excitation of the nucleus. 
\end{quotation} 
The fifty years elapsed since the time of the exposure of his ideas have  
contradicted the last sentence: it is indeed possible to associate solutions 
of the Bohr hamiltonian with nuclear spectra, giving an immediate explanation 
of many nuclear properties.  
The collective model, that was subsequently developed in collaboration  
with Ben R. Mottelson and that has taken the name of Bohr-Mottelson model,  
should be regarded, as Bohr was warning, as a complementary approach to  
the shell model and should shed light only 
on some aspects of the still dim complexity of nuclear spectra. 
 
The innovative idea underlying the already cited work was 
\begin{quotation} 
...to consider various properties of a nucleus described in terms of a 
deformable surface coupled to the motion of individual nucleons. This combined 
model may be referred to as the quasi-molecular model. 
\end{quotation} 
It is strange that the name proposed by the author has been lost. He found that 
his model bore a resemblance to the theory of molecules in the very same way 
in which the single particle shell model bore a resemblance to the  
atomic theory. 
 
\begin{center} 
\begin{figure}[!t] 
\begin{center} 
\begin{picture}(200,220)(0,0) 
\psset{unit=0.85pt}    
\pscircle[linewidth=1pt](120,120){50} 
\psarc[arcsepB=2pt]{->}(120,120){60}{20}{40} 
\rput(180,150){\small $\gamma$} 
\rput(185,120){\small $\beta$} 
\psline{-}(60,120)(180,120)  
\psline{-}(83,70.5)(157,169.5)  
\psline{-}(157,70.5)(83,169.5) 
\psline{-}(240,85)(240,170) 
\psline{-}(0,95)(0,165)  
\rput(240,120){\epsfig{file=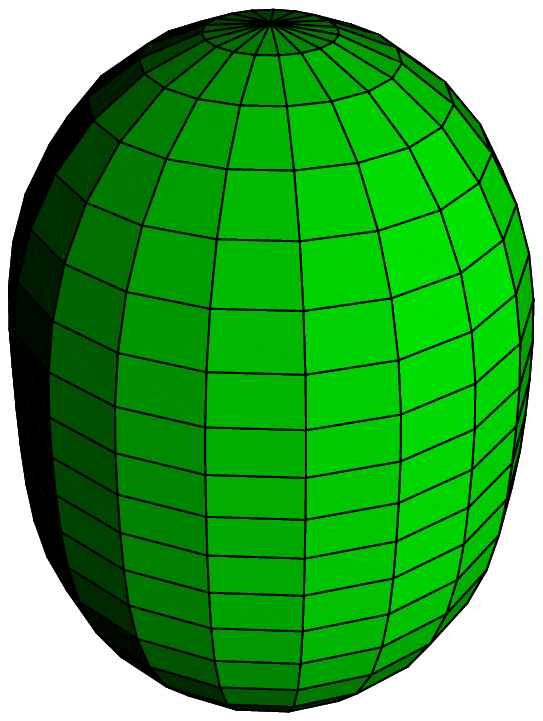,scale=0.3}} 
\rput(0,120){\epsfig{file=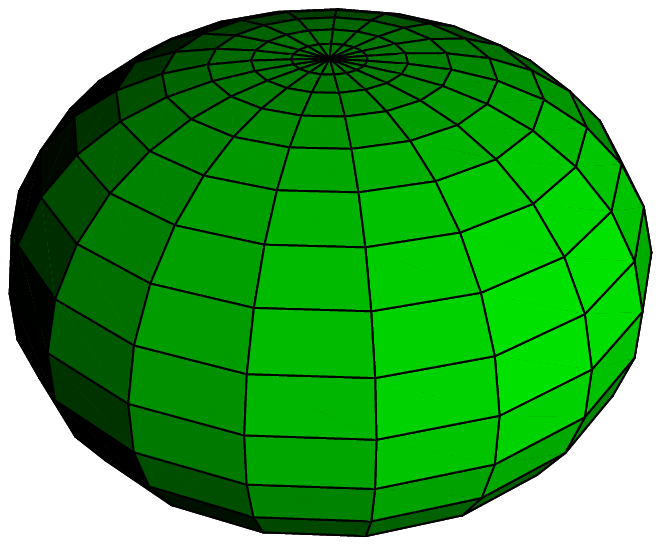,scale=0.3}} 
\rput(245,168){\small z} 
\rput(5,163){\small z} 
\psline{-}(164,188)(215,225) 
\psline{-}(20,14)(88,58)  
\rput(190,202.5){\epsfig{file=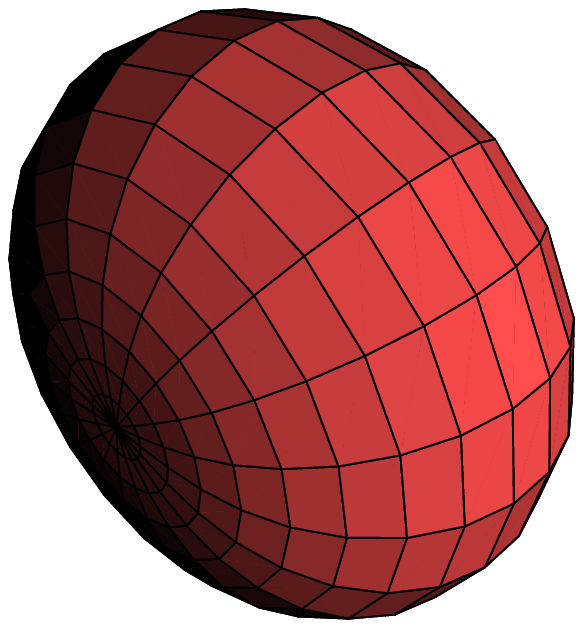,scale=0.3}} 
\rput(54,31.5){\epsfig{file=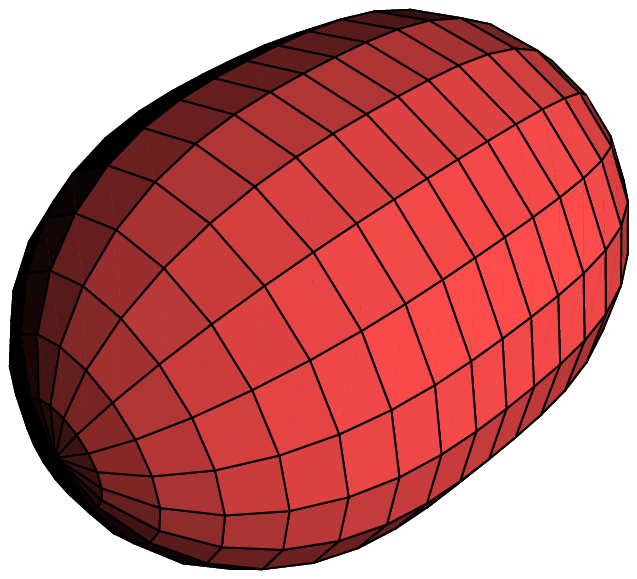,scale=0.3}} 
\rput(220,225){\small y} 
\rput(26,11){\small y} 
\psline{-}(22,226)(95,190) 
\psline{-}(175,60)(232,28)  
\rput(56,202.5){\epsfig{file=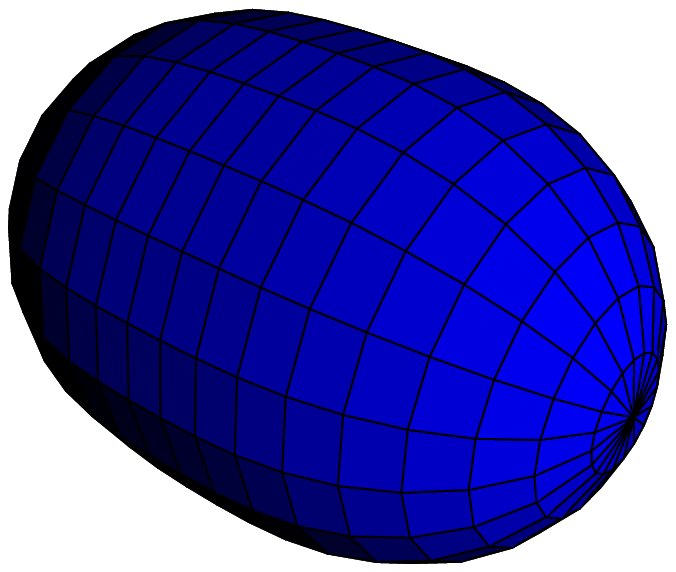,scale=0.3}} 
\rput(200,35.5){\epsfig{file=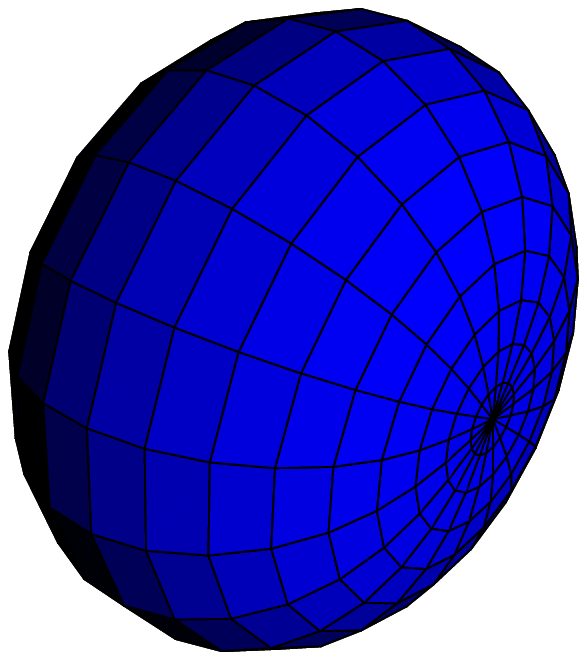,scale=0.3}} 
\rput(19,222){\small x} 
\rput(228,24){\small x} 
\end{picture} 
\caption{Hill-Wheeler coordinates $\{\beta,\gamma\}$. The quadrupole deformed  
shapes corresponding to $\beta = 0.4$ and $\gamma = n\pi/3$  
(with $n=0,...,5$) are shown for reference.  Different colors are  
connected with different principal axes of symmetry  
({\green green} for z, {\red red} for y and {\blue blue} for x). }  
\label{Hill} 
\end{center} 
\end{figure} 
\end{center}

Bohr's paper was confined to the treatment of a single nucleon interacting 
with the nuclear surface. The radius of the latter was expressed,  
in polar coordinates, as an expansion in spherical harmonics: 
\be 
R(\theta,\phi)\= R_0 \Bigl(1+\sum_{\lambda,\mu}\alpha_{\lambda,\mu} 
Y_{\lambda,\mu}(\theta, \phi) \Bigr) \,,
\label{expa} 
\ee 
where $R_0$ is the radius of the nucleus when it has the spherical  
equilibrium shape. The expansion parameters $\alpha_{\lambda,\mu}$ are the  
coordinates that define a multidimensional space, whose points represent a  
deformed surface. The requirement of reality for the nuclear radius implies 
\be 
\alpha_{\lambda,\mu}\=(-1)^\mu \alpha_{\lambda,-\mu}^* \,.
\ee 
The consistency of expansion (\ref{expa}) with the 'atomicity' of the 
nucleus,  
i.e. the fact that it is composed by smaller particles, the nucleons, 
is assured by realizing that the idea of a continuous surface loses its 
meaning 
if one takes into consideration surface elements whose dimensions are  
comparable with the average distances between the constituents. This happens 
whenever $\lambda$ is larger or equal to the linear dimensions of the nucleus, 
$A^{1/3}$.\\ 
In the following we will confine to the quadrupole degree of freedom  
($\lambda=2$) leaving aside the discussions about other multipolarities:
in fact quadrupole states are the most fundamental collective type of 
low-lying excitations in nuclei and deserve a thorough analysis. 
If only small values of $\alpha$'s are taken into account  
the quadrupole deformed surface is an ellipsoid randomly oriented in space. 
The five coordinates $\{\alpha_{2,\mu}\}$ may thus be mapped onto a set  
of other five variables $\{a_0,a_2,\theta_1,\theta_2,\theta_3 \}$,  
three of which describe angular orientations of the ellipsoid, the other two 
parameters being related to the extent of the deformation. This mapping  
from the frame of reference $R$ to a second frame of reference $R'$, called  
'intrinsic', may be achieved via a unitary transformation: 
\begin{eqnarray} 
a_{2,\nu}&\=&\sum_{\mu=-2}^{2} \alpha_{2,\mu} \WD_{\mu,\nu} (\theta_i)\\ 
\alpha_{2,\mu}&\=&\sum_{\mu=-2}^{2}a_{2,\nu}  \WD_{\mu,\nu}^* (\theta_i) \,,
\label{transf} 
\end{eqnarray} 
where $\WD_{\mu,\nu} (\theta_i)$ are the Wigner rotation functions  
for spherical 
harmonics with $\lambda\=2$.  The set of three angles $\{\theta_i\}$  
describes the relative orientation of the two frames of reference.
From now on we will drop the index $2$ since, as already said, we will 
deal only with quadrupole deformations. 
The transformation (\ref{transf}) may be conveniently chosen so that the  
axes of the frame of reference $R'$ coincide with the principal axes of the 
ellipsoid. In this case  
\be 
a_1=a_{-1}=0 ~~\mbox{~~~and~~~}~~ a_2=a_{-2}\,. 
\ee 
The transformation of coordinates described above is however not unambiguous  
because of the possible different order of labeling of the axes and of the  
choice of the direction of the versors. This amounts to 48 sets of tranformed 
coordinates which correspond to the initial set (24 as long as
right handed coordinate systems only are considered). The symmetry 
properties of the wave functions 
will be affected by the arbitrariness of the choice of the transformed  
coordinate system and this fact implies certain symmetry relations. 
Another substitution 
is required to reach a new set of coordinates $\{\beta,\gamma,\theta_i \}$  
called the Hill-Wheeler coordinates, in which the subset  $\{\beta,\gamma\}$
defines a two-dimensional polar coordinate system within the 
five-dimensional quadrupole deformation space, namely: 
\begin{eqnarray} 
a_0&\=&\beta\cos{\gamma}\\  
a_2&\=& (\beta/\sqrt{2})\sin{\gamma} \,,
\end{eqnarray} 
with $\sum_\mu \mid \alpha_\mu \mid^2 \= \beta^2$. 
The total deformation of the nucleus is measured by $\beta$ in such a way that 
$\beta=0$ represents a sphere and $\beta\ne 0$ is an ellipsoid. The larger 
the value of $\beta$, the more deformed the surface. The parameter $\gamma$ 
describes the deviations from rotational symmetry: whenever  
$\gamma\= n{\pi\over 3}$, with $n\in\mathbb{Z}$, two of the three semi-axes  
are of equal length. This is depicted in fig. \ref{Hill}. 
To each (solid-line) radius in the polar plot of this figure corresponds an
axially symmetric ellipsoid with a well-specified symmetry axis and a 
given character for what concerns prolateness or oblateness. 
Every point in the areas within the radii corresponds to triaxial shapes,
that are characterized by three different semi-axes.  

A convenient way to deal with quadrupole surfaces is to rewrite eq.  
(\ref{expa}) in terms of the new variables. The three radii defining the  
ellipsoid in the body-fixed frame are thus
\be 
R_k\=R_0 \Bigl[ 1+{5\over 4\pi} \beta \cos{\bigl( \gamma-{2\over 3} \pi  
k\bigr)} \Bigr] 
\ee 
with $k=1,2,3$ . 
 
Now we are ready to introduce the Bohr hamiltonian, that is the hamiltonian 
built with generalized coordinates and momenta in quadrupole deformation 
space: 
\be 
H_B\= T+V\=\sum_{\mu} \Bigl\{ {1\over 2B_2}\mid\pi_{\mu} \mid^2 + {C_2\over 2} 
\mid\alpha_\mu \mid^2   \Bigr\} 
\label{hb_alpha} 
\ee  
where $B_2$ (that will be called $B_m$ i n the following) and $C_2$ are 
the mass and stiffness parameters of the 
liquid drop model for the quadrupole multipolarity. Here $\pi_\mu$ are the
conjugate momenta associated to $\alpha_\mu$. 
We wish to call 'Bohr hamiltonian' the set of more general expressions 
in which the potential term is a generic function of the parameters.  
We separate the Bohr hamiltonian into three terms: 
\be 
H_B\= T_{vib} + T_{rot} + V. 
\label{hb} 
\ee 
All the above terms are in general functions of $\beta$ and $\gamma$. 
The first term is the kinetic energy term related with shape vibrations  
with fixed orientation in space, the second instead is the kinetic energy of 
a rotational motion of the nuclear surface without any change of shape. The  
third term is the restoring potential in the shape parameters. 
Without entering into the details of the derivation \cite{Jean}  
we merely restate the  
definitions of the two kinetic terms using the ${\beta,\gamma}$ variables: 
\begin{eqnarray} 
T_{vib}&=& -{\hbar^2\over 2B_m} \Biggl\{{1\over \beta^4}{\partial \over 
 \partial \beta} \beta^4{\partial \over \partial  
\beta}+{1\over \beta^2}{1\over \sin{3\gamma}}{\partial \over \partial\gamma}  
\sin{3\gamma}{\partial \over \partial\gamma}\Biggr\}\nonumber \\ 
&&\label{tvib}\\ 
T_{rot}&=& {\hbar^2\over 2B_m}{1\over 4\beta^2} \sum_{k=1}^3 {Q_k^2 \over  
\Bigl[\sin{(\gamma -{2\pi\over 3}k)}\Bigr]^2} \label{trot} \,. 
\end{eqnarray} 
The vibrational part is usually divided into a $\beta$ and a $\gamma$ part
(corresponding to the two terms in parenthesis in (\ref{tvib})), though
the two variables are actually mixed by the second term.
We will consider potential energies that are only function of the internal  
coordinates $\beta$ and $\gamma$.  
The wave functions $\Psi(\beta,\gamma,\theta_i)$ are solutions of the  
eigenvalue equation $H_B\Psi\= E\Psi$ for the hamiltonian (\ref{hb}). 
\par 
It may be thought that, in some sense essentially, the equation to solve 
is nothing but the Sch\"odinger equation and that the record of cases that  
have already been known in the quantum mechanical context,
apply to the present situation as well. This is  
partly true, but there are anyway a number of major differences that must  
be underlined and for which the collective hamiltonian deserves special cares: 
 the Bohr equation is richer being expressed in terms of two variables 
(read two out of five), its 'natural' space is five  
dimensional instead of three dimensional and these affects not only  
asymptotic behaviours and boundary conditions (that would be trivial), 
but also the group structure that we can identify with the Bohr hamiltonian. 
 
The main part of the following will be 
concerned with $\gamma-$unstable solution, that is to say, exact solutions 
with a potential that is independent of $\gamma$. At the end of this  
enumeration we will treat approximate solutions with potentials that have 
also some dependence on $\gamma$.

\section{$\gamma-$unstable cases} 
Whenever the potential energy is only a function of $\beta$ the hamiltonian is  
separable \cite{Wilet}. Setting  
\be 
\Psi(\beta,\gamma,\theta_i)\=f(\beta) \Phi(\gamma,\theta_i) 
\ee 
we may write two equations, one for the $\beta$ variable,
\be 
\Biggl\{ {\hbar^2\over 2B_m} \Bigl(-{1\over \beta^4}{\partial \over 
 \partial \beta}\beta^4{\partial \over \partial \beta}+ 
{\Lambda^2 \over \beta^2}\Bigr) +V(\beta) \Biggr\}  f(\beta)  
\= E f(\beta) 
\label{betaeq0} 
\ee 
that may alternatively be expressed in the canonical form as 
\be 
\Biggl\{ {\hbar^2\over 2B_m} \Bigl(-{\partial^2 \over 
 \partial \beta^2}+{(\tau+1)(\tau+2)\over \beta^2}\Bigr) + 
V(\beta) \Biggr\} (\beta^2 f(\beta))\= E (\beta^2 f(\beta)) 
\label{betaeq} 
\ee 
and  the other for the $\gamma$ variable
\be
\Biggl\{ -{1\over \sin{3\gamma}}{\partial \over \partial\gamma}  
\sin{3\gamma}{\partial \over \partial\gamma}+ {1\over 4}  
\sum_{k=1}^3 {{\hat Q}^2_k \over (\sin{(\gamma -{2\pi\over 3}k)})^2}\Biggr\} 
\Phi(\gamma,\theta_i) 
=\Lambda \Phi(\gamma,\theta_i) \,,
\label{gammaeq} 
\ee 
where $\Lambda\=\tau(\tau+3)$ is the separation parameter with  
$\tau\=0,1,2,...$~. 
Since the potential is only a function of $\beta$, the angular momentum  
${\hat {\bf Q}}^2$ and its third component ${\hat Q}_z$  
are constants of motion and the quantum numbers associated with them, 
$L$ and $M$, are thus good quantum numbers.  
The so-called $\gamma-$angular part of the wavefunction may be written as 
\be 
\Phi(\gamma,\theta_i)\= \sum_{K=-L}^L g_K^{L,\tau,\nu} \WD_{M,K}^L (\theta_i) 
\,,
\ee 
where the rotation functions are eigenfunctions of the two operators 
${\hat Q}_z$ and ${\hat Q}_3$, respectively  
the third component of the angular momentum vector along the z axis in  
the fixed frame of reference (eigenvalue M) and the third component of the   
angular momentum vector along the z' axis in  
the intrinsic frame of reference of the nucleus (eigenvalue K). 
The functions $g$, that have the property $g_K\= g_{-K}$, are given explicitely
by B\'es \cite{Bes} and in addition to $L$ and $M$, are also labeled by  
other two quantum numbers $\tau$ and  
$\nu$. The former is the quantum number that comes from the solution of the 
eigenvalue equation of the Casimir operator of the SO(5) group (also 
called SO(5) seniority), while the latter is an empirical label needed in  
order to distinguish between the occurrence of multiple set of the same $L,M$ 
within a given SO(5) IR and takes the values \cite{Corr} :
\be 
\nu\=0,1,2,~...~,[\tau/3] \,,
\ee 
where square brackets indicates the integer part. 
Here we cannot omit to underline a fact that is worthy of remark: Bohr  
correctly denoted with  
$\tau$ the set of these two quantum numbers, but sometimes in the literature  
the $\tau$ quantum number has been taken as the label of SO(5) (as we do) and  
the $\nu$ quantum number has been left out (or forgot!). We think that the  
reason of this omission is that the importance of $\nu$ is seen only when one 
takes into account states with $\tau \geq 6$. In appendix A we give a 
detailed discussion of the determination of the sequence of repetitions.\\ 
The $K$ quantum number takes the value $\tau-3\nu$ and for each $K$ one 
has the following list of possible $L$'s: 
\be 
L=2K,2K-2,2K-3,\cdots,K 
\ee 
that is to say all the integers between $K$  and $2K$ except for $2K-1$. 
The resulting spectrum displays a typical degeneracy pattern in energy 
that is summarized in table I. 
 
\begin{table}[!t] 
\begin{tabular}{cccc} 
~~~~$\tau$~~~~&~~~~$\nu$~~~~&~~~~$K$~~~~&~~~~$L$~~~~\\ \hline\hline 
0&0&0&0\\ \hline 
1&0&1&2\\ \hline 
2&0&2&2,4\\ \hline 
3&0&3&3,4,6\\  
~&1&0&0\\ \hline 
4&0&4&4,5,6,8\\ 
~&1&1&2\\ \hline 
5&0&5&5,6,7,8,10 \\ 
~&1&2&2,4\\ \hline 
6&0&6&{\red 6},7,8,9,10,12\\ 
~&1&3&3,4,{\red 6}\\ 
~&2&0&0 \\ \hline 
\end{tabular} 
\caption{Quantum numbers for the $\gamma-$unstable states. Every 
group of states with the same $\tau$ has the same energy. The lowest  
possible multiple occurrence of states with the same $L$ within a given IR of  
SO(5) is marked in red (See appendix for details).} 
\label{ta1} 
\end{table} 
 
The sets of quantum numbers and the $\Phi$ eigenfunctions of (\ref{gammaeq}) 
discussed above are common to all $\gamma-$unstable problems of this section. 
The main differences in the spectra are thus to be searched in eq.  
(\ref{betaeq}), where in principle every possible potential function may 
be chosen. It is our aim to give here a discussion, as complete as possible,
of the known cases.

\subsection{Bohr's harmonic oscillator solution} 
\begin{figure}[!t] 
\epsfig{file=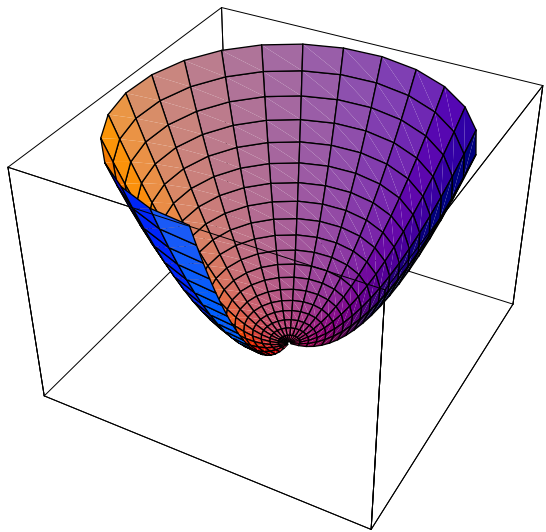,scale=0.7} 
\caption{Plot $V(\beta,\gamma)$ of the harmonic oscillator  
potential (\ref{vb}) discussed by Bohr with $C=1$ and $0<\gamma <3\pi/2$. } 
\label{harmo} 
\end{figure}

The solution given by A.Bohr \cite{Bohr1} was historically the first one.  
He put from the beginning an oscillator potential of the form 
\be 
V(\beta) \={1\over 2}C\sum_{\mu=-2}^2 \mid\alpha_\mu\mid^2  
\={1\over 2}C\beta^2 
\label{vb} 
\ee 
in eq. (\ref{betaeq}), obtaining the harmonic oscillator hamiltonian 
in a five  
dimensional space. The potential $V(\beta,\gamma)$ is depicted in fig. 
\ref{harmo}. 
The spectrum is given by 
\be 
E=(N+5/2) \hbar\omega 
\ee  
with $\omega=\sqrt{C/B}$ and  
\be 
N\= \sum_\mu n_\mu \=2n_\beta + \tau 
\ee 
with  
$N=0,1,2,3,...$\,. The numbers $n_\mu$ represent the number of phonons with a 
given $\mu \hbar$ component of angular momentum along the z axis. 
The last relation evidences the fact that the $\gamma-$unstable case, with a 
harmonic oscillator spectrum and a   
minimum in $\beta=0$, has a further degeneracy with respect to the case 
discussed above: to a given $N$ may correspond different sets  
of $(n_\beta,\tau)$. 
\bc 
\begin{table}[!t] 
\begin{tabular}{ccc} 
~~~~$N$~~~~&~~~~$(n_\beta,\tau)$~~~~&~~~~$L$~~~~\\ \hline\hline 
0&(0,0)&0\\ \hline 
1&(0,1)&2\\ \hline 
2&(0,2)&2,4\\  
~&(1,0)&0\\ \hline 
3&(0,3)&0,3,4,6\\ 
~&(1,1)&2 \\ \hline 
4&(0,4)&{\red 2},{\red 4},5,6,8\\ 
~&(1,2)&{\red 2},{\red 4}\\ 
~&(2,0)&0\\ \hline 
\end{tabular} 
\caption{Quantum numbers for the $\gamma-$unstable states of the  
harmonic oscillator potential. Every group of states with the same 
$N$ has the same energy. The first repetitions are marked in red.  
See appendix A and text for details on the sequence of repetitions.} 
\label{ta2} 
\end{table} 
\ec 
 With the content of Table I in mind, we list in Table II 
the quantum numbers of the first few states. 
States with the same $N$ have the same energy. We perform also in this case 
the analysis of the number of repetitions, that can be found in appendix A.  

It is customary to normalize the energies in such a way that the lowest state 
is at 0, and the first excited is at 1. This normalization is equivalent to 
set the overall energy scale and the relative energy scale respectively.
It turns out, obviously, that  
the energy of the second group of excited states in this energy scale 
is at 2.  
The ratio $R_{4/2}=E_{4^+}/E_{2^+}$, or better 
$(E_{4^+}-E_{0^+})/(E_{2^+}-E_{0^+})$, is the standard 
reference point for all the solutions of the  
collective hamiltonian and for the comparison with experimental data. 
Thus, for the harmonic oscillator, we have $E_{4^+}/E_{2^+}\=2$. 
The indexes are 
referred to the labeling of the states from the bottom to the top. 
This is unambiguous for the present case, but in the following we will
encounter many situations in which the degeneration typical of the harmonic  
oscillator will 
be removed and we will have two indexes, in order to distinguish between 
different bands (or families) of states and between different states 
within a given family. 
 
It is useful to introduce the reduced energy $\epsilon= E {2B_m\over \hbar^2}$ 
and reduced potential $v(\beta)= V(\beta) {2B_m\over \hbar^2}$ in eq.  
(\ref{betaeq}). 
In the case we are discussing $v(\beta)=k\beta^2$ with $k=CB/\hbar^2$ and  
hence 
\footnotetext[1]{Alternatively one can take the standard form 
$$ \Biggl\{ {\partial^2 \over 
 \partial \beta^2}+\epsilon -{(\tau+1)(\tau+2)\over \beta^2} - 
k\beta^2 \Biggr\} \chi(\beta) = 0 $$ 
with $\chi ( \beta ) = \beta^2 f(\beta )$. It is indeed always possible to 
write a linear second order differential equation in its canonical form.} 
\be 
\Biggl\{ -{1\over \beta^4}{\partial \over 
 \partial \beta} \beta^4{\partial \over 
 \partial \beta} -\epsilon +{(\tau+1)(\tau+2)\over \beta^2} + 
k\beta^2 \Biggr\} f(\beta)\= 0 
\ee 
with normalized solution 
\be 
f_n^\tau (\beta)=\Biggl[ {2n! \over \Gamma (n+\tau+5/2)}\Biggr]^{1/2}  
\beta^\tau L_n^{\tau+3/2}(\beta^2)e^{-\beta^2/2} 
\ee 
that contains associated Laguerre polynomials. 

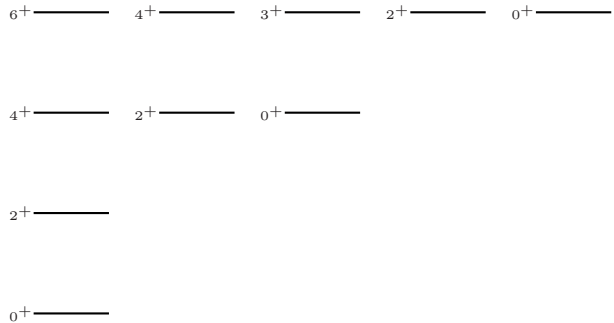
\begin{figure}[t] 
\bc 
\begin{picture}(280,160)(0,0) 
\psset{unit=.95pt} 
\psline{-}(20,20)(50,20)\rput(15,20){\tiny $0^+$}
\psline{-}(20,60)(50,60)\rput(15,60){\tiny $2^+$}
\psline{-}(20,100)(50,100)\rput(15,100){\tiny $4^+$}
\psline{-}(20,140)(50,140)\rput(15,140){\tiny $6^+$}
\psline{-}(70,100)(100,100)\rput(65,100){\tiny $2^+$}
\psline{-}(70,140)(100,140)\rput(65,140){\tiny $4^+$}
\psline{-}(120,100)(150,100)\rput(115,100){\tiny $0^+$}
\psline{-}(120,140)(150,140)\rput(115,140){\tiny $3^+$}
\psline{-}(170,140)(200,140)\rput(165,140){\tiny $2^+$}
\psline{-}(220,140)(250,140)\rput(215,140){\tiny $0^+$}
\end{picture} 
\caption{Spectrum of the harmonic oscillator potential.  
The energy scale ($\varepsilon'$) is chosen by fixing $\hbar \omega =1$.
See text and table II for explanations.} 
\label{ee1}   
\ec 
\end{figure}

We display in fig. \ref{ee1} the lowest energy states for the harmonic 
oscillator. 

\subsection{Wilets and Jean's solutions} 
After a very clear and concise introduction to the subject,  
Wilets and Jean \cite{Wilet} gave the solution in a couple of cases: 
the infinite square well and the displaced harmonic oscillator. 
 
\subsubsection{'Anharmonic oscillator'} 
Their aim was to discuss the addition of anharmonicities to the $\beta$  
potential and they took as a limiting case what they called, with a somewhat  
misleading designation, 'anharmonic oscillator'.  
In reality they were the first to treat the case  
of the infinite square well  in the form 
\be 
V(\beta)\= \left\{  
\begin{array}{ccc} 
 const &,& \beta < \beta_w \\ 
\infty &,& \beta > \beta_w 
 \end{array} \right.  
\label{square} 
\ee 
where $\beta_w$ takes a non-null positive real value.
They gave the eigenenergies in terms of zeros of the Bessel functions and 
they found that the ratio $R_{4/2}$ is 2.20, 
but they said (cited work): 
\begin{quotation}  
..., it represents the extreme case, not realizable in nature. 
\end{quotation} 
It seems thus that they underestimated the importance of this solution.  
In fact, only recently, Iachello \cite{Iac1} realized that this potential 
may furnish a good description for the shape phase transition between 
spherical and $\gamma-$unstable nuclei, although the real form of the  
potential at the critical point is, to the leading order, a quartic  
oscillator: the infinite square well approximates very well the   
behaviour of the $\beta^4$ potential, being very flat around the origin, and 
displays a qualitative agreement with the quickly rising asymptotic  
behaviour of the quartic potential at infinity. 
We will discuss later his solution, 
that provides not only the spectrum, but also eigenfunctions and transition  
rates. We will thus describe the complete solution for this potential as it is 
given in \cite{Iac1}, in the appropriate section.

\subsubsection{Displaced harmonic oscillator} 
\begin{figure}[!t] 
\epsfig{file=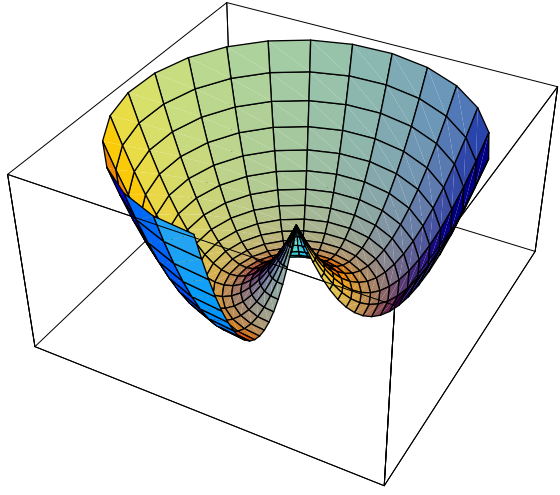,scale=0.7} 
\caption{Plot  $V(\beta,\gamma)$ of the displaced  
harmonic oscillator potential (\ref{vwj}) discussed by Wilets and Jean  
with $C=2$ and $\beta_0=0.6$ (these values are chosen merely for purpose of 
illustration) and $0<\gamma<3\pi/2$. } 
\label{anharmo} 
\end{figure} 
The same authors treated the modification of the harmonic potential, 
called displaced harmonic potential, whose expression is given by 
\be 
V(\beta)\= {1\over 2}C (\beta-\beta_0)^2. 
\label{vwj} 
\ee 
It was introduced to describe a situation where the minimum along the $\beta$ 
direction is located at some fixed value $\beta_0$ (see fig. \ref{anharmo}). 
With the substitutions: 
$$ 
x=\sqrt{B\omega\over \hbar \beta}, \quad \epsilon={E\over \hbar \omega},  
\quad \varphi (x)=x^2f(\beta) 
$$ 
one can recast equation (\ref{betaeq}) in the following form 
\be 
{1\over 2}\Biggl({\partial^2\over \partial x^2} -{(\tau+1)(\tau+2)\over x^2}-(x-x_0)^2 
\Biggr) \varphi(x)=\epsilon  \varphi(x) \,.
\ee 
Wilets and Jean considered the potential formed by the displaced harmonic  
oscillator plus the $1/x^2$ term and expanded it in Taylor series around  
the minimum $x'$ obtaining 
\be 
v(x)\equiv  {1\over 2}\Biggl( {(\tau+1)(\tau+2)\over x^2}+(x-x_0)^2 
\Biggr) \= v(x')+{\omega'^2\over 2}(x-x')^2+O\bigl((x-x')^3\bigr) \,.
\ee 
Neglecting terms of order three the effective potential is approximated
by a harmonic oscillator and therefore the spectrum may be written as 
\be 
\epsilon\=(n_\beta+1/2)\omega'+v(x') \,,
\ee 
with $n_\beta\=0,1,2,...$. The spectrum looks like the one in fig. \ref{ee1}
with some proper scaling and shift of the energies. The eigenfunctions read 
\be 
\varphi(x)\=h_{n_\beta}\bigl((x-x')\sqrt{\omega'} \bigr) 
e^{-(x-x')^2\omega/2} \,,
\ee 
where $h$ denotes Hermite polynomials. 
This is not an exact solution, but it is worth saying that it is a very good  
approximation for large values of $x_0$.

\subsection{Elliott-Evans-Park's solution for the Davidson potential.} 
Davidson \cite{Dav} introduced a potential of the form $a r^2+b/r^2$  
for the interaction between 
the constituents of a diatomic molecule. Elliott and collaborators 
\cite{Elli,Elli2} 
employed (though they were not historically the firsts, see section on the 
Warsaw solution!) a similar form in the context of quadrupole  
deformations and Rowe and Bahri \cite{Rowe1} gave a detailed algebraic  
discussion of this potential both for molecular and nuclear spectra (see 
section VI). 
\begin{figure}[!t] 
\epsfig{file=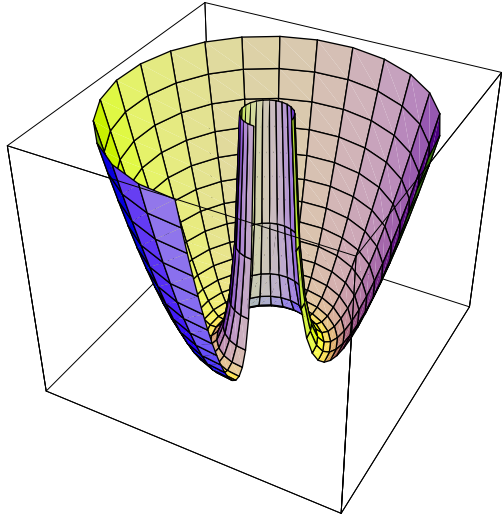,scale=0.7} 
\caption{Plot $V(\beta,\gamma)$ of the Davidson potential  
(\ref{vdav}) discussed analytically by Elliott {\it et al.} 
\cite{Elli,Elli2} and algebraically  
by Rowe and Bahri \cite{Rowe1} with all the constants taken equal to 1 and  
$0<\gamma<3\pi/2$. } 
\label{ellio} 
\end{figure} 
We will confine here to the direct analytic solution of the Bohr equation  
with the potential: 
\be 
V(\beta)\=A\Bigl( {\beta\over \beta_0}-{\beta_0\over\beta} \Bigr)^2  
\label{vdav} 
\ee 
that presents a minimum in $\beta_0$ and $A$ is related to its steepness. 
Then the differential equation in the 'radial' variable is  
\be
\Biggl\{-{\hbar^2\over 2B_m}\Bigl[ {1\over \beta^4}{\partial \over \partial  
\beta} \beta^4{\partial \over \partial \beta} -{\tau(\tau+3)\over\beta^2}  
\Bigr]+A\Bigl( {\beta\over \beta_0}-{\beta_0\over\beta} \Bigr)^2 \Biggl\}  
f(\beta) \= E f(\beta) \,.
\ee 
By regrouping the terms coming from the potential $V(\beta)$ 
it is possible to rewrite the eigenvalue equation as the equation for the 
harmonic oscillator as follows,
\be
\Biggl\{-{\hbar^2\over 2B_m}\Bigl[ {1\over \beta^4}{\partial \over \partial  
\beta} \beta^4{\partial \over \partial \beta} -{p(p+3)\over\beta^2}  
\Bigr]+A{\beta^2\over \beta_0^2}-2A \Biggl\}  
f(\beta)\= E f(\beta) \,,
\ee 
where $p(p+3)\= \tau(\tau+3)+2AB_m\beta_0^2/\hbar^2$. 
The solution for the five dimensional oscillator is then  known to be given 
in terms of associated Laguerre polynomials 
\be 
f(\beta)\=\sqrt{2(n!)\over\Gamma(n+p+5/2)} {\beta^p \over b^{-p-5/2} } 
L_n^{p+3/2}(\beta^2/b^2) e^{-{\beta^2\over(2b^2)}} 
\ee 
where we have set $b=(\hbar^2\beta_0^2/2AB_m)^{1/4}$. 
The expression of the spectrum is 
\be 
E\= \hbar \omega (2n+p+5/2) -2A\,, 
\ee 
where the non-integer quantum number $p$ is introduced for convenience.
Here $n$ should be interpreted as $n_\beta$ and $\hbar\omega=b^{-2}$. By
expressing $p$ as a function of $\tau$ and by introducing a parameter 
$G$ we can express the spectrum as (see \cite{Rowe1}):
\be
E\=\hbar \omega (2n+1+\sqrt{(\tau+3/2)^2+G}) -2A \,,
\ee
where we can forget the energy shift and we can set $\hbar\omega$ to 1.
The spectrum exhibits a number of interesting features: for $G=0$ we obtain
again the harmonic oscillator spectrum with the degeneracy of the 
$(n=0, \tau=2)$ state with the  $(n=1, \tau=0)$ state and so on.  
For $G\rightarrow\infty$ the ground state band
tends to obey the rule $\tau(\tau+3)$ typical of the Wilets-Jean model. 
The intermediate situation may be used to describe $\gamma-$unstable 
situations where the $0^+$ of the second band typically lies at higher 
energy with respect to the $2^+$ of the ground state band.

\begin{figure}[t] 
\bc 
\begin{picture}(280,320)(0,-10) 
\psset{unit=.95pt} 
\rput(15,0){\small $n=0$} \rput(30,-10){\small $G=0$}
\psline{-}(5,10)(25,10)\rput(15,13){\tiny $\tau=0$}
\psline{-}(5,40)(25,40)\rput(15,43){\tiny $\tau=1$}
\psline{-}(5,70)(25,70)\rput(15,73){\tiny $\tau=2$}
\psline{-}(5,100)(25,100)\rput(15,103){\tiny $\tau=3$}
\rput(45,60){\small $n=1$}
\psline{-}(35,70)(55,70)\rput(45,73){\tiny $\tau=0$}

\rput(85,0){\small $n=0$}\rput(100,-10){\small $G=20$}
\psline{-}(75,10)(95,10)\rput(85,13){\tiny $\tau=0$}
\psline{-}(75,40)(95,40)\rput(85,43){\tiny $\tau=1$}
\psline{-}(75,80.99)(95,80.99)\rput(85,84){\tiny $\tau=2$}
\psline{-}(75,130.1)(95,130.1)\rput(85,133.1){\tiny $\tau=3$}
\rput(115,147){\small $n=1$}
\psline{-}(105,157.6)(125,157.6)\rput(115,160.6){\tiny $\tau=0$}

\rput(155,0){\small $n=0$}\rput(170,-10){\small $G=100$}
\psline{-}(145,10)(165,10)\rput(155,13){\tiny $\tau=0$}
\psline{-}(145,40)(165,40)\rput(155,43){\tiny $\tau=1$}
\psline{-}(145,83.96)(165,83.96)\rput(155,87){\tiny $\tau=2$}
\psline{-}(145,140.78)(165,140.78)\rput(155,143.78){\tiny $\tau=3$}
\rput(185,306){\small $n=1$}
\psline{-}(175,316)(195,316)\rput(185,319){\tiny $\tau=0$} 

\rput(215,0){\small $n=0$}\rput(230,-10){\small $G\rightarrow\infty$}
\psline{-}(205,10)(225,10)\rput(215,13){\tiny $\tau=0$}
\psline{-}(205,40)(225,40)\rput(215,43){\tiny $\tau=1$}
\psline{-}(205,84.88)(225,84.88)\rput(215,87.88){\tiny $\tau=2$}
\psline{-}(205,144.55)(225,144.55)\rput(215,147.55){\tiny $\tau=3$}
\rput(245,306){\small $n=1$}
\psline{-}(235,320)(255,320)\psline{->}(245,320)(245,330) 

\end{picture} 
\caption{Spectrum of the Davidson potential. The energy scale is chosen 
by fixing $\hbar \omega =1$. Notice that this figure is inspired to the 
corresponding fig. 1 in \cite{Rowe1}, although we have kept the energy 
normalization consistent for different values of the parameter $G$ and we 
have added the rightmost case to illustrate the SO(6) limit.} 
\label{ee3}   
\ec 
\end{figure}
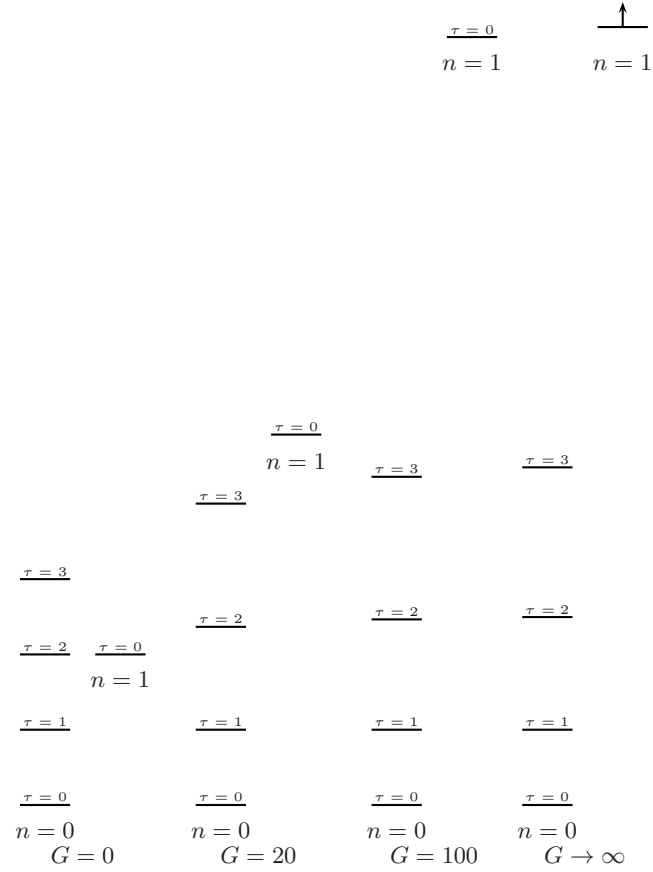  
These features are illustrated in fig. \ref{ee3}, where the spectrum of the
Davidson potential is shown for a few values of the parameter G.

\subsection{Iachello's infinite square well solution or E(5)} 
\begin{figure}[t] 
\bc 
\begin{picture}(280,230)(0,0) 
\psset{unit=.95pt} 
\psline{-}(10,10)(30,10)\rput(5,10){\tiny $0^+$}
\psline{-}(10,40)(30,40)\rput(5,40){\tiny $2^+$}
\psline{-}(10,76)(30,76)\rput(5,76){\tiny $4^+$}
\psline{-}(45,76)(65,76)\rput(40,76){\tiny $2^+$}
\psline{-}(10,117.7)(30,117.7)\rput(5,117.7){\tiny $6^+$}
\psline{-}(45,117.7)(65,117.7)\rput(40,117.7){\tiny $4^+$}
\psline{-}(80,117.7)(100,117.7)\rput(75,117.7){\tiny $3^+$}
\psline{-}(115,117.7)(135,117.7)\rput(110,117.7){\tiny $0^+$}

\psline{->}(20,40)(20,10)\rput(12,25){\tiny 100}
\psline{->}(20,76)(20,40)\rput(12,58){\tiny 168}
\psline{->}(20,117.7)(20,76)\rput(12,95){\tiny 221}
\psline{->}(55,76)(22,40)\rput(46,58){\tiny 168}
\psline{->}(55,117.7)(22,76)\rput(36,103){\tiny 105}
\psline{->}(55,117.7)(55,76)\rput(48,92){\tiny 116}
\psline{->}(90,117.7)(57,76)\rput(70,101){\tiny 157}
\psline{->}(125,117.7)(59,76)\rput(110,101){\tiny 221}
\psline{->}(160,100.9)(35,40)\rput(95,65){\tiny 86}
\psline{->}(160,154)(160,100.9)\rput(155,127){\tiny 75}
\psline{->}(160,213.4)(160,154)\rput(154,184){\tiny 165}
\psline{->}(195,213.4)(162,154)\rput(175,190){\tiny 165}

\psline{-}(150,100.9)(170,100.9)\rput(145,100.9){\tiny $0^+$}
\psline{-}(150,154)(170,154)\rput(145,154){\tiny $2^+$}
\psline{-}(150,213.4)(170,213.4)\rput(145,213.4){\tiny $4^+$}
\psline{-}(185,213.4)(205,213.4)\rput(180,213.4){\tiny $2^+$}

\psline{-}(220,237.4)(240,237.4)\rput(215,237.4){\tiny $0^+$}

\rput(20,0){\small $\xi=1$}
\rput(160,90){\small $\xi=2$}
\rput(230,227){\small $\xi=3$}
\psline[linestyle=dotted]{-}(20,130)(20,180)
\psline[linestyle=dotted]{-}(160,220)(160,240)
\end{picture} 
\caption{Spectrum of the five-dimensional infinite square well. Energies
are scaled shifting the ground state to zero and fixing the lowest 
excited $2^+$ state to lie at $1$. Transition rates are also 
reported fixing $B(E2,2_{1,1}\rightarrow 0_{1,0})=100$. Adapted from \cite{Iac1}} 
\label{ee2}   
\ec 
\end{figure}
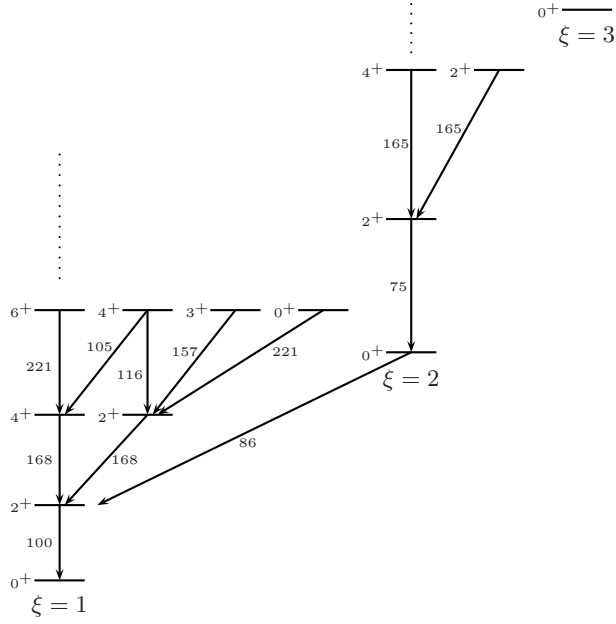

As previously remarked, Iachello \cite{Iac1} brought renewed attention on the  
square well solution, proposing the square well as a  
convenient substitute for the description of  the critical point in the 
U(5)-SO(6) shape phase transition between the vibrator and the  
$\gamma-$unstable rotor. 
\begin{figure}[!t] 
\epsfig{file=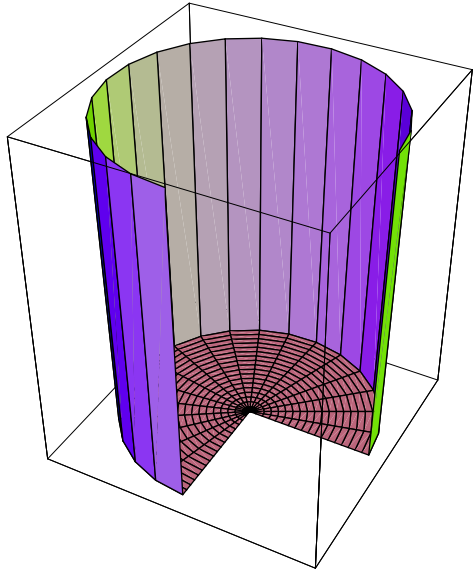,scale=0.7} 
\caption{Plot $V(\beta,\gamma)$ of the infinite well potential  
(\ref{square}) discussed analytically by Wilets and Jean and  
and by Iachello with $0<\gamma<3\pi/2$. } 
\label{bucainf} 
\end{figure} 
With the potential (\ref{square}) and setting $\varphi(\beta)\=\beta^{3/2}  
f(\beta)$, $z=\beta k$,  
$k=\sqrt{\varepsilon}$ and $\nu\= \tau+3/2$, the $\beta-$part of the  
Bohr equation may be written as a Bessel equation whose solutions 
(regular in the origin) are Bessel $J$ function 
\be 
\varphi''+{\varphi'\over z} +\Bigl[1-{(\tau+3/2)^2\over z^2}\Bigr] 
\varphi \= 0  
\label{iacheq}
\ee 
with Dirichlet boundary condition ($\varphi(\beta_w)\=0$). This fact  
determines the spectrum as a function of the $\xi$th zero, $x_{\xi,\tau}$, 
of the $J_{\tau+3/2}(z)$ function, namely:
\be 
\varepsilon_{\xi,\tau} \= \Biggl( {x_{\xi,\tau}\over \beta_w}\Biggr)^2 \,.
\ee 
We display in fig. (\ref{ee2}) the lowest part of the spectrum.
The wave functions are  
\be 
f_{\xi,\tau}(\beta)\= c_{\xi,\tau}\beta^{-3/2}J_{\tau+3/2} 
\bigl({x_{\xi,\tau} \beta \over \beta_w} \bigr) \,.
\ee 
The normalization constant may be found analytically from the  
normalization condition $\int d\beta \beta^4 f^2(\beta)\=1$. The 
result, containing an hypergeometric function, reads 
\be 
{1\over c_{\xi,\tau}^2}\= {k_{\xi,\tau}^{3+2\tau}\beta_w^2 \over 2^{4+2\tau}} 
{ _1F_2 \bigl( 2+\tau; 7/2+\tau, 4+2\tau; -k^2 \bigr) \over  
\Gamma(5/2+\tau)\Gamma(7/2+\tau)  } 
\label{mianorma} 
\ee 
and, although rather complicated, may furnish an alternative to direct 
numerical computation. 
 
This extraordinary simple, but nonetheless very successful, solution has been 
labeled E(5), as the euclidian group in the five dimensional  
space of the quadrupole variables. This group label may be interpreted in a
very straightforward way by noticing that the $\beta-$part eq. 
(\ref{betaeq0}), and hence eq. (\ref{iacheq}), may be written as
\be
[\pi^2+u(\beta)-\varepsilon]f(\beta)=0\,
\label{conjham}
\ee 
where $\hat \pi^\mu$ are the conjugate momenta with respect to the 
five quadrupole deformation variables, $\hat \alpha_\mu$.
The hamiltonian (\ref{conjham}) is invariant with respect to rotations in 
the five dimensional Hilbert space associated with the variables defined
above, and in the special case of the square well either $u(\beta)=0$ or 
$u(\beta)=\infty$, thus the hamiltonian in the relevant region is written 
as a function of $\pi$ only (plus a constant). Therefore the hamiltonian is
also invariant with respect to translations in the five dimensional space.
The present solution is in summary invariant with respect to the 
transformations induced by the group E(5) that is the semidirect sum of 
the five dimensional translation and rotation groups.

\subsection{Caprio's finite square well solution} 
\begin{figure}[!t] 
\epsfig{file=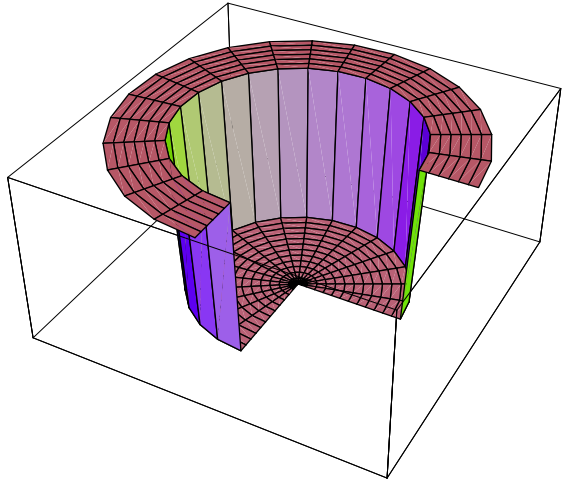,scale=0.7} 
\caption{Plot $V(\beta,\gamma)$ of the finite well potential  
(discussed  by Caprio \cite{Cap}), with $0<\gamma<3\pi/2$. } 
\label{bucafin} 
\end{figure} 
The previous analytical solution is a successful approximation of the physics 
of an entire class of the so-called $\gamma-$unstable nuclei, based on the  
infinite well potential. Caprio \cite{Cap} has investigated whether a  
more realistic behaviour  
of the potential well, taken as a finite square well, alters this view.  
The finite square well potential is depicted in fig. \ref{bucafin} 
and has the following expression 
\be 
V(\beta)\= \left\{  
\begin{array}{ccc} 
 V_0 &,& \beta \le \beta_w \\ 
 0 &,& \beta > \beta_w 
 \end{array} \right.  
\label{fsquare} 
\ee 
where $\beta_w$ is the position of the wall or step.
Eigensolutions may be found: inside the well they are again Bessel functions,  
while outside 
the well, taking into account the correct asymptotic behaviour at infinity,  
the solution may  
be written in terms of modified spherical Bessel functions. The complete  
solution is 
\be 
f_{\xi,\tau}(\beta) \= \left\{  
\begin{array}{ccc} 
 A_{\xi,\tau} \beta^{-1} j_{\tau+1}\big[(\epsilon_{\xi,\tau}-v_0)^{1/2}\beta  
\big] &,&  
\beta \le \beta_w \\ 
 B_{\xi,\tau} \beta^{-1} k_{\tau+1}\big[(-\epsilon_{\xi,\tau})^{1/2}\beta  
\big] &,&  
\beta > \beta_w 
 \end{array} \right.  
\label{finsquare} 
\ee 
where $v_0=(2B_m/\hbar^2)V_0$ is the reduced potential and  
$\epsilon_{\xi,\tau}$ are the  
reduced eigenvalues, that are obtained by requiring the continuity of the  
wave function  
and of its first 
derivative at the position of the step $\beta_w$. One can define a  
dimensionless energy variable 
\be 
\eta(\epsilon) \equiv \Bigl(1-{\epsilon\over v_0}\Bigr)^{1/2} 
\ee 
and the parameter  
\be 
x_0\equiv \sqrt{-v_0} \beta_w 
\ee 
and substitute into the matching condition to obtain a transcendental equation 
 that must be solved numerically to determine the eigenvalues. 
\begin{figure}[!t] 
\epsfig{file=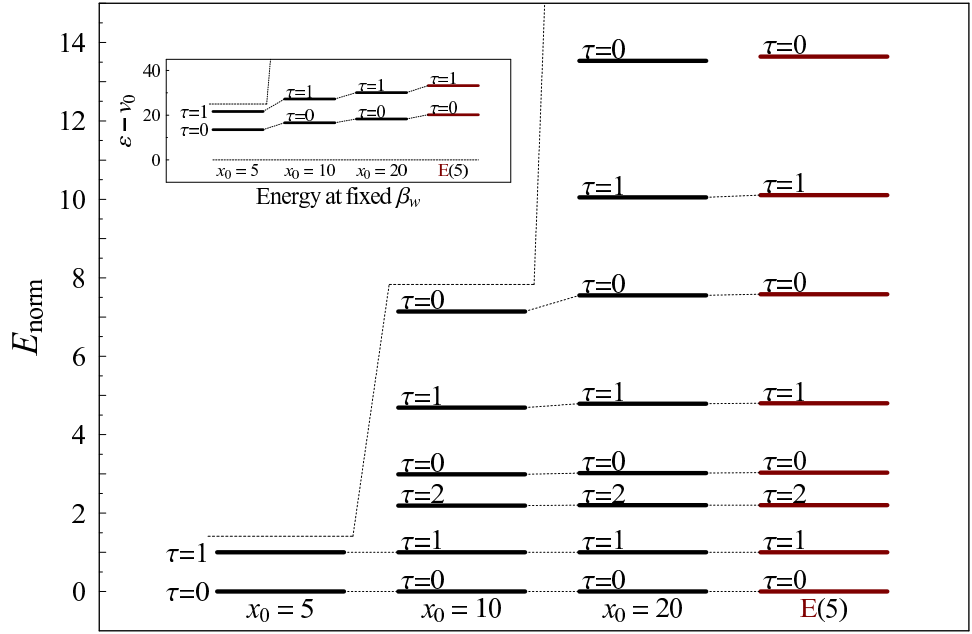,scale=0.7} 
\caption{Evolution of excitation energies as a function of $x_0$ normalized
to the first excited state (absolute energies in the inset). The threshold 
between bound and unbound states is indicated with a dashed line. 
Adapted from \cite{Cap}, courtesy of the author.}
\label{mar_evo}
\end{figure} 
Unlike the previous case the spectrum, displayed in fig. \ref{mar_evo} 
has a finite number of discrete eigenvalues,  
but only those that are lying close to the threshold are appreciably modified  
with respect to the infinite well solution. 
For the most part the spectrum does not differ substantially from the E(5)  
case and the same is true for electromagnetic transition rates.

\subsection{Solutions for the Coulomb-like and Kratzer-like potentials} 
\label{cke5}
\begin{figure}[!t] 
\epsfig{file=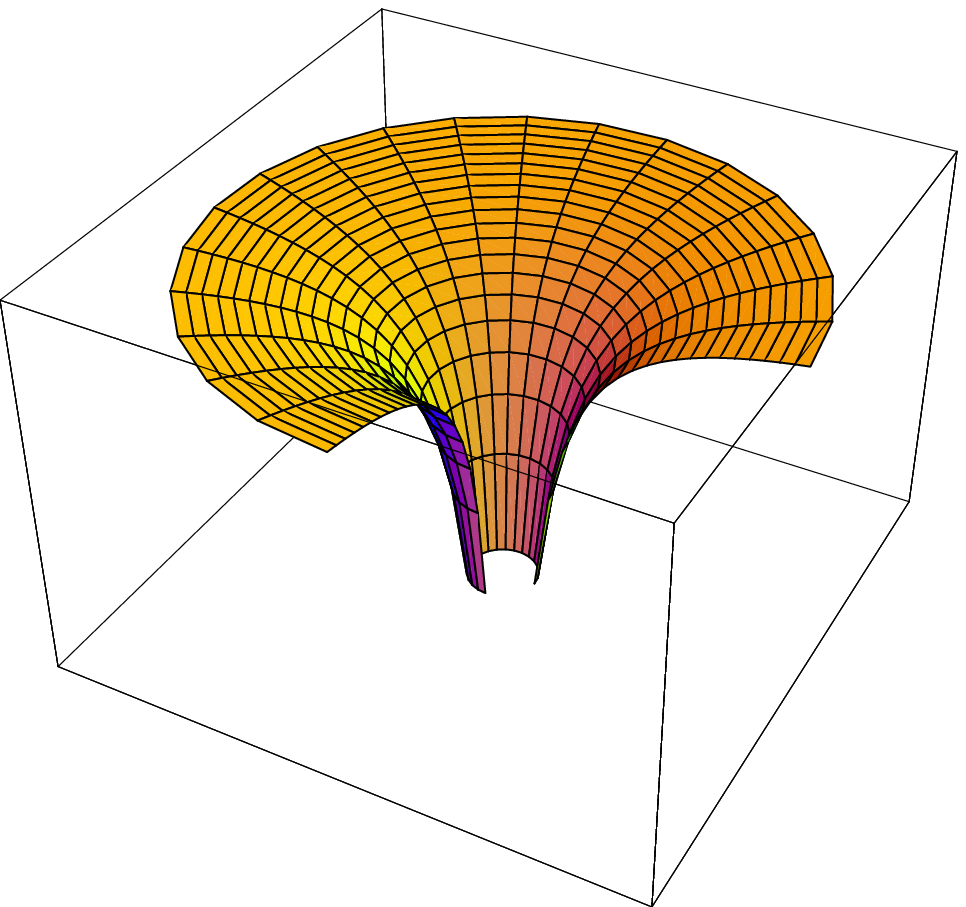,scale=0.7} a) 
\epsfig{file=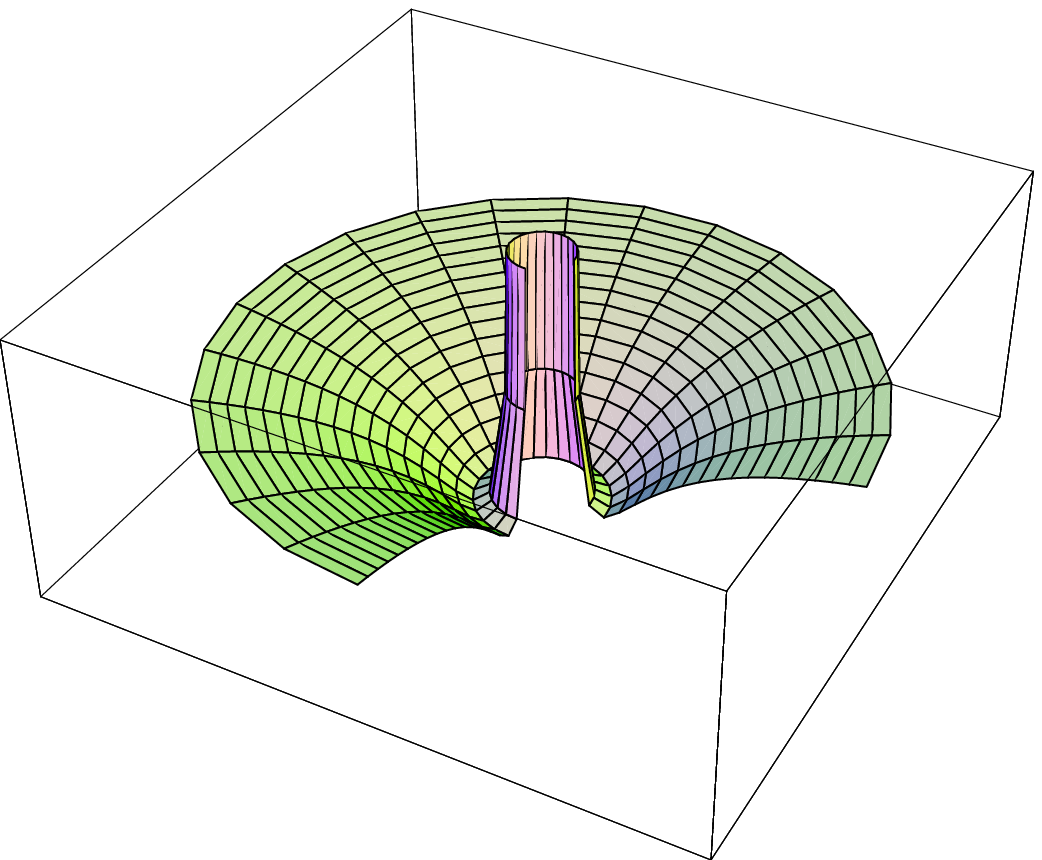,scale=0.7}b) 
\caption{Plot $V(\beta,\gamma)$ of the Coulomb-like (a) and Kratzer-like (b)  
potentials (\cite{FV1}) discussed by Fortunato and Vitturi,  
with $0<\gamma<3\pi/2$. } 
\label{forfigs} 
\end{figure} 
The analytic solution of the problem with a Coulomb-like potential (fig.  
\ref{forfigs}, a)  
and with a Kratzer-like potential (fig. \ref{forfigs}, b) has been discussed  
in ref. \cite{FV1}. 
For these two potentials it is possible to identify the Bohr's equation with  
the Whittaker's  
equation, whose solution is known in terms of Whittaker's functions. 
Namely, after having introduced reduced energies and potential as in the  
harmonic oscillator case, we can rewrite eq. (\ref{betaeq}) 
in its standard form  
with the transformation $\chi (\beta) = \beta^2 f(\beta)$. The canonical  
form for the Bohr equation is thus 
\be 
\chi''(\beta) +\Bigl\{ \epsilon-u(\beta)-{(\tau+3/2)^2\over \beta^2}+ 
{1\over 4\beta^2} \Bigl\} \chi(\beta) \=0 
\label{stdform} 
\ee 
and will be used in the following to derive analytic solutions in the  
two cases cited above. 

The Coulomb-like potential reads: 
\be 
u_C(\beta)= -{A\over \beta} \,,
\ee 
with $A>0$. With the substitutions 
$\varepsilon =-\epsilon$,  $x=2 \sqrt{\varepsilon}\beta$,  
$k=A/(2\sqrt{\varepsilon})$ and  
$\mu=\tau+3/2$, equation (\ref{stdform}) 
 takes the Whittaker's standard form \cite{Bat}: 
\be 
\chi''(x) +\Bigl\{-{1\over 4}+{k\over x}+{(1/4-\mu^2)\over x^2} 
\Bigl\} \chi(x) \=0 
\ee 
and its regular solution for negative energies 
is (as in \cite{Bat}) expressed in terms of the Whittaker's function 
$M_{k,\mu}(x)$ : 
\be 
 \chi_{k,\mu}(x)\= {\cal N}_{\tau,\xi} 
x^{(2\mu+1)\over 2} e^{-x/2}  
~_1F_1 \bigl(\mu+1/2-k,2\mu+1; x\bigr) \,.
\ee 
When the first parameter $\mu+1/2-k = \tau+2-A/(2\sqrt{\varepsilon})$ is  
a negative integer, that we call $-\xi$, the Whittaker's function reduces to 
an associated Laguerre polynomial and the wave function may be written as 
\be 
\chi(x)_{\tau,\xi}={\cal N}_{\tau,\xi}~ x^{2\mu+1\over 2} 
e^{-x/2} {\xi!\over (2\mu+1)_\xi}  
L_\xi^{(2\mu)}(x) 
\ee 
where the denominator is a Pochhammer symbol. 
The condition given above fixes analytically the spectrum  
\be 
\varepsilon_{\tau,\xi}={A^2/4\over (\tau+\xi+2)^2}. 
\ee 
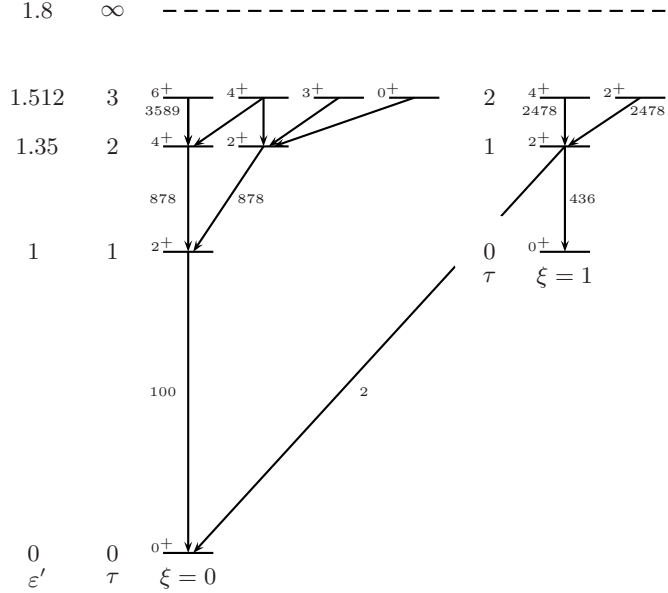
\begin{figure}[t] 
\bc 
\begin{picture}(280,260)(0,0) 
\psset{unit=.95pt} 
\rput(70,0){\small $\xi=0$} 
\rput(40,0){\small $\tau$} 
\rput(40,10){\small $0$} 
\rput(40,130){\small $1$} 
\rput(40,172){\small $2$} 
\rput(40,191.4){\small $3$} 
\rput(40,226){\small $\infty$} 
\rput(10,0){\flushleft\small $\varepsilon'$} 
\rput(10,10){\flushleft\small 0 } 
\rput(10,130){\flushleft\small 1 } 
\rput(10,172){\flushleft\small 1.35} 
\rput(10,191.4){\flushleft\small 1.512} 
\rput(10,226){\flushleft\small  1.8} 
\rput(220,120){\small $\xi=1$} 
\psline{-}(60,10)(80,10) \rput(60,13){\tiny $0^+$} 
\psline{-}(60,130)(80,130) \rput(60,133){\tiny $2^+$} 
\psline{-}(60,172)(80,172) \rput(60,175){\tiny $4^+$} 
\psline{-}(60,191.4)(80,191.4) \rput(60,194.4){\tiny $6^+$} 
\psline{-}(90,172)(110,172) \rput(90,175){\tiny $2^+$} 
\psline{-}(90,191.4)(110,191.4) \rput(90,194.4){\tiny $4^+$} 
\psline{-}(120,191.4)(140,191.4) \rput(120,194.4){\tiny $3^+$} 
\psline{-}(150,191.4)(170,191.4) \rput(150,194.4){\tiny $0^+$} 
\psline{-}(210,130)(230,130) \rput(210,133){\tiny $0^+$} 
\psline{-}(210,172)(230,172)\rput(210,175){\tiny $2^+$} 
\psline{-}(210,191.4)(230,191.4)\rput(210,194.4){\tiny $4^+$} 
\psline{-}(240,191.4)(260,191.4)\rput(240,194.4){\tiny $2^+$} 
\psline{->}(70,130)(70,10) 
\psline{->}(70,191.4)(70,172) 
\psline{->}(70,172)(70,130) 
\psline{->}(100,172)(72,130) 
\psline{->}(100,191.4)(72,172) 
\psline{->}(130,191.4)(102,172) 
\psline{->}(160,191.4)(104,172) 
\psline{->}(100,191.4)(100,172) 
\psline{->}(70,191.4)(70,172) 
\psline{->}(220,172)(220,130) 
\psline{->}(220,191.4)(220,172) 
\psline{->}(250,191.4)(221,172) 
\psline{->}(220,172)(72,10) 
\rput(60,74){\tiny 100} 
\rput(60,151){\tiny 878} 
\rput(60,186){\tiny 3589} 
\rput(95,151){\tiny 878} 
\rput(227,151){\tiny 436} 
\rput(210,186){\tiny 2478} 
\rput(253,186){\tiny 2478} 
\rput(140,74){\tiny 2} 
\pscircle*[linecolor=white](190,130){15} 
\rput(190,120){\small $\tau$} 
\rput(190,130){\small $0$} 
\rput(190,172){\small $1$} 
\rput(190,191.4){\small $2$} 
\psset{linestyle=dashed} 
\psline{-}(60,226)(260,226) 
\end{picture} 
\caption{Spectrum of the Coulomb-like potential.  
The energy scale ($\varepsilon'$) is chosen by fixing the  
 energy of the first two states respectively to 0 
and 1. The transition rate for $2_{0,1}^+ \rightarrow 
0_{0,0}^+$ has been fixed to 100. 
Some selected quadrupole transitions are shown in the figure for simplicity. } 
\label{Hyd}   
\ec 
\end{figure}  
A portion of this spectrum is shown in fig. \ref{Hyd}.  
Its most distinctive features are the position of the $(4^+,2^+)$  
doublet at an energy of $1.35$ and the presence of  
a threshold at $1.8$ that corresponds to an infinite quantum number. 
Note also that each group of $(\xi,\tau)$ states is degenerate with any 
$(\xi+k,\tau-k)$ group of states with $k=1,...,\tau$. 
 
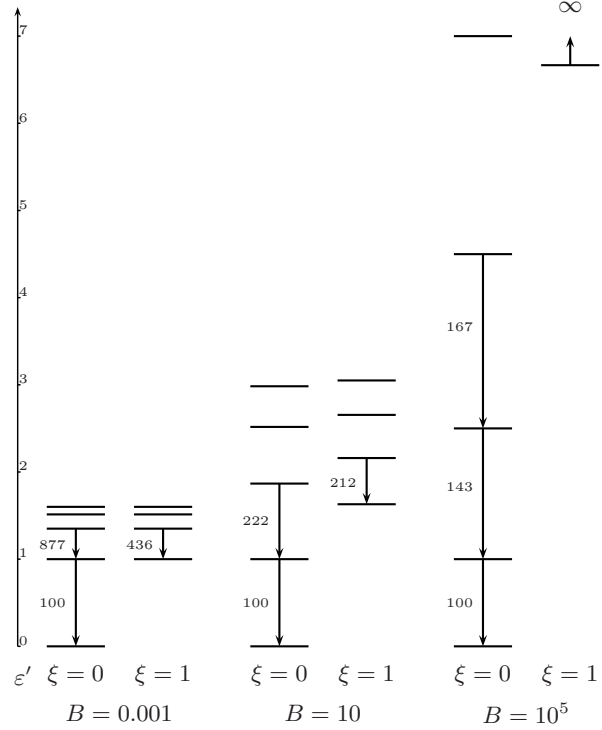
\begin{figure}[t] 
\bc 
\begin{picture}(220,270)(0,0) 
\psset{unit=1.1pt} 
\rput(2,10){\small $\varepsilon'$} 
\psline[linewidth=0.5]{->}(0,20)(0,240) 
\psline[linewidth=0.5]{-}(0,20)(1,20) 
\psline[linewidth=0.5]{-}(0,50)(1,50) 
\psline[linewidth=0.5]{-}(0,80)(1,80) 
\psline[linewidth=0.5]{-}(0,110)(1,110) 
\psline[linewidth=0.5]{-}(0,140)(1,140) 
\psline[linewidth=0.5]{-}(0,170)(1,170) 
\psline[linewidth=0.5]{-}(0,200)(1,200) 
\psline[linewidth=0.5]{-}(0,230)(1,230) 
\rput(2,22){\tiny 0} 
\rput(2,52){\tiny 1} 
\rput(2,82){\tiny 2} 
\rput(2,112){\tiny 3} 
\rput(2,142){\tiny 4} 
\rput(2,172){\tiny 5} 
\rput(2,202){\tiny 6} 
\rput(2,232){\tiny 7} 
\rput(20,10){\small $\xi=0$} 
\psline{-}(10,20)(30,20)  
\psline{-}(10,50)(30,50)   \psline{->}(20,50)(20,20) 
\psline{-}(10,60.5)(30,60.5) \psline{->}(20,60.5)(20,50) 
\psline{-}(10,65.4)(30,65.4) 
\psline{-}(10,68)(30,68) 
\rput(50,10){\small $\xi=1$} 
\psline{-}(40,50)(60,50) 
\psline{-}(40,60.5)(60,60.5) 
\psline{-}(40,65.4)(60,65.4)  \psline{->}(50,60.5)(50,50) 
\psline{-}(40,68)(60,68) 
\rput(12,35){\tiny 100} 
\rput(12,55){\tiny 877} 
\rput(42,55){\tiny 436} 
\rput(90,10){\small $\xi=0$} 
\psline{-}(80,20)(100,20) 
\psline{-}(80,50)(100,50)     \psline{->}(90,50)(90,20) 
\psline{-}(80,76.01)(100,76.01) \psline{->}(90,76.01)(90,50) 
\psline{-}(80,95.51)(100,95.51) 
\psline{-}(80,109.5)(100,109.5) 
\rput(120,10){\small $\xi=1$} 
\psline{-}(110,68.9)(130,68.9) 
\psline{-}(110,84.8)(130,84.8)   \psline{->}(120,84.8)(120,68.9) 
\psline{-}(110,99.65)(130,99.65) 
\psline{-}(110,111.5)(130,111.5) 
\rput(82,35){\tiny 100} 
\rput(82,63){\tiny 222} 
\rput(112,76){\tiny 212} 
\rput(160,10){\small $\xi=0$} 
\psline{-}(150,20)(170,20)    \psline{->}(160,50)(160,20) 
\psline{-}(150,50)(170,50)   \psline{->}(160,95)(160,50) 
\psline{-}(150,95)(170,95)     \psline{->}(160,154.97)(160,95) 
\psline{-}(150,154.97)(170,154.97) 
\psline{-}(150,230)(170,230) 
\rput(190,10){\small $\xi=1$} 
\psline{-}(180,220)(200,220) 
\psline{->}(190,220)(190,230) 
\rput(190,240){\small $\infty$} 
\rput(152,35){\tiny 100} 
\rput(152,75){\tiny 143} 
\rput(152,130){\tiny 167} 
\rput(35,-3){\small $B=0.001$} 
\rput(105,-3){\small $B=10$} 
\rput(175,-3){\small $B=10^5$} 
\end{picture} 
\caption{Evolution of spectra with fixed $A=20$ and increasing  
$B$ for the two limiting situation ($B=0$, Coulomb-like and $B\rightarrow  
\infty$, O(6) limit). The first two bands ($\xi=0,1$) are displayed with  
their lowest states ($\tau=0,1,2,..$). The various substates are not  
displayed for sake of simplicity. Some transition rates are indicated.} 
\label{evoB}   
\ec 
\end{figure}

The Kratzer-like potential may be thought as a modification of the  
former that may be shaped, adjusting the parameters, in such a way to  
display a pocket at some fixed point. It has the  
following expression 
\be 
u_K(\beta)= -{A\over \beta}+{B\over \beta^2} = -2{\cal D} \Biggl( {\beta_0  
\over \beta} -{1\over 2}{\beta_0^2 \over \beta^2}  \Biggr) \,,
\ee 
where we have shown two possible ways to parameterize this potential. 
The first  
is the easiest to use as it will be clear from the formula for the spectrum,  
while the second is related to the geometrical shape of the potential:  
${\cal D}$ is the depth of pocket and $\beta_0$ is the position of the minimum. 
These two sets of parameters are connected by simple relations that may be  
deduced from the equation above. In the following we shall make use of the  
former notation. 
Equation (\ref{stdform}) with the Kratzer potential and the substitutions  
$\varepsilon =-\epsilon$,  $x=2 \sqrt{\varepsilon}\beta$,  
$k=A/(2\sqrt{\varepsilon})$ and  
$\mu^2=(\tau+3/2)^2+B$, takes again the Whittaker's standard form. 
The solutions are the same as above with the new  
substitutions and the same arguments apply for the properties of convergence. 
Now $\mu+1/2-k = \sqrt{(\tau+3/2)^2+B}+1/2- 
A/(2\sqrt{\varepsilon})$ must be a negative integer, $-\xi$.  
The spectrum is in this case:  
\be 
\varepsilon_{\tau,\xi}  
\={A^2/4 \over (\sqrt{(\tau+3/2)^2+B}+1/2+\xi)^2} 
\label{spectrum} 
\ee 
with $\xi=0,1,2,..$ ~to label different families. 
In reference \cite{FV1} the evolution of the spectra with $\beta_0$ 
and ${\cal D}$ is studied in detail in order to establish a connection  
between shape of the potential and spectrum. Note, however, that if we 
use the two parameters $A$ and $B$, we can realize that $A$ does not play 
any role in determining the scaled eigenvalues and therefore the 
various spectra depend only on $B$, ranging from the Coulomb-like limit  
($B=0$) to the O(6) limit ($B\rightarrow \infty$). 
This is illustrated in fig. (\ref{evoB}). 
The threshold varies with $B$ from $1.8$ to infinity and the position of the  
$4^+$ varies from $1.35$ to $10/3$. This makes the Kratzer-like solution  
very flexible.
The parameter $B$ not only fixes the position 
of the $4_0^+$, but also all the other excited states including all the 
$\beta-$bands.

\subsection{L\'evai and Arias' sextic potential solution} 
\begin{figure}[!t] 
\epsfig{file=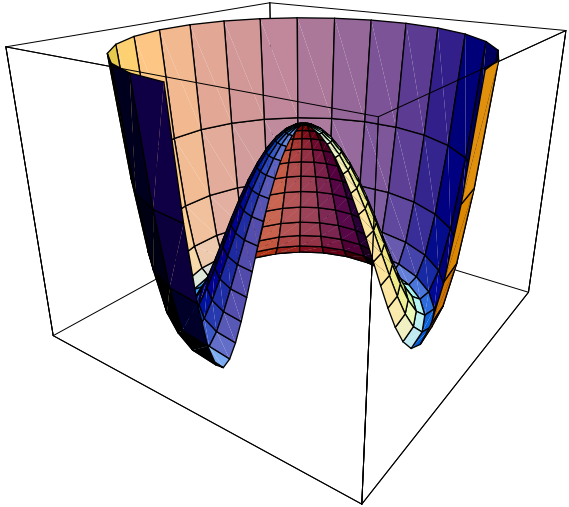,scale=0.7} 
\caption{Plot $V(\beta,\gamma)$ with $u^+(\beta)$ of the sextic potential 
discussed by L\'evai and Arias \cite{Pepe}
with $0<\gamma<3\pi/2$ for $\tau=0$, $M=1$ and a set of parameters 
chosen for the purpose of illustration. } 
\label{sexpo} 
\end{figure}
J.M.Arias and G.L\'evai \cite{Pepe} proposed the sextic oscillator  
as an example 
of quasi-exactly solvable potential for the Bohr hamiltonian. Only a  
certain number of eigenvalues may be obtained in closed form for this  
class of potentials. In the present case this is possible for the lowest 
few values of the principal quantum number. The sextic oscillator has the 
following expression 
\be 
u^\pi(\beta)\= (b^2-4ac^\pi)\beta^2+2ab\beta^4+a^2\beta^6+u_0^\pi \,,
\ee 
where the parameters $a,b$ are used to determine, with an ample choice, 
the shape of the potential. The index $\pi\=\pm$ is used to distinguish 
between potentials for even and odd values of $\tau$. The constants $c^\pi$ 
are obtained setting $\tau=2s-5/2$ (that must be a non-negative integer)  
and requiring that  
\be 
s+M+{1\over 2}\equiv c =const. 
\ee 
being $M$ a non-negative integer. For each $M$ it is possible 
to get analytically $M+1$ of its solutions.  
The last condition is needed in order to keep constant the quadratic  
term of the potential, otherwise the shape of the potential will change  
with $s$. This implies also that when $M$ is increased by one unity,  
the value of $\tau$ must correspondingly decrease of two units.  
The constants $u_0^\pi$ may be used to shift the minima of the two potentials 
at the same energy. There are various considerations that can be made here,  
for which the reader is referred  to the cited work. For our purpose it will 
suffice to summarize the solutions in table \ref{tapepe}, where the  
formulae for the lowest eigenvalues are given. 
\bc 
\begin{table}[!t] 
\begin{tabular}{cccc} 
~~~$M$~~~&~~~$\tau$~~~&~~~$n$~~~&~~~$E_{n+1,\tau}$ \\ \hline 
1&0&$\left. \begin{array}{c} 0 \\1 \end{array} \right.$& 
$\left. \begin{array}{c} E_{1,0}=7b-2\sqrt{b^2+10a}+u_0^+ \\ 
 E_{2,0}=7b+2\sqrt{b^2+10a}+u_0^+ \end{array} \right. $ \\\hline 
0&2&0&$E_{1,2}=9b+u_0^+$\\\hline 
1&1&$\left. \begin{array}{c} 0 \\1 \end{array} \right.$& 
$\left. \begin{array}{c} E_{1,1}=9b-2\sqrt{b^2+14a}+u_0^- \\ 
 E_{2,1}=9b+2\sqrt{b^2+14a}+u_0^- \end{array} \right. $ \\ \hline 
0&3&0&$E_{1,3}=11b+u_0^-$\\ \hline 
\end{tabular} 
\caption{Summary of the lowest eigenvalues of the sextic oscillator. 
For fixed parameters $a,b$ the energies for which a close form may be 
obtained are labeled by $n$, the number of nodes of the radial wavefunction, 
and $\tau$.} 
\label{tapepe} 
\end{table} 
\ec 
We give in Fig. (\ref{sexpo}) an example of the shape of the potential surface
in order to help the reader to visualize the sextic potential.
This potential is a rather flexible one, since the parameters may be 
adjusted in order to yield a minimum at $\beta=0$ or at $\beta>0$.
In the latter case it might also have a local maximum at some
$\beta>0$ before reaching the global minimum. These features may
be very useful for an accurate description of shape phase transitions.

The  unnormalized solutions of the canonical form of the differential  
equation (having set $\phi(\beta)=\beta^2f(\beta)$) may be written as 
\be 
\phi_n(\beta)\= P_n(\beta^2)\beta^{2s-1/2}exp\Bigl( -{a\beta^4\over 4} 
-{b\beta^2\over 2} \Bigr) \,,
\ee 
where $P_n$ are polynomials of order $n$. The property of normalizability 
imposes $a\ge 0$.

\subsection{Linear potential} 
\begin{figure}[!t] 
\epsfig{file=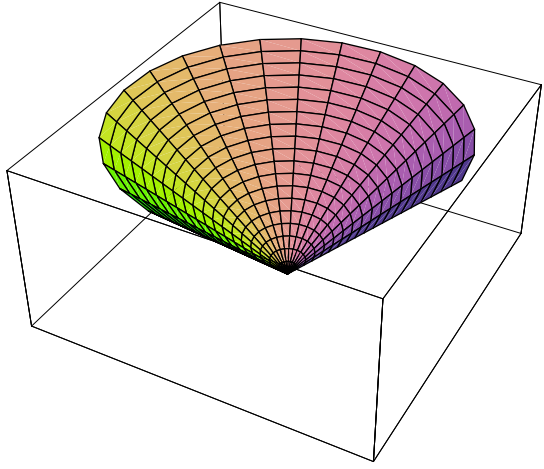,scale=0.7} 
\caption{Plot $V(\beta,\gamma)$ of the linear potential 
with $0<\gamma<3\pi/2$. } 
\label{linpo} 
\end{figure} 
We give in this paragraph an approximated treatment {\it \`a la Wilets et  
Jean} that may be applied to the linear potential  
\be 
u(\beta) = A\beta 
\ee 
being $A$ a positive constant. This is depicted in fig. \ref{linpo} 
 
As far as we know this potential 
has not been treated in any paper concerning the subject.  
Regrouping all 
the terms except the energy in the second term of eq. (\ref{stdform}),  
we can define an effective potential 
\be 
V_{eff}(\beta)= A\beta +{(\tau+3/2)^2\over \beta^2}-{1\over 4\beta^2} 
\ee 
that can be expanded as a Taylor series around the minimum that is located 
at $\beta_0 \= (2(\tau+1)(\tau+2)/A)^{1/3}$. Neglecting terms of cubic or 
higher orders, we are now in the conditions to solve the differential equation
(analogous to the displaced harmonic oscillator) 
 \be 
\chi''(\beta) +\Bigl\{ \epsilon - {3\over 2}A\beta_0 - {3A\over 2\beta_0} 
(\beta-\beta_0)^2 \Bigl\} \chi(\beta) \=0 
\ee 
that gives the following spectrum: 
\be 
\epsilon_{n_\beta,\tau}= {3\over 2}A\beta_0 + (n_\beta+ {1\over 2}) 
\sqrt{3A\over \beta_0}. 
\ee 
The wave functions are expressed in terms of Hermite polynomials $H$ of order 
$n_\beta$ 
\be 
\chi(\beta) \= H_{n_\beta}\Bigl((\beta-\beta_0)\sqrt[4]{3A/ \beta_0} \Bigr)  
e^{-{(\beta-\beta_0)^2/2\sqrt{3A/\beta_0}}} 
\ee 
When one measures, as usual, the energies starting 
from the ground state in units of the 
energy of the first $2^+$, the spectrum does not depend on the $A$ parameter. 
 
\begin{table}[!t] 
\begin{tabular}{c|c|c||c|c} 
$J^\pi$&$n$&$\tau$ &  approximation &    numeric \\ \hline 
$0^+$&0&  0    &    0           &      0  \\ 
$2^+$&0&  1    &    1           &      1  \\ 
$2,4^+$&0&  2    &    1.885       &      1.911 \\ 
$3,4,6^+$&0&  3    &    2.697       &      2.755 \\ 
$4,5,6,8^+$&0&  4    &    3.456       &      3.548 \\ \hline 
$0^+$&1&  0    &    1.466       &      1.734 \\ 
$2^+$&1&  1    &    2.220       &      2.568 \\ 
\end{tabular} 
\caption{Comparison between approximated eigenvalues calculated 
with the method explained in the text, and the numerical values courtesy of 
M.Caprio \cite{CapPri}. The result is satisfactory for the first family (1 
to 3 \% errors), while it gets worse, as expected, when $n$ growths.} 
\label{ta3} 
\end{table} 
In table \ref{ta3} we report a list of the lowest eigenvalues calculated in  
the present approach (by the author) and numerically by Caprio \cite{CapPri}. 
He had performed numerical  
calculations to check the validity of this approximation as a premise 
to the so-called sloped wall potential, that is flat from the center to 
a given point after which it has a linear dependence 
(see Sect. (\ref{sloped})).

\subsection{Other cases} 
The cases treated in the previous pages do not exhaust completely  
the list of known (analytic or approximated) solution, but, to remain  
loyal to the statements made in the introduction, we preferred to  
distinguish between the former cases and the present subsection. 
Solutions of the Bohr hamiltonian have been derived, for instance, from  
the classical limit of the interacting boson model (IBM). Other potentials, 
that are interesting for a phenomenological modelization, have been  
treated numerically. We will briefly examine here some of these cases. 
 
\subsubsection{Ginocchio's anharmonic potential solution} 
The idea in Ginocchio's paper \cite{Gino}, that we will resume without  
going into the details because it would be a task beyond our aims,  
is to bridge the IBM and the collective model by means of coherent boson  
states. Starting from an IBM hamiltonian with anharmonicities (and with a  
fixed number of bosons) and pairwise interactions he derived a second order 
differential operator which has a spectrum that contains as the lowest levels 
the IBM eigenenergies. The radial profile of the potential connected 
with this solution is very interesting since it is negative in the origin 
and is very flat near it; it grows rather steadily towards zero at 
some given point and has a small tail that asymptotically approaches zero.
It may be written as
\be
V_A(\beta)=V_0 {b\beta^2\over 1+b\beta^2}\,,
\ee
where the strength of the potential is connected to the number of bosons
and $b$ measures the anharmonicity of the system.
The eigenfunctions are given in terms of Jacobi polynomials:
\be
\varphi_{n,\tau}(\beta)=N_{n\tau} (1-x)^{\tau/2}(1+x)^{1+\alpha'/2}
P_m^{\alpha,\alpha'}(x)\,,
\ee
where $m=(n-\tau)/2$, $\alpha=\tau+3/2$, $\alpha'=\bar N-n$ and
\be
x={1-b\beta^2\over 1+b\beta^2}\,.
\ee
The normalization constant is given by
\be
N_{n,\tau}^2={b^{5/2}\alpha' m!\Gamma(\alpha'+\alpha+m+1) \over 
2^{\alpha+\alpha'-1/2} \Gamma(\alpha+m+1)\Gamma(\alpha'+m+1)}\,.
\ee
The reader is referred to the original paper and to Ref. \cite{Gino2}
for details on the normalization,
on the volume element, on the eigenenergies, on the additional 
definitions and on the quantum numbers that find their roots in the IBM.

\subsubsection{Warsaw solution}
In ref. \cite{War1} an interesting $\gamma-$unstable case
is solved numerically and compared with data. In particular the so-called
Myers-Swiatecki potential:
\be
V(\beta^2) \= {1\over 2} V_c \beta^2 +G[e^{-(\beta/a)^2}-1]
\ee
is solved. With this potential the authors say that they get practically
the same results that can be obtained with a modification of 
the Davidson potential. They also give the analytic solution in this case
(about ten years before the work by Elliott and collaborators).
This correspondence 
can be understood, developing the exponential function, from the 
interplay between the $\beta^2$ and $\beta^{-2}$ terms.

In ref. \cite{War2} the generalized Bohr hamiltonian is employed (we
do not go into details, see \cite{GenBo}) with a potential of the 
following form:
\be
V(\beta) \= {1\over 2} C_2 \beta^2  +C_8 \beta^8+G [e^{-(\beta/a)^2}-1 ]\,.
\ee
It is interesting to note that the comparison with the spectrum of $^{134}$Ba
that they obtained is not less valuable that the one obtained in the context
of the critical point symmetry E(5). We would like to point out that, if 
(again) a series expansion for the transcendental function is used, 
the result is that we have to deal with a generalization of the Davidson 
potential with even powers of $\beta$. This has been treated in \cite{Bona2}
where the authors show how the solution of such a potential tend to the 
infinite square well solution when the leading power increases. They have
reasons to believe that a term like $\beta^8$ is enough to have a good 
correspondence.
\label{corrispondenza}

\subsubsection{Quartic potential}
The quartic oscillator in $\beta$ is expected to 
play an important role in the shape
phase transition from harmonic oscillator, U(5), to $\gamma-$unstable cases
, O(6).  In fact it can be shown \cite{Pepe2} that the large N limit of
the IBM at the critical point and the (numerical) solution of the Bohr 
differential equation with a $\beta^4$ potential lead to the same results 
and these results are fundamentally different from the analytic solution 
obtained from the solution of 
 the Bohr equation with an infinite square well, the so called E(5)
symmetry. Thus it should be stressed that E(5) symmetry is the {\it exact}
mathematical solution for an {\it approximate} physical problem. 
In table \ref{ta4} we give numerical values for the excitation energies of the
$\beta^4$ potential and we refer to \cite{Pepe2} for a thorough comparison
with E(5).
\begin{table}[!t]
\begin{tabular}{c|cccc|}
&$\xi=1$&$\xi=2$&$\xi=3$&$\xi=4$\\\hline
$\tau=0$&0.00&2.39&5.15&8.20\\
$\tau=1$&1.00&3.63&6.56&9.75\\
$\tau=2$&2.09&4.92&8.01&11.34\\
$\tau=3$&3.27&6.26&9.50&12.95\\\hline
\end{tabular}
\caption{Excitation energies for a $\beta^4$ potential relative to the energy
of the first excited state (from \cite{Pepe2}).}
\label{ta4}
\end{table}
They obtain a geometrical limit of IBM by using a coherent state formalism. 
Considering two different IBM hamiltonian they obtain energy surfaces showing
that a $\beta^4$ potential is associated with the critical point. It is also
argued that the IBM could provide the finite N correction to the predictions
of the simple collective model. This is of interest for an identification
of nuclei at the critical point.

\subsubsection{Bonatsos' {\it et al.} solution}
\label{sec_bon1}
A study that deserves comments is a generalization of 
the harmonic oscillator and quartic potentials  that has been
discussed in \cite{Bona1}. The authors solve numerically the Bohr hamiltonian
with the sequence of potential $\beta^{2n}$ with $n$ positive non-null 
integer. 

The harmonic oscillator case occurs for $n=1$, while the square well may be 
considered as the limit for $n\rightarrow \infty$. The structure of the 
spectra of all these potentials changes smoothly from one extreme to the 
other, thus bridging the two solutions. This model allows thus for
a large variety of different spectra with the $R_{4/2}$ ratio varying from $2$
to $2.20$.

\subsubsection{Caprio's sloped wall potential}
\label{sloped}
The E(5) and X(5) solutions generated not only an experimental effort 
aimed at the recognition of new patterns in nuclear spectra, but also 
a theoretical effort aimed at the assessment of its most important 
characteristics. The solution
for the finite square well and the solution for the sloped wall potential 
are parts of this effort. In the former case the conclusion was that the
results of the E(5) model are quite robust with respect to the introduction
of a finite depth. In the latter case a flat potential with a sloped wall
was investigated \cite{Capslo}
to establish if the results of the E(5) and X(5) symmetries 
are robust with respect to (small) inclinations of the wall of the infinite 
potential well.  
The sloped-wall potential reads:
\be
V(\beta)\= \left\{  
\begin{array}{ccc} 
 V_0 &,& \beta \le \beta_w \\ 
 C(\beta-\beta_w) &,& \beta > \beta_w 
 \end{array} \right.   \,.
\label{slo} 
\ee
The solution in the internal region may be found analytically in terms
of Bessel function, while in the external region an analytic treatment 
is not possible. The author notice that when the term with a $\beta^{-2}$
dependence is not present, the equation in $\beta$ reduces to the Airy
equation for which the solution is possible.
The problem is then solved numerically using the Airy functions as an
efficient basis for diagonalization. The eigenvalues are determined 
by the condition of continuity of the logarithmic derivative at the 
matching point. 
The eigenvalues are lowered relative to the respective 
infinite square well cases, and the lowering is more effective for
high lying states. On the whole the spectrum is very sensitive
to the stiffness of the wall, and the deviation from the X(5) predictions
are able to generate a closer agreement with nuclear spectra in the
$N\sim 90$ region. \footnotetext[2]{This subsection has been inserted here
because of its numerical character, although a better place would have been
at the end of the next section.}

\section{Axial $\gamma-$stable cases.}
The present section deals mainly with potentials of the type 
$u(\beta,\gamma)\= u(\beta)+v(\gamma)$, where the term in $\gamma$ is 
taken as an harmonic oscillator. 
This clearly violates the property of periodicity  
in $\gamma$ that our problem possesses. However if we consider only a narrow 
interval of $\gamma$'s around zero it is possible to give an analytic  
solution to the differential equations using the expansion of the  
periodic functions around $\gamma=0$ (see for example \cite{Wilet} for a  
thorough analysis). Although this is not the only  
possibility to obtain solvable, or approximatively solvable, Bohr  
hamiltonians, we will discuss it in detail for its importance in connection  
with the issue of critical point symmetries and shape phase transitions. 
This was first discussed in ref. \cite{Iac2}, where the approximations 
that will be used throughout were introduced. Within this class of potentials 
one may study approximate solutions for a $\beta-$soft, $\gamma-$soft rotor. 
In subsection E a different method, that leads to an exact separation,
is used.

It is perhaps worth saying that when the $\gamma$ degree of freedom 
comes into play, one usually have to consider $\gamma$ phonons.
The $\beta$ phonon shares the same total angular momentum ($\lambda=2$) with
the $\gamma$ phonon, but while the former has a null component along 
the quantization axis, the latter has a non null ($\mu=\pm 2$) component.
The rules to assign quantum numbers to the various states, for a given
number of quanta in $\gamma$, are:
$$
j=0,...,n_\gamma $$
$$
K=2n_\gamma-4j$$
\be
L=\left\{
\begin{array}{rr}
0,2,4,6,8,...~~~~~~~~&K=0\\
K,K+1,K+2,...& K\ne 0
\end{array}
\right.
\label{regole}
\ee
When $K$ is negative one must consider its absolute value.

\subsection{Iachello's square well solution or X(5)} 
This important 
case gave the starting signal to a number of experimental works about 
the so-called X(5) symmetry. Iachello proposed an approximate separation 
of variables for the Bohr hamiltonian that consists of two steps. 
The first point is to realize that around $\gamma\=0^o$ the rotational 
kinetic energy, eq. (\ref{trot}), may be written as 
\be 
\sum_{k=1}^3 {{\hat Q}_k^2 \over \sin{\bigl(\gamma -{2\pi\over 3}k}\bigr)^2} 
\simeq  
{4\over 3} (\underbrace{{\hat Q}_1^2+{\hat Q}_2^2+{\hat Q}_3^2}_{L^2})+
{\hat Q}_3^2 
\Bigl({1\over \sin{\gamma}^2}-{4\over 3} \Bigr)\,.
\label{trk1}  
\ee 
In this case the problem is no longer $\gamma-$unstable and the separation 
of the wavefunction {\it \'a la Bes} do not hold anymore. We should instead 
look 
for solutions of the type $\Psi(\beta,\gamma,\theta_i)\=\varphi_K^L(\beta, 
\gamma)\WD_{M,K}^L(\theta_i)$, where $\WD$ is a Wigner function, 
eigenfunction of the square of total angular momentum and of its third  
component. The action of $L^2$ and ${\hat Q}_3^2$ leaves the following 
equation 
\begin{widetext} 
\be 
\Biggl\{-{1\over \beta^4}{\partial \over 
 \partial \beta} \beta^4{\partial \over \partial  
\beta}-{1\over \beta^2}{1\over \sin{3\gamma}}{\partial \over \partial\gamma}  
\sin{3\gamma}{\partial \over \partial\gamma}+{1\over 4\beta^2}\Biggl[  
{4\over 3}L(L+1) +K^2\Bigl({1\over \sin{\gamma}^2}-{4\over 3} \Bigr) 
\Biggr]+u(\beta,\gamma) \Biggr\} \varphi_K^L(\beta,\gamma)\=\epsilon 
\varphi_K^L(\beta,\gamma)\,.
\label{iach-wide} 
\ee 
Considering now a potential of the type $u(\beta,\gamma)\= u(\beta)+v(\gamma)$ 
the above equation may be approximately separated into the following set 
\be 
\Biggl[ -{1\over \beta^4}{\partial \over 
 \partial \beta}\beta^4{\partial \over 
 \partial \beta}+{L(L+1) \over 3\beta^2} + 
u(\beta) \Biggr] \xi_L(\beta) \= \epsilon_\beta \xi_L(\beta) 
\label{x5-uno} 
\ee 
\be 
\Biggl[ -{1\over \langle \beta^2\rangle\sin{3\gamma}}{\partial \over  
\partial\gamma} \sin{3\gamma}{\partial \over \partial\gamma}+ {1\over  
4\langle \beta^2\rangle}K^2\Biggl({1\over \sin{\gamma}^2}-{4\over 3} \Biggr) 
+v(\gamma) \Biggr]\eta_K (\gamma)\= \epsilon_\gamma \eta_K (\gamma) \,,
\label{x5-due} 
\ee 
\end{widetext} 
where $ \langle \beta^2\rangle $ is the average of $\beta^2$ over 
$\xi(\beta)$. Here we have $\epsilon \simeq\epsilon_\beta+\epsilon_\gamma$ 
and $\varphi_K^L(\beta,\gamma)\simeq \xi_L(\beta)\eta_K (\gamma)$.  
Some authors noticed that a different set of equations, in which the term 
$K^2/(3 \langle \beta^2\rangle)$ is kept in the first equation, may have 
been derived as well. In this case, that we will not take into account, the  
solution of the $\beta$ equation would carry also the $K$ quantum number. 
 
Insofar the procedure has been carried for a general potential. 
Iachello treated a square well potential in $\beta$ combined with 
a harmonic oscillator in $\gamma$. 
Introducing the square well potential (\ref{square}) into eq. (\ref{x5-uno}) 
and setting $\tilde \xi(\beta)=\beta^{3/2}\xi(\beta)$,  
$\epsilon_\beta=k_\beta^2$ and $z=\beta k_\beta$, he obtained a 
Bessel equation  
\be 
\tilde \xi''+{\tilde \xi' \over z} +\Biggl[ 1 - {\nu^2\over z^2} \Biggr] 
\tilde \xi\=0 \,,
\ee 
where  
\be 
\nu\=\Bigl( {L(L+1)\over 3} +{9\over 4}\Bigr)^{1/2}~~. 
\ee 
The solution of this equation is therefore 
expressed in terms of Bessel function with  
a non-integer index, $\nu$,
\be 
\xi_{s,L}\=c_{s,L} \beta^{-3/2}J_\nu(k_{s,L}\beta) \,,
\ee 
where the set of quantum numbers refers to the $s$th zero of the Bessel  
function $J_\nu(z)$ and to the total angular momentum, $L$. 
In fact the boundary  
condition at the wall of the well, $\tilde \xi (\beta_w)\=0$, 
determines the eigenvalues, because the wave function has a node if  
and only if the Bessel function has a zero, that we call $x_{s,L}$.  
The spectrum is thus 
\be 
\epsilon_{s,L}\=(k_{s,L})^2\,,
\ee 
where $k_{s,L}\= {x_{s,L}\over \beta_w}$ and $\beta_w$ is the position of 
the wall.

\begin{figure}[!h]
\epsfig{file=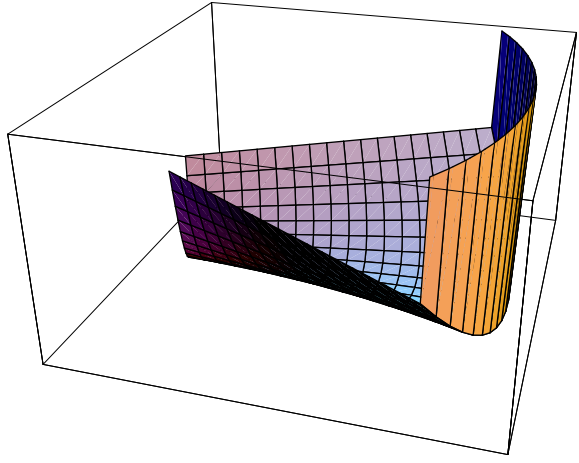,scale=0.65}
\caption{Plot $V(\beta,\gamma)$ of the square well potential in $\beta$
combined with a harmonic oscillator in $\gamma$
discussed by Iachello, with $-\pi/4<\gamma<\pi/4$. }
\label{x5fig}
\end{figure}
 
We give here the solution of the $\gamma-$part for the harmonic oscillator,
as it was given by Iachello. The following subsections 
\ref{subck}~-~\ref{subpg} will share the same treatment for 
the $\gamma-$variable.
A trigonometric expansion, strictly valid in a small sector around the origin 
($\sin{\gamma}\sim \gamma$), is made in eq. 
(\ref{x5-due}) with a harmonic oscillator potential, obtaining formally 
the radial equation of a two dimensional harmonic oscillator:
\be
\Bigl[-{1\over \langle \beta^2\rangle}{1\over \gamma}
{\partial\over \partial \gamma}\gamma{\partial\over \partial \gamma}+
{K^2\over 4 \langle \beta^2\rangle}{1\over \gamma^2}+(3a)^2{\gamma^2\over 2} 
\Bigr]\eta_{K}(\gamma)\={\tilde \varepsilon}_\gamma \eta_{K}(\gamma)\,,
\ee
with ${\tilde \varepsilon}_\gamma\=\varepsilon_\gamma+
{K^2\over 3\langle \beta^2\rangle}$. The solution is given by
\be
{\tilde \varepsilon}_\gamma\={3a\over \sqrt{\langle \beta^2\rangle}}
(n_\gamma+1)
\ee
and
\be
\eta_{n_\gamma,K}(\gamma)\=c_{n,K} \gamma^{\mid K/2\mid}
e^{-3a\gamma^2/2}L_n^{\mid K \mid}(3a\gamma^2)\,,
\ee
where $n=(n_\gamma-\mid K \mid)/2$ with $n_\gamma=0,1,2,\ldots~$ and 
$L_n^{\mid K \mid}$ are Laguerre polynomials. The parameter $a$ is the 
strength of the harmonic oscillator potential, or in other words its
amplitude.

\subsection{Solutions for Coulomb- and Kratzer-like potentials}  
\label{subck}
\begin{figure}[!t] 
\epsfig{file=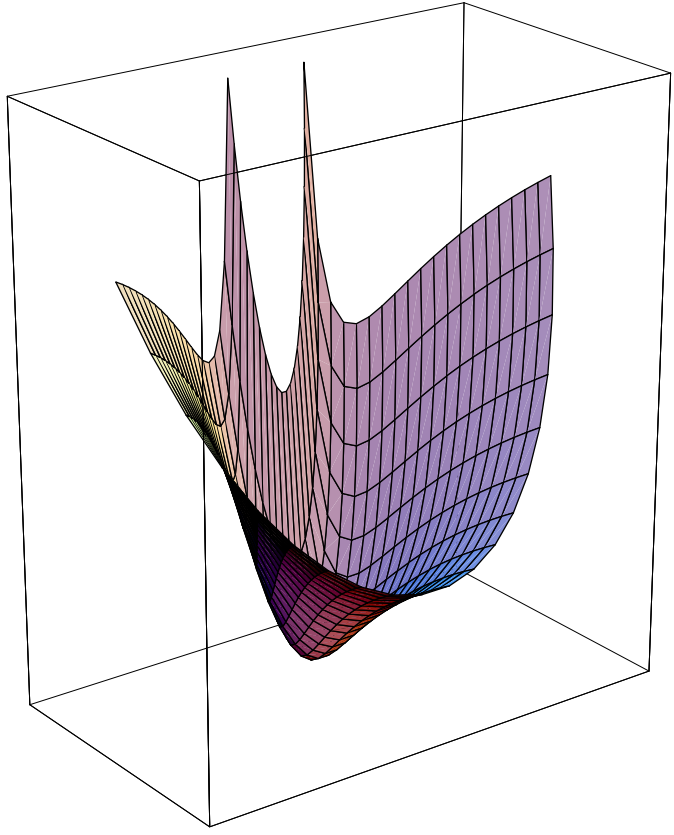,scale=0.65} 
\caption{Plot $V(\beta,\gamma)$ of the Kratzer-like  
potential (\cite{FV2}) combined with a harmonic oscillator in $\gamma$ 
discussed by Fortunato and Vitturi, with $-\pi/6<\gamma<\pi/6$. } 
\label{forfig2} 
\end{figure} 
 With the substitution $\xi_L(\beta)\= \chi_L(\beta)\beta^{-2}$,
equation (\ref{x5-uno}) may be simplified to its standard form:
\be
{\partial^2 \over \partial \beta^2}\chi_L(\beta) +\Biggl\{ 
\varepsilon^{(\beta)}
-u(\beta)-\Biggl({2+{L(L+1)\over 3} \over \beta^2} \Biggr)\Biggr\}\chi_L(\beta)
\= 0 \,.
\label{general}
\ee
By inserting the Coulomb-like
potential $u(\beta)\=-A/\beta$ with $A>0$ and recasting the problem in the 
variable $x\=2\beta\sqrt{\epsilon}$, with the further substitutions 
$\varepsilon\=-\epsilon$,  $k={A\over2\sqrt{\epsilon}}$ and $\mu^2=
\bigl({9\over 4}+{L(L+1)\over 3}\bigr)$, we obtain the Whittaker's equation
\be
{\partial^2 \over \partial x^2}\chi_L(x) +\Biggl\{ -{1\over 4}
+{k\over x}+{1/4-\mu^2\over x^2} \Biggr\}\chi_L(x)\=0\,.
\ee
Its regular solution is the Whittaker function $M_{k,\mu}(x)$ that reads:
\be
\chi_L(x)\= {\cal N}e^{-x/2} x^{{1\over 2}+\mu} 
~_1F_1 \bigl({1\over 2}+\mu-k,1+2\mu,x\bigr)
\label{wave}
\ee 
which is, in general, a multivalued function. The constant ${\cal N}$ is 
determined from the normalization condition. We adopt usual conventions 
and thus the function is analytic on the real axis.
The hypergeometric series is an infinite series and to recover a good 
asymptotic behaviour we must require that it terminates, i.e. that it becomes 
a polynomial. This happens when the first argument is a negative integer, 
$-v$, that must thence be regarded as an additive quantum number. 
This condition fixes unambiguously the $\beta$ part of the spectrum:
\be
\epsilon_{v,L}^{(\beta)} \= {A^2/4 \over\Biggl(\sqrt{{9\over 4}+
{L(L+1)\over 3}}+{1\over 2}+v \Biggr)^2}\,.
\label{spectr1}
\ee
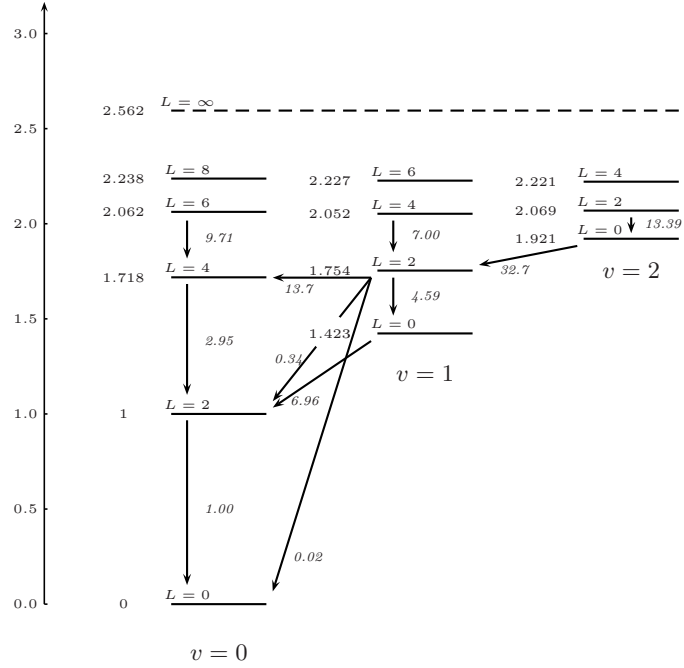
\begin{figure}[!h]
\bc
\vspace{0.6cm}
\begin{picture}(230,300)(0,0)
\psset{unit=1.2pt}
\psline[linewidth=0.5]{->}(0,20)(0,210)
\psline[linewidth=0.5]{-}(0,20)(1,20)\rput(-6,20){\tiny $0.0$}
\psline[linewidth=0.5]{-}(0,50)(1,50)\rput(-6,50){\tiny $0.5$}
\psline[linewidth=0.5]{-}(0,80)(1,80)\rput(-6,80){\tiny $1.0$}
\psline[linewidth=0.5]{-}(0,110)(1,110)\rput(-6,110){\tiny $1.5$}
\psline[linewidth=0.5]{-}(0,140)(1,140)\rput(-6,140){\tiny $2.0$}
\psline[linewidth=0.5]{-}(0,170)(1,170)\rput(-6,170){\tiny $2.5$}
\psline[linewidth=0.5]{-}(0,200)(1,200)\rput(-6,200){\tiny $3.0$}
\rput(55,5){\small $v=0$}
\psline{-}(40,20)(70,20)
\rput(25,20){\tiny $0$}\rput(45,23){\tiny $L=0$}
\psline{-}(40,80)(70,80)
\rput(25,80){\tiny $1$}\rput(45,83){\tiny $L=2$}
\psline{-}(40,123.1)(70,123.1)
\rput(25,123.1){\tiny $1.718$}\rput(45,126.1){\tiny $L=4$}
\psline{-}(40,143.75)(70,143.75)
\rput(25,143.75){\tiny $2.062$}\rput(45,146.75){\tiny $L=6$}
\psline{-}(40,154.25)(70,154.25)
\rput(25,154.25){\tiny $2.238$}\rput(45,157.25){\tiny $L=8$}
\psline[linestyle=dashed]{-}(40,175.7)(200,175.7)
\rput(25,175.7){\tiny $2.562$}\rput(45,178.7){\tiny $L=\infty$}
\psline{->}(45,78)(45,26)\rput(55,50){\tiny \it 1.00}
\psline{->}(45,121)(45,86)\rput(55,103){\tiny \it 2.95}
\psline{->}(45,141)(45,129)\rput(55,135){\tiny \it 9.71}

\rput(120,93){\small $v=1$}
\psline{-}(105,105.4)(135,105.4)
\rput(90,105.4){\tiny $1.423$}\rput(110,108.4){\tiny $L=0$}
\psline{-}(105,125.23)(135,125.23)
\rput(90,125.23){\tiny $1.754$}\rput(110,128.23){\tiny $L=2$}
\psline{-}(105,143.15)(135,143.15)
\rput(90,143.15){\tiny $2.052$}\rput(110,146.15){\tiny $L=4$}
\psline{-}(105,153.6)(135,153.6)
\rput(90,153.6){\tiny $2.227$}\rput(110,156.6){\tiny $L=6$}
\psline{->}(110,123)(110,111)\rput(120,117){\tiny \it 4.59}
\psline{->}(110,141)(110,131)\rput(120,136){\tiny \it 7.00}

\rput(185,125.4){\small $v=2$}
\psline{-}(170,135.27)(200,135.27)
\rput(155,135.27){\tiny $1.921$}\rput(175,138.27){\tiny $L=0$}
\psline{-}(170,144.15)(200,144.15)
\rput(155,144.15){\tiny $2.069$}\rput(175,147.15){\tiny $L=2$}
\psline{-}(170,153.27)(200,153.27)
\rput(155,153.27){\tiny $2.221$}\rput(175,156.27){\tiny $L=4$}
\psline{->}(185,142)(185,137)\rput(195,140){\tiny \it 13.39}

\psline{->}(168,133)(137,127) \rput(148,125){\tiny \it 32.7}
\psline{->}(103,103)(72,82) \rput(82,84){\tiny \it 6.96}

\psline{->}(103,123)(72,24)\rput(83,35){\tiny \it 0.02}
\psline{->}(103,123)(72,84)\rput(77,97){\tiny \it 0.34}
\psline{->}(103,123)(72,123)\rput(80,119){\tiny \it 13.7}

\pscircle*[linecolor=white](90,105.4){6}
\rput(90,105.4){\tiny $1.423$}

\end{picture}
\caption{$\beta-$spectrum of the Coulomb-like case ($n_\gamma=0$). 
The vertical scale is expressed in units of the energy of the 
$(v=0,L=2)$ state. Some B(E2) values are displayed in units of the
lowest transition of the first band (numbers in italics). 
Notice the presence and energy of 
a threshold and the absence of degeneracy. From \cite{FV2}.}
\label{ffff}  
\ec
\end{figure}

Fig. \ref{ffff} shows the spectrum of the axial rotor with coulomb-like
potential.

The case of the Kratzer-like potential is treated in a very similar way. 
Inserting in eq. (\ref{general}) the potential 
\be
u(\beta)=-A/\beta+B/\beta^2
\label{potenKr}
\ee 
and setting $x\=2\beta\sqrt{\epsilon}$, 
$\varepsilon\=-\epsilon$,  $k={A\over2\sqrt{\epsilon}}$ and $\mu^2=
\bigl(B+{9\over 4}+{L(L+1)\over 3}\bigr)$, 
we obtain again the Whittaker's equation whose
solutions may be written as in eq. (\ref{wave}) and we may repeat the 
whole procedure of the previous section, the only major difference being the 
definition of $\mu$. The spectrum assumes in this case the form
\be
\epsilon_{v,L}^{(\beta)} \= 
{A^2/4 \over\Biggl(\sqrt{B+{9\over 4}+{L(L+1)\over 3}}+{1\over 2}+v \Biggr)^2}
 ~.
\label{ep2}
\ee
We recall (see sect. \ref{cke5}) that the two parameters used here, 
$A$ and $B$, have an 
immediate translation into the position, $\beta_0$, and depth of 
the minimum of the potential, ${\cal D}$, being valid the following relations: 
$A\=2\beta_0{\cal D}$ and $B\=\beta_0^2{\cal D}$. Consequently $\beta_0\=2B/A$
and ${\cal D}\=A^2/(4B)$.

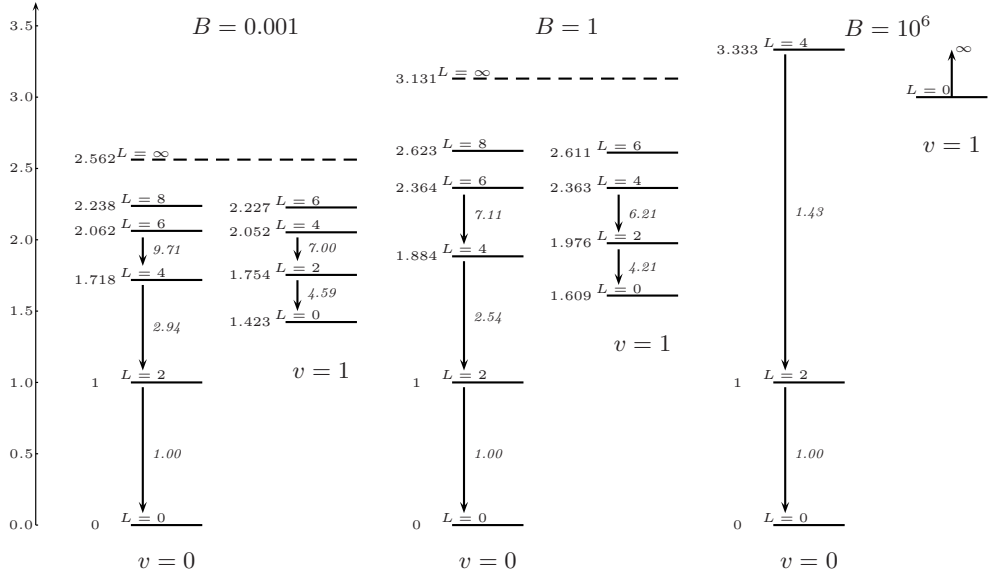
\begin{figure}[!h]
\bc
\begin{picture}(300,250)(25,0)
\psset{unit=0.9pt}
\psline[linewidth=0.5]{->}(0,20)(0,240)
\psline[linewidth=0.5]{-}(0,20)(1,20)\rput(-6,20){\tiny $0.0$}
\psline[linewidth=0.5]{-}(0,50)(1,50)\rput(-6,50){\tiny $0.5$}
\psline[linewidth=0.5]{-}(0,80)(1,80)\rput(-6,80){\tiny $1.0$}
\psline[linewidth=0.5]{-}(0,110)(1,110)\rput(-6,110){\tiny $1.5$}
\psline[linewidth=0.5]{-}(0,140)(1,140)\rput(-6,140){\tiny $2.0$}
\psline[linewidth=0.5]{-}(0,170)(1,170)\rput(-6,170){\tiny $2.5$}
\psline[linewidth=0.5]{-}(0,200)(1,200)\rput(-6,200){\tiny $3.0$}
\psline[linewidth=0.5]{-}(0,230)(1,230)\rput(-6,230){\tiny $3.5$}
\rput(55,5){\small $v=0$}
\psline{-}(40,20)(70,20)
\rput(25,20){\tiny $0$}\rput(45,23){\tiny $L=0$}
\psline{-}(40,80)(70,80)
\rput(25,80){\tiny $1$}\rput(45,83){\tiny $L=2$}
\psline{-}(40,123.1)(70,123.1)
\rput(25,123.1){\tiny $1.718$}\rput(45,126.1){\tiny $L=4$}
\psline{-}(40,143.75)(70,143.75)
\rput(25,143.75){\tiny $2.062$}\rput(45,146.75){\tiny $L=6$}
\psline{-}(40,154.25)(70,154.25)
\rput(25,154.25){\tiny $2.238$}\rput(45,157.25){\tiny $L=8$}
\psline[linestyle=dashed]{-}(40,173.72)(135,173.72)
\rput(25,173.72){\tiny $2.562$}\rput(45,176.72){\tiny $L=\infty$}
\rput(120,85){\small $v=1$}
\psline{-}(105,105.4)(135,105.4)
\rput(90,105.4){\tiny $1.423$}\rput(110,108.4){\tiny $L=0$}
\psline{-}(105,125.23)(135,125.23)
\rput(90,125.23){\tiny $1.754$}\rput(110,128.23){\tiny $L=2$}
\psline{-}(105,143.15)(135,143.15)
\rput(90,143.15){\tiny $2.052$}\rput(110,146.15){\tiny $L=4$}
\psline{-}(105,153.6)(135,153.6)
\rput(90,153.6){\tiny $2.227$}\rput(110,156.6){\tiny $L=6$}
\psline{->}(45,78)(45,25)\rput(55,50){\tiny \it 1.00}
\psline{->}(45,121)(45,85)\rput(55,103){\tiny \it 2.94}
\psline{->}(45,141)(45,129)\rput(55,135){\tiny \it 9.71}
\psline{->}(110,123)(110,110)\rput(120,117){\tiny \it 4.59}
\psline{->}(110,141)(110,131)\rput(120,136){\tiny \it 7.00}
\rput(90,230){\small $B=0.001$ }

\rput(190,5){\small $v=0$}
\psline{-}(175,20)(205,20)
\rput(160,20){\tiny $0$}\rput(180,23){\tiny $L=0$}
\psline{-}(175,80)(205,80)
\rput(160,80){\tiny $1$}\rput(180,83){\tiny $L=2$}
\psline{-}(175,133.04)(205,133.04)
\rput(160,133.04){\tiny $1.884$}\rput(180,136.04){\tiny $L=4$}
\psline{-}(175,161.84)(205,161.84)
\rput(160,161.84){\tiny $2.364$}\rput(180,164.84){\tiny $L=6$}
\psline{-}(175,177.4)(205,177.4)
\rput(160,177.4){\tiny $2.623$}\rput(180,180.4){\tiny $L=8$}
\psline[linestyle=dashed]{-}(175,207.8)(270,207.8)
\rput(160,207.8){\tiny $3.131$}\rput(180,210.8){\tiny $L=\infty$}
\rput(255,96.5){\small $v=1$}
\psline{-}(240,116.54)(270,116.54)
\rput(225,116.54){\tiny $1.609$}\rput(245,119.54){\tiny $L=0$}
\psline{-}(240,138.56)(270,138.56)
\rput(225,138.56){\tiny $1.976$}\rput(245,141.56){\tiny $L=2$}
\psline{-}(240,161.8)(270,161.8)
\rput(225,161.8){\tiny $2.363$}\rput(245,164.8){\tiny $L=4$}
\psline{-}(240,176.66)(270,176.66)
\rput(225,176.66){\tiny $2.611$}\rput(245,179.66){\tiny $L=6$}
\rput(225,230){\small $B=1$ }

\psline{->}(180,78)(180,25)\rput(190,50){\tiny \it 1.00}
\psline{->}(180,131)(180,85)\rput(190,107){\tiny \it 2.54}
\psline{->}(180,159)(180,138)\rput(190,150){\tiny \it 7.11}
\psline{->}(245,136)(245,121)\rput(255,128){\tiny \it 4.21}
\psline{->}(245,159)(245,143)\rput(255,151){\tiny \it 6.21}

\rput(325,5){\small $v=0$}
\psline{-}(310,20)(340,20)
\rput(295,20){\tiny $0$}\rput(315,23){\tiny $L=0$}
\psline{-}(310,80)(340,80)
\rput(295,80){\tiny $1$}\rput(315,83){\tiny $L=2$}
\psline{-}(310,219.98)(340,219.98)
\rput(295,219.98){\tiny $3.333$}\rput(315,222.98){\tiny $L=4$}
\rput(360,230){\small $B=10^6$ }
\psline{-}(370,200.)(400,200.)\psline{->}(385,200)(385,220)
\rput(375,203){\tiny $L=0$}\rput(390,220){\tiny $\infty$}
\rput(385,180){\small $v=1$}

\psline{->}(315,78)(315,25)\rput(325,50){\tiny \it 1.00}
\psline{->}(315,218)(315,85)\rput(325,151){\tiny \it 1.43}  
\end{picture}
\caption{
Evolution of the $\beta-$spectrum of the Kratzer-like rotor with
the parameter $B$. The vertical scale is expressed in units of the energy 
of the $(v=0,L=2)$ state. The B(E2) values, calculated with formula 
(\ref{be2}), between the lowest states are indicated in italics 
beside the corresponding downward arrows. Intraband transitions are 
not marked for the sake of simplicity. From \cite{FV2}.}
\label{ffff2}  
\ec
\end{figure}

In fig. \ref{ffff2} we report a study of the evolution of the spectrum 
and transition rates for the Kratzer-like rotor as a function of the
parameter $B$. When $B$ is large a typical rotational spectrum is recovered.

\subsection{Bonatsos' {\it et al.} solution}              
As in sec. \ref{sec_bon1}, we summarize here briefly another work by Bonatsos
and collaborators \cite{Bona2} that deals with the numerical solution of a 
sequence of potentials interpolating between the U(5) and X(5) models of 
the form:
\be
u_{2n}(\beta,\gamma)\={\beta^{2n}\over 2}+c\gamma^2 \,,
\ee
where the harmonic dependence on $\gamma$ is the same of the X(5) case,
while the potential in $\beta$ may be considered as a generalization of
other cases. For $n=1$ one gets an exactly soluble model that is called 
X(5)$-\beta^2$, while for 
$n\rightarrow \infty$ the infinite square well potential is recovered.
As expected energy ratios and  transition rated change smoothly from
one limiting case to the other and it is shown that the X(5) results are 
already well approximated for $n\sim 4$.

\subsection{Pietralla and Gorbachenko's solution or CBS-model}
\label{subpg}
\begin{figure}[!t] 
\epsfig{file=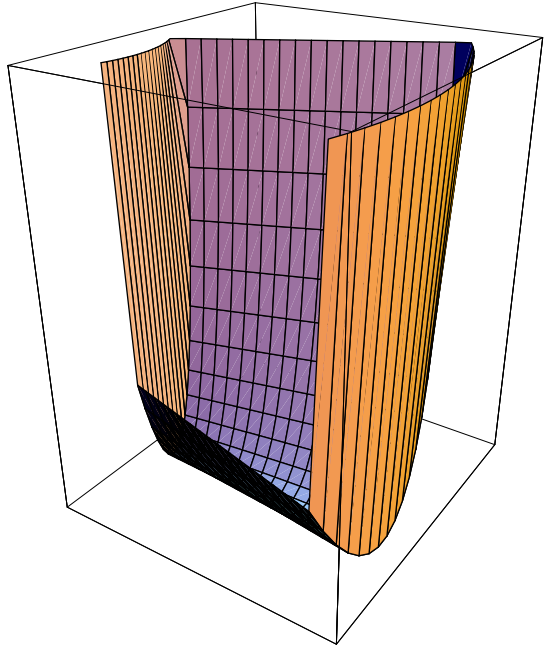,scale=0.65} 
\caption{Plot $V(\beta,\gamma)$ of the two-sided infinite square well 
potential in $\beta$ combined with a harmonic oscillator in $\gamma$ 
discussed by Pietralla and Gorbachenko \cite{Pietr}, 
with $-\pi/12<\gamma<\pi/6$. } 
\label{pietra} 
\end{figure} 
An extension of the X(5) model has been recently proposed \cite{Pietr} 
with the aim 
to study the evolution of the first excited $0^+$ state in transitional
nuclei ($N\sim 90$). The authors consider the same problem discussed by 
Iachello for axially symmetric prolate nuclei, but they take as a potential
in $\beta$ the infinite square well with two boundaries:
\be 
V(\beta)\= \left\{  
\begin{array}{ccc} 
 \infty &,& \beta < \beta_m \\ 
 0 &,& \beta_m \le \beta < \beta_M \\
   \infty &,& \beta \ge \beta_M
 \end{array} \right.  
\label{twosquare} 
\ee 
The number $r_\beta=\beta_m/\beta_M$ may be used to parameterize the 
stiffness of the potential.
The X(5) solution is a special case of this model for $r_\beta=0$ as well
as the rigid rotor is another special case for $r_\beta=1$.  
Eq. (\ref{x5-uno}) may be transformed into the Bessel equation with the 
substitutions $z=\sqrt{E/(\hbar^2/2B_m)}\beta$ and $\tilde \xi(z)= 
\beta^{3/2} \xi_L(\beta)$. The solution is a linear combination of Bessel's
$J_\nu(z)$ and $Y_\nu(z)$ functions with index $\nu=\sqrt{L(L+1)/3+9/4}$.
The authors give the complete solution in the following form:
\be
\xi_{L,s}(\beta)={c_{L,s}\over \beta^{3/2}}\Bigl[J_\nu\bigl(z_{L,s}^{r_\beta}
\beta/\beta_M \bigr)+\gamma_Y Y_\nu\bigl(z_{L,s}^{r_\beta}
\beta/\beta_M \bigr) \Bigr]\,,
\ee
where $c_{L,s}$ is a normalization constant.
The spectrum is found imposing the condition that the wave functions in 
$\beta$ must have nodes at the two boundaries:
$$
\xi(\beta_m)=\xi(\beta_M)=0
$$
or
\be
\tilde \xi(r_\beta z_m)=\tilde \xi(z_M)=0\,,
\ee
where $z_M=\sqrt{E/(\hbar^2/2B_m)}\beta_M$. The quantization condition
is thus
\be
J_\nu(z_M)Y_\nu(r_\beta z_M)-J_\nu(r_\beta z_M)Y_\nu(z_M)=0\,,
\ee
whose $s$th zero, $z_{L,s}^{r_\beta}$, must be calculated numerically.
The eigenvalues are found as
\be
E_{L,s}= {\hbar^2\over 2B_m \beta_M^2} (z_{L,s}^{r_\beta})^2\,.
\ee
This solution is rather interesting because there is one more
degree of freedom with respect to X(5) and moving $r_\beta$ from 0 to 1
has not only the effect of increasing the position of the $\beta$ bands,
but also to invert the trend of the ratio $R_{4/2}(s)$: while for $r_\beta=0$
the ratio $R_{4/2}$ of the ground state band is higher than for any 
excited band, for greater values of $r_\beta$ the opposite happens.
Another interesting feature of this model is the possibility to study the 
effect of the 'centrifugal stretching': the term with $\beta^{-2}$ power
acts as a 'centrifugal stretch' (although for the sake of clarity the variable
here is not the radius of the system, but its deformation) in the sense that
the wave functions of states labeled by high values of the SO(6) quantum 
number will be pushed towards higher deformations, decreasing their 
probability amplitudes in the region close to the origin. Within this model
one actually cut the region close to the origin with the internal wall. 
As a consequence only states with appreciable amplitude in the classically 
forbidden area will be more affected: the overall result is that states
with high $\tau$ are not much affected by the presence of the internal wall,
while states with low $\tau$ are shifted up in energy. The spectrum will 
therefore show smaller energy distances among low-lying states with respect   
to X(5).

The authors named their model as "confined $\beta-$soft" rotor model (CBS)
by analogy with the Wilets-Jean $\gamma-$soft model (see Appendix B).

\subsection{Prolate and oblate axial rotor: new cases}
\label{new-sol}
We propose in this section new solutions that, as far as we know, 
have not been published anywhere else, although they are pretty simple.
To obtain them we have done nothing but collecting knowledge coming from
the various solutions that have been devised so far and apply known 
approximations. In doing so we made use of some suggestions
due to D.J.Rowe \cite{RoPR}. The novelty here is represented by
the transformation (\ref{keyt}) that allows a better level of 
approximation for the goniometric functions. This is reflected in four new
analytically solvable cases.

The Schr\"odinger equation for the Bohr hamiltonian
may be profitably simplified for the prolate
and oblate soft axial rotors (respectively around $\gamma\sim 0$ and
around $\gamma \sim \pi$) using the approximation (\ref{trk1}) 
devised by Iachello. Alternatively one can treat the 
$\gamma\sim \pi/3$ case by noticing that in that case the projection 
of the angular momentum on the intrinsic x-axis is a good quantum number 
(see fig. \ref{Hill}). 
We will seek solutions of the type 
\be
\Psi(\beta,\gamma\theta_i)=
\xi_L(\beta)\eta_K(\gamma){\cal D}^L_{M,K}(\theta_i) \,.
\ee
Since the rotational part is standard and the action of the square
of the angular momentum and of its third component are trivially
evaluated, we will restrict to the $\beta-\gamma$ part, as done in 
\cite{Iac2}. The result is formula (\ref{iach-wide}), that we employ 
here as a starting point for new solutions.

Contrary to what was done in the X(5) model \cite{Iac2}, we may separate 
exactly the Bohr equation as explained in section \ref{exa-sep}, 
using a potential of the form $u(\beta, \gamma)\= u_1(\beta)+ 
u_2(\gamma)/ \beta^2$, obtaining the following two equations
\begin{widetext}
\be
\Biggl\{-{1\over \beta^4}{\partial \over 
\partial \beta}\beta^4{\partial \over \partial \beta}+u_1(\beta)-\epsilon+
{\omega+ {L(L+1)-K^2\over 3}\over \beta^2}\Biggr\} \xi_{L,K}(\beta) \=0 
\label{cia1}
\ee
\be
\Biggl\{-{1\over \sin{(3\gamma)}}{\partial \over \partial 
\gamma}\sin{(3\gamma)}{\partial \over \partial \gamma}
+{K^2\over 4 \sin{(\gamma)}^2} 
+u_2(\gamma)-\omega\Biggr\} 
\eta_K(\gamma)  \=0
\label{cia2}
\ee
\end{widetext}
where $\omega$ is a separation constant and the energy, $\epsilon$, is
contained only in the first equation. Notice that one could have 
alternatively retained one or both of the terms $L(L+1)/3$ and $K^2/3$
in the equation for $\gamma$. We prefer to keep them in the $\beta-$part
because this part is solved exactly. The $\gamma-$part usually experiences 
a second turn of approximations.

In the second equation one can adopt the same
trigonometric simplifications ($\sin{\gamma}\sim\gamma$ and $\cos{\gamma}\sim
 1$) that are implicitly used in (\ref{trk1}) or may
try a more sophisticated approach. In fact eq. (\ref{cia2}) may be written as
\be
\Biggl\{ {\partial^2 \over \partial\gamma^2}+3\cot{3\gamma}
{\partial \over \partial \gamma}+\omega-u_2(\gamma)-
{K^2\over 4\sin^2{(\gamma)}} \Biggr\}\eta_K(\gamma)=0
\ee
and with the transformation 
\be
\eta_K(\gamma)={\rho_K(\gamma)\over \sqrt{\sin{3\gamma}}}
\label{keyt}
\ee
it may be brought in the standard form
\be
\Biggl\{ {\partial^2 \over \partial\gamma^2}+\omega-u_2(\gamma)
+{9\over 2}-{K^2\over 4\sin^2{\gamma}}+{9\over 4}\cot^2{(3\gamma)}
\Biggr\}\rho_K(\gamma)=0\,,
\ee
where the term with the first derivative has been eliminated.
It is worth noticing that now a better approximation may be used
for the trigonometric functions, without loosing the possibility
to solve exactly the equation. The series expansion for the 
cotangent, including the second order, may be used 
\be
{9\over 4}\cot^2(3\gamma)\sim {1\over 4\gamma^2}-{3\over 2}+
{27\over 20}\gamma^2\,.
\ee
Therefore we can recast the differential equation above as
\be
\Biggl\{ {\partial^2 \over \partial\gamma^2}+
\underbrace{\omega +3}_{q^2}
+{27\over 20}\gamma^2-u_2(\gamma)+{1-K^2\over 4\gamma^2} \Biggr\}
\rho_K(\gamma)=0\,.
\ee
This expression represents a better level of approximation with respect to 
any equation used so far for the $\gamma$ part of the problem.
The 'extra' term with dependence $\gamma^{-2}$ has the same behaviour of 
a 'centrifugal' term and the differential equation can be solved exactly 
with harmonic or Davidson potentials for $u_2(\gamma)$.

\begin{figure}[!t]
\begin{picture}(250,180)(-5,0)
\psset{unit=1 pt}
\psline{-}(0,0)(240,0)
\psline{-}(0,0)(0,175)
\rput(-5,0){\tiny 0}\rput(-5,20){\tiny 2}\rput(-5,40){\tiny 4}
\rput(-5,60){\tiny 6}\rput(-5,80){\tiny 8}\rput(-5,100){\tiny 10}
\rput(-5,120){\tiny 12}\rput(-5,140){\tiny 14}
\psline[linewidth=1.3pt]{-}(10,20)(30,20)
\rput(20,15){\tiny $n_\gamma=0$}
\rput(20,25){\tiny $K=0$}
\psline[linestyle=dotted](20,35)(20,55)
\psline[linewidth=1.3pt]{-}(40,40)(60,40)
\rput(50,35){\tiny $n_\gamma=1$}
\rput(50,45){\tiny $K=2$}
\psline[linestyle=dotted](50,55)(50,75)
\psline[linewidth=1.3pt]{-}(70,80)(90,80)
\rput(80,75){\tiny $n_\gamma=1$}
\rput(80,85){\tiny $K=2$}
\psline[linestyle=dotted](80,95)(80,115)
\psline[linewidth=1.3pt]{-}(100,100)(120,100)
\rput(110,95){\tiny $n_\gamma=2$}
\rput(110,105){\tiny $K=0$}
\psline[linestyle=dotted](110,115)(110,135)
\psline[linewidth=1.3pt]{-}(130,100)(150,100)
\rput(140,95){\tiny $n_\gamma=2$}
\rput(140,105){\tiny $K=4$}
\psline[linestyle=dotted](140,115)(140,135)
\psline[linewidth=1.3pt]{-}(160,140)(180,140)
\rput(170,135){\tiny $n_\gamma=2$}
\rput(170,145){\tiny $K=4$}
\psline[linestyle=dotted](170,155)(170,175)
\psline[linewidth=1.3pt]{-}(190,80)(210,80)
\rput(200,75){\tiny $n_\gamma=3$}
\rput(200,85){\tiny $K=6$}
\psline[linestyle=dotted](200,95)(200,115)
\end{picture}
\caption{Graphical representation of the values of $\omega_{n_\gamma,K}$
for the lowest states of a few bands, obtained from formula (\ref{2ome}) 
with $\lambda=1$ (the shift, $-3$, has been omitted). The assignation of
quantum numbers to each band follows from (\ref{regole}). }
\label{om-band}
\end{figure}
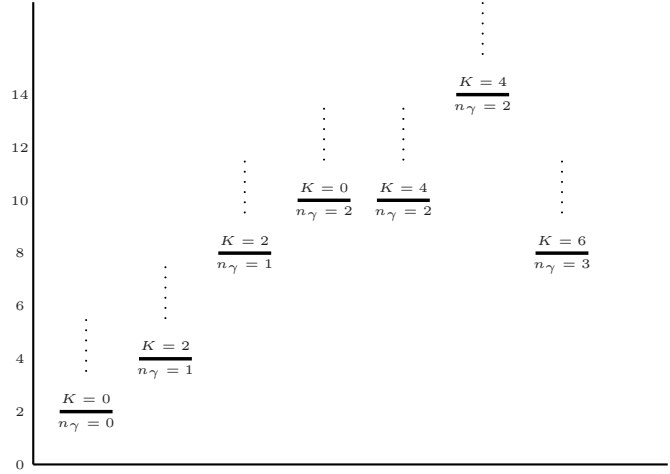

With $u_2(\gamma)=C\gamma^2$, defining $\mu=q^2/(2\lambda)$ and 
$\lambda^2=C-{27\over 20}$ 
and $t(t+1)=(K^2-1)/4$ we obtain
\be
\rho_K(\gamma) = {\cal N} \gamma^{t+1}e^{-{\lambda\over 2}\gamma^2}
~_1F_1 \Bigl({t+3/2-\mu\over 2}, t+3/2;\lambda\gamma^2\Bigr)\,,
\ee
where the confluent hypergeometric function is not divergent only when
it is a polynomial: this happens if the first argument is equal to 
$-n_\gamma$ with $n_\gamma=0,1,2,3,...$. This requirement sets the formula 
for the spectrum
\be
\omega_{n_\gamma,K}=\sqrt{C-{27\over 20}}\Bigl(2t+3+4n_\gamma\Bigr)-3\,.
\ee
Notice the constraint on the value of $C$ that comes from the square root.
With
\be
t={-1\pm\mid K\mid \over 2}
\ee
we can write 
\be
\omega_{n_\gamma,K}= \sqrt{C-{27\over 20}}\Bigl( \pm\mid K\mid
+2+4n_\gamma\Bigr) -3\,.
\label{2ome}
\ee
The structure of the various bands predicted by this formula 
is depicted in fig. \ref{om-band}. We shifted the energies eliminating the
term $-3$ and we used a constant strength ($\lambda$) equal to unity for 
purpose of illustration. 

The expression for $\omega$ must be now used inside the 
equation for the $\beta$ part.
Choosing for example the harmonic oscillator: $u_1(\beta)=k\beta^2$
one has:
\be
\epsilon=\sqrt{k}\bigl(2n_\beta+\tau+5/2 \bigr)
\ee
with $\tau$ is determined through the relation
\be
\tau(\tau+3)=\omega_{n_\gamma,K}+{L(L+1)-K^2\over 3}
\ee
(one can employ the Davidson potential in a very similar way).
Alternatively we may consider the Coulomb potential, $u_1(\beta)=-A/\beta$, 
or the Kratzer potential, $u_1(\beta)=-A/\beta+B/\beta^2$.
In the latter more general case we obtained
\be
\epsilon={A^2/4\over \Bigl(\sqrt{1/4+B+\omega_{n_\gamma,K}
+{L(L+1)-K^2\over 3}}+1/2+n_\beta\Bigr)^2}\,,
\ee
where one should insert the values of $\omega$ previously found in 
formula (\ref{2ome}). The most relevant implication of this model is the 
difference between bands with opposites signs of $K$.

\section{Triaxial cases}
We divided this section from the previous one, although they both
deal with $\gamma-$stable cases, because in triaxial nuclei the 
minimum of the potential along $\gamma$ is not located at $n{\pi\over 3}$,
 with $n\in \mathbb{Z}$. This implies the absence of axial symmetry.

In triaxial nuclei the eigenvalue of the projection of the third component 
of the angular momentum is no more a good quantum number (this can be seen
already at the classical level). In the special case of $\gamma=30^o$ the
projection of the angular momentum on the intrinsic y-axis, called sometimes
$R$, is a conserved quantity, but in the regions $0^o<\gamma<30^o$ and 
$30^o<\gamma<60^o$ is not possible to use $K$ or $R$ to label bands.
Notice that in the axial case the lowest member of a group of states with
the same $L$ has always the lowest possible value of $K$, while at 
$\gamma=30^o$ the lowest state with a given $L$ has the highest possible 
value of $R$ (this is a trivial consequence of formula (\ref{meyer-form})).

\subsection{Davydov's classical solution or rigid rotor model}
Traditionally the name of Davydov is associated with a series of papers 
\cite{Dav1,Dav2,Dav3,Dav4} in collaboration with other various researchers.
Davydov and Filippov proposed the existence of triaxial nuclei,
the properties of which  have been investigated in the
adiabatic approximation, which assumes rotation of the nucleus without 
change of the intrinsic state. The equilibrium shape is a triaxial 
ellipsoid, whose hamiltonian, frequently called the rigid rotor hamiltonian,
 may be written as:
\be
H={1\over 2} \sum_{\kappa=1}^3 {{\hat Q}_\kappa^2 \over \sin{(\gamma-
{2\over3}\pi\kappa)}^2}\,,
\label{rigrot}
\ee
where ${\hat Q}_\lambda$ are the projections of the total angular 
momentum operator on the axes of the intrinsic system.
The wave functions are expanded in terms of (axial) rotational 
wave functions
\be
\Psi_{JM}=\sum_K \mid JK\rangle A_K\,,
\ee
with
\be
\mid JK \rangle =\sqrt{2J+1\over 16\pi^2(1+\delta_{K0})}
\Bigl({\cal D}_{M,K}^J+(-)^J{\cal D}_{M,-K}^J \Bigr)\,,
\ee
where $K=0,2,4,...$ and 
\be
J= \left\{ 
\begin{array}{lc}
K,K+1,K+2,... & K\ne 0 \\
0,2,4,... & K=0
\end{array} 
\right. \,.
\ee
Within this rigid-rotor approach analytic expressions may be derived 
for a few rotational levels with low total angular momentum, although
this model completely neglects the shape vibrational degrees of freedom. 

A more complicated version, containing the kinetic energy in the $\beta$
variable along with the above rigid rotor hamiltonian and a potential term
$V(\beta)$ may be employed \cite{Dav3}. The corresponding Schr\"odinger 
equation for this $\gamma-$rigid, $\beta-$soft model is 
separable and leads to lengthy expressions for the energy of collective 
nuclear states. We give here a simplified version of the spectrum:
\be
E_{J,K,n_\beta}=(n_\beta+1/2)E_\beta+c_1(J(J+1)-K^2)+c_2K^2\,,
\ee
where $n_\beta=0,1,2,..$ is the number of $\beta$ phonons.

Davydov proposed also a solution that takes into account $\beta$ and 
$\gamma$ vibrations. When in the Bohr hamiltonian a potential of the type
\be
V(\beta, \gamma)={1\over 2} C(\beta-\beta_0)^2 +{1\over 2} C_\gamma \beta_0^2
(\gamma-\gamma_0)^2 
\ee
is employed, we can look for solutions that are product of $f(\beta)$ and
$\Phi(\gamma,\theta_i)$ functions, obeying the following equations:
\be
\Bigl( -{1\over \sin{3\gamma}}{\partial\over \partial \gamma}\sin{3\gamma}
{\partial\over \partial \gamma}+{1\over 4}
\sum_{\kappa=1}^3{{\hat Q_\kappa}^2\over \sin{(\gamma-{2\over3}\pi\kappa)}^2}
+D(\gamma-\gamma_0)^2-\Lambda \Bigr)
\Phi(\gamma,\theta_i)=0
\ee
and
\be
\Bigl[ -{\hbar^2\over 2B_m}{d^2\over d\beta^2}+{1\over 2}C(\beta-\beta_0)^2 +
{\hbar^2(\Lambda+2)\over 2B\beta^2} -E\Bigr] \beta^2 f(\beta)=0\,,
\ee
where $C_\gamma\=\omega_\gamma^2B$ contains the frequency of the $\gamma$
vibration.

The above equations can be solved exactly for a spherical nucleus, 
reproducing known results. Davydov proposed also a solution for
non-axial nuclei when small oscillations around the equilibrium position 
are considered. The main point is to approximate the effective potential,
that consists of the displaced harmonic oscillator and the 'centrifugal' 
term, expanding it as a power series as Wilets and Jean did for the 
$\gamma-$unstable case. In the present approach (as in the previous one,
that has only $\beta$ vibrations) a rather complicated expression for the 
energy levels is found that accounts for rotations as well as $\beta$ and
$\gamma$ vibrations.
Although we do not enter into the details, it should be said that this
approach had considerable success and applications to nuclear spectra.

\subsection{Meyer-ter-vehn formula}
In a work focused on the spectrum of an odd nucleon coupled with a 
rotating triaxial core \cite{MTV}, Meyer-ter-vehn introduced also a simple 
analytic formula for the spectrum of the rigid rotor at $\gamma=30^o$.
At that angle two of the three moment of inertia are equal (albeit
the axis of the ellipsoid does not have equal length) and the hamiltonian 
(\ref{rigrot}) is axially symmetric around the intrinsic y-axis. 
The energy spectrum is easily
written (apart from a factor) in the following form
\be
E_{J,{ R}}= J(J+1)-{3\over4}{ R}^2
\label{meyer-form}
\ee
where ${R}$ is the sharp projection of the angular momentum on 
the y-axis. The wave functions
\be
\Psi_{J,M,{ R}}(\theta_i) 
=\sqrt{2J+1\over 16\pi^2
(1+\delta_{{ R}0})}\Bigl({\cal D}_{M,{ R}}^J(\theta_i)+
(-)^J{\cal D}_{M,-{ R}}^J(\theta_i) \Bigr)
\ee
do not depend on the $\gamma$ variable, but only on the Euler angles.
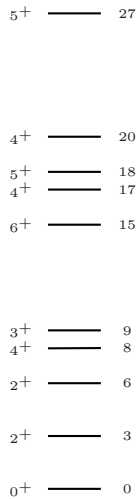
\begin{figure}[!t]
\bc
\begin{picture}(40,190)(0,0)
\psset{unit=1.pt}
\psline{-}(10,0)(30,0) \rput(0,0){\tiny $0^+$} \rput(40,0){\tiny $0$}
\psline{-}(10,20)(30,20)\rput(0,20){\tiny $2^+$} \rput(40,20){\tiny $3$}
\psline{-}(10,40)(30,40)\rput(0,40){\tiny $2^+$} \rput(40,40){\tiny $6$}
\psline{-}(10,53.2)(30,53.3) \rput(0,53.3){\tiny $4^+$} 
\rput(40,53.3){\tiny $8$}
\psline{-}(10,60)(30,60)\rput(0,60){\tiny $3^+$} \rput(40,60){\tiny $9$}
\psline{-}(10,100)(30,100)\rput(0,100){\tiny $6^+$} \rput(40,100){\tiny $15$}
\psline{-}(10,113.3)(30,113.3)\rput(0,113.3){\tiny $4^+$} 
\rput(40,113.3){\tiny $17$}
\psline{-}(10,120)(30,120)\rput(0,120){\tiny $5^+$} \rput(40,120){\tiny $18$}
\psline{-}(10,133.3)(30,133.3)\rput(0,133.3){\tiny $4^+$} 
\rput(40,133.3){\tiny $20$}
\psline{-}(10,180)(30,180)\rput(0,180){\tiny $5^+$} \rput(40,180){\tiny $27$}
\end{picture}
\ec
\caption{Spectrum of the rigid triaxial rotor at $\gamma=30^o$, according to 
the formula by Meyer-ter-Vehn. This spectrum is also predicted by the Davydov 
rigid model, although only a few low-lying states may be calculated 
analytically. Quantum numbers and energies are reported on the left and 
right respectively. Notice that here the first excited states lies at $3$.}
\label{mtvf}
\end{figure}

We illustrate in fig. \ref{mtvf} the spectrum predicted by eq. 
(\ref{meyer-form}).

\subsection{Iachello's solution or Y(5)}
\begin{figure}[!t] 
\epsfig{file=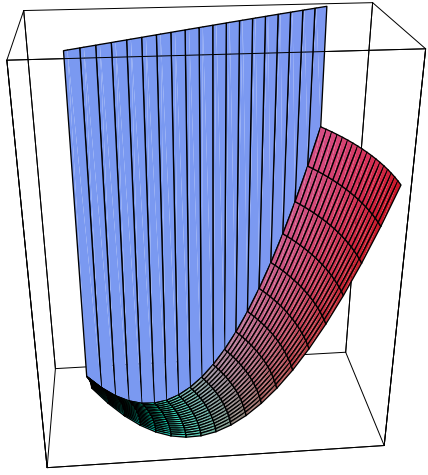,scale=0.65} 
\caption{Plot $V(\beta,\gamma)$ of the potential used in the Y(5) symmetry
discussed by Iachello (\cite{Iac3}): 
an infinite square well in $\gamma$ with $0<\gamma<\gamma_w=\pi/6$ 
combined with a displaced harmonic oscillator in $\beta$. } 
\label{y5figura} 
\end{figure} 
The solution called Y(5) was introduced in Ref. \cite{Iac3} with the aim
to describe the critical point of the axial to triaxial shape phase 
transition. The starting point is again the set of differential equations
(\ref{x5-uno}, \ref{x5-due}) obtained inserting in the Bohr hamiltonian 
a potential 
of the type $V(\beta,\gamma)=u(\beta)+v(\gamma)$ that allows an
approximate separation of variables. In the present case a displaced harmonic
oscillator potential in the variable $\beta$ is considered, namely
\be
u(\beta)={u_0\over 2}(\beta-\beta_0)^2\,,
\ee
while an infinite square well potential is considered in the asymmetry 
variable $\gamma$, as follows
\be 
v(\gamma)\= \left\{  
\begin{array}{ccc} 
 0 &,& \gamma < \gamma_w \\ 
\infty &,& \gamma > \gamma_w 
 \end{array} \right.  
\label{square-gamma} 
\ee
This particular choice of the potential, depicted in fig. (\ref{y5figura}), 
is an approximation to a more
general expression, $-\cos(3\gamma)+\xi\cos(3\gamma)^2$, that changes 
smoothly from an axial to a triaxial minimum, passing through a critical
point when $\xi=1/2$. When at the critical point the shape of this 
potential is flat around $\gamma=0$ and eventually rises steadily at some
point: this can be roughly approximated by a square well. Although this is 
a rather crude approximation, it has the advantage of generating an exact
solution, that may again be used as a benchmark. When $\gamma << 60^o$ one
may also take $\sin(\gamma)\sim \gamma$ and eq. (\ref{x5-due}) may be written
as
\be
\Bigl[ -{1\over \gamma}{\partial \over \partial \gamma}\gamma
{\partial \over \partial \gamma}+  \Bigl({K\over 2}\Bigr)^2
\Bigl({1\over \gamma^2}-{4\over 3}\Bigr)  \Bigr]\eta(\gamma) =
{\bar \epsilon}_\gamma \eta(\gamma)\,,
\ee
with $K=0,\pm 2,\pm 4,...~$. This equation is the Bessel equation
with solutions given in terms of Bessel functions as
\be
\eta_{s,K}(\gamma) = c_{s,K} J_{K/2}(k_{s,K}\gamma)\,,
\ee
with $k_{s,K}=x_{s,K}/\gamma_w$. Here $x_{s,K}$ is the $s$th zero of the
$J_{K/2}$ Bessel function. The eigenfunction must be zero at the 
wall ($\gamma_w$) of the infinite square well and this fixes the spectrum
to depend upon the zeros of the Bessel function as
\be
{\bar \epsilon}_{\gamma,s,K}=k_{s,K}^2-{K^2/2}\,.
\ee 
The full solution is obtained solving also the $\beta$ part of the problem.

\subsection{Solution around $\gamma\sim \pi/6$}\label{exa-sep}
\begin{figure}[!t] 
\epsfig{file=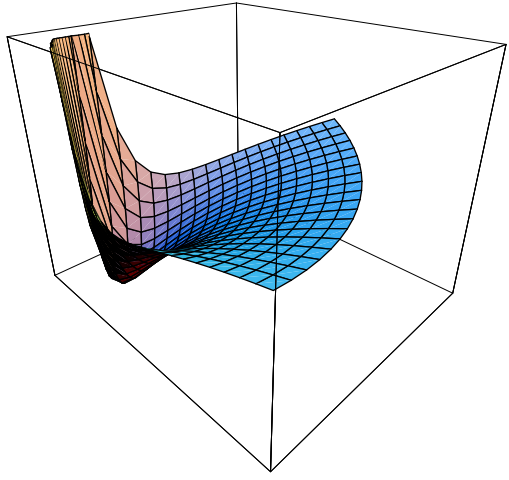,scale=0.65} 
\caption{Plot $V(\beta,\gamma)$ of the potential used the solution
of the $\gamma\sim \pi/6$ soft triaxial rotor
discussed by Fortunato (\cite{FL}): 
a harmonic oscillator in $\gamma$ with $0<\gamma<\pi/3$ 
combined with a Kratzer potential in $\beta$ (both taken as a very 
shallow for purpose of illustration).} 
\label{triax30} 
\end{figure} 
It has been shown in \cite{FL} that whenever the potential is chosen as 
\be
V(\beta, \gamma)\= V_1(\beta)+{V_2(\gamma)\over \beta^2}\,,
\label{pote}
\ee
the Schr\"odinger equation 
$H_B \Psi(\beta,\gamma, \theta_i)\=E  \Psi(\beta,\gamma, \theta_i)$
is separable \cite{Wilet}. The set of second order differential equations
that comes from this separation contains the separation constant $\Omega$
and reads:
\be
\Bigl(-{ \hbar^2\over 2B_m}{1\over \beta^4}{\partial \over 
\partial \beta}\beta^4{\partial \over \partial \beta}+V_1(\beta)-
E+{\Omega\over \beta^2}\Bigr) f(\beta)  \=0
\label{kk1}
\ee
\be
\Bigl(-{\hbar^2\over 2B_m}{1\over \sin{(3\gamma)}}{\partial \over \partial 
\gamma}\sin{(3\gamma)}{\partial \over \partial \gamma}+V_2(\gamma)-\Omega+
+{\hbar^2\over 8B_m}\sum_{\kappa=1,2,3} {{\hat Q}_\kappa^2 \over 
\sin^2(\gamma-2\pi\kappa/3)} \Bigr) 
\Phi(\gamma,\theta_i) \=0\,, \label{kk2}
\ee
with  $\Psi(\beta,\gamma, \theta_i) \= f(\beta) \Phi(\gamma,\theta_i)$. 
The set above may describe a $\beta-$soft, $\gamma-$soft triaxial 
rotor with a potential that has a minimum located in 
$\beta=\beta_0$ and $\gamma=\gamma_0 \= \pi/6$. 
The same considerations about the privileged role of the projection
of the angular momentum along the intrinsic y-axis apply in the present case.
Restricting ourselves to a small region around $\pi/6$, we multiply
the two equations above by $2B_m/\hbar^2$ and we define reduced energy and 
potentials. The new set of equations reads
\be
\Bigl(-{1\over \beta^4}{\partial \over 
\partial \beta}\beta^4{\partial \over \partial \beta}+u_1(\beta)-
E+{\Omega\over \beta^2}\Bigr) f(\beta) \=0 
\label{b-a}
\ee
\be
\Bigl(-{1\over \sin{(3\gamma)}}{\partial \over \partial 
\gamma}\sin{(3\gamma)}{\partial \over \partial \gamma}+u_2(\gamma)-\Omega
+{1\over 4}\sum_{\kappa=1,2,3} {{\hat Q}_\kappa^2 \over 
\sin^2(\gamma-2\pi\kappa/3)} \Bigr) 
\Phi(\gamma,\theta_i)  \=0 \,.
\label{g-a}
\ee
The spectrum is determined by the solution of the first differential
equation in which $\omega$, that it is found from the solution of the 
second differential equation, plays the role of the coefficient
of a 'centrifugal' term and, as we will show, yields a 
non-trivial expression for the energy levels.
Around $\pi/6$, setting $\gamma \= \pi/6+x$, the 
rotational part of the Bohr hamiltonian becomes
\be
\sum_{\kappa=1,2,3} {{\hat Q}_\kappa^2 \over \sin^2(\gamma-2\pi\kappa/3)}
\= 4(\underbrace{{\hat Q}_1^2+{\hat Q}_2^2+{\hat Q}_3^2)}_{\displaystyle 
{\hat L}^2}+ {\hat Q}_1^2\Biggl({1\over  \cos^2(x)}-4\Biggr)\,.
\label{sum}
\ee 
Changing variable in eq. (\ref{g-a}), introducing the harmonic dependence
 of the potential, $u_2(x)\= Cx^2$,
 and using the simplifications
\be
\sin{3\gamma}=\cos{3x}\sim 1 \qquad \,\,\cos{x}\sim 1 
\ee
together with relation (\ref{sum}) leads to a simplified equation
\be
\Biggl(
 -{\partial^2 \over \partial x^2}+Cx^2-\omega+
{\hat{\vec Q}}^2-{3\over 4}{\hat{Q}_1}^2 \Biggr) 
\Phi(x,\theta_i)\=0\,.
\label{demonio}
\ee
The wave function $\Phi(x,\theta_i)$ may be written, following
\cite{MTV}, as
\be
\sqrt{2L+1 \over 16\pi^2(1+\delta_{R,0})} 
\eta_{n_\gamma,L,R}(x)\bigl[{\cal D}^{(L)}_{M,R}(\theta_i)+(-1)^L
{\cal D}^{(L)}_{M,-R}(\theta_i)\bigr]
\ee
where the angular part is written in terms of Wigner functions labeled 
by the projection of the total angular momentum on the intrinsic y-axis, $R$, 
that is a good quantum number, while the functions $\eta_{n_\gamma,L,R}(x)$ 
are eigenfunctions of the one dimensional Schr\"odinger equation for the 
harmonic oscillator. The index $n_\gamma$ is the quantum number 
associated with the vibrations in the $\gamma$ degree of freedom.
The spectrum is therefore written as
\be
\omega_{L,R,n_\gamma} = \sqrt{C}(2n_\gamma+1) + L(L+1)-{3\over 4}R^2 
\label{omega}
\ee
where the first term corresponds to the $\gamma-$vibration, while
the other two terms correspond to the Meyer-ter-vehn 
formula \cite{MTV} that accounts for the rotational energy of the 
$\gamma=\pi/6$ rigid triaxial rotor.
By imposing $f(\beta)\=\beta^{-2} \chi(\beta)$, we can write the eigenvalue 
equation in the $\beta$ variable as
\be
\chi''(\beta)+\Bigl( \varepsilon -u_1(\beta) -{\omega+2\over \beta^2}\Bigr)
\chi(\beta) =0\,.
\label{equa}
\ee
The potential $u_1(\beta)$ is taken of a Kratzer-like form 
$-A/\beta+B/\beta^2$  \cite{FV1,FV2}. Setting 
$z=2\beta\sqrt{\epsilon}$, $\epsilon=-\varepsilon$, $k=A/(2\sqrt{\epsilon})$
 and $\mu^2=9/4+B+\omega$, eq. (\ref{equa}) may be rewritten as the 
Whittaker's differential equation:
\be
\Bigl\{{d^2\over dz^2} -{1\over 4} +{k\over z}+{1/4-\mu^2\over z^2} \Bigr\}
\chi(z)=0\,.
\ee
The reduced eigenvalues are
\be
\epsilon(n_\gamma,n_\beta,L,R)={A^2/4 \over \Bigl(\sqrt{9/4+B+
\omega_{L,R,n_\gamma}} +1/2+n_\beta \Bigr)^2}\,,
\label{spec1}
\ee
where $\omega$ comes from eq. (\ref{omega}). \\

Setting as usual the energy of the ground state to zero and the unit 
of the energy scale to be the energy of the lowest excited $2^+$ state 
of the ground state band leads to a spectrum that does not depend on $A$.
The interplay between the two parameters $B$ and $C$ may be exploited
to fit experimental data. Notice that $B$ and $C$ do not separately
fix the position of the respective $\beta$ and $\gamma$ bands, but they 
both take part in a non-trivial way to determine the energy levels. 
It is clear that, while the solution of the $\gamma-$angular part
of the problem gives a straightforward extension of the rigid rotor formula
in which a simple harmonic term for the $\gamma$ degree of freedom appears,
the full spectrum is rather a more complicated function that essentially
depends on the choice of the potential in $\beta$.

\begin{figure}[!t]
\epsfig{file=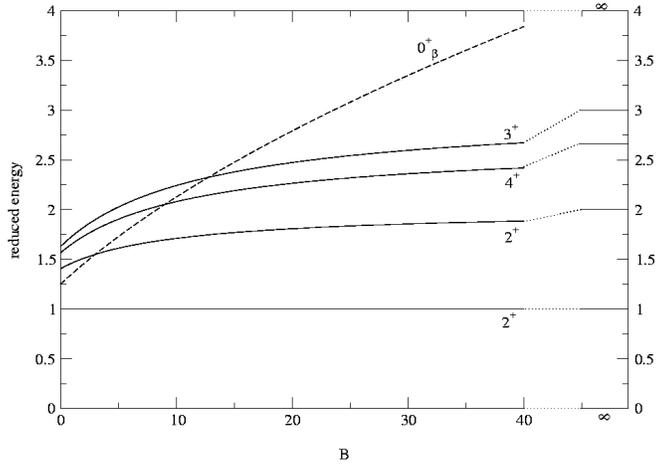,width=0.48\textwidth}
\caption{Reduced energies of the lowest state of the $\beta-$band
(dashed line) and of a few lowest states of the ground state band
(solid lines) as a function of $B$. The limits for the energy levels
when $B\rightarrow \infty$, that correspond to the rigid triaxial rotor
energies, are reported in the right side. Here we fixed $C=1$.}
\label{pi6f1}
\end{figure}
\begin{figure}[!t]\vspace{10mm}
\epsfig{file=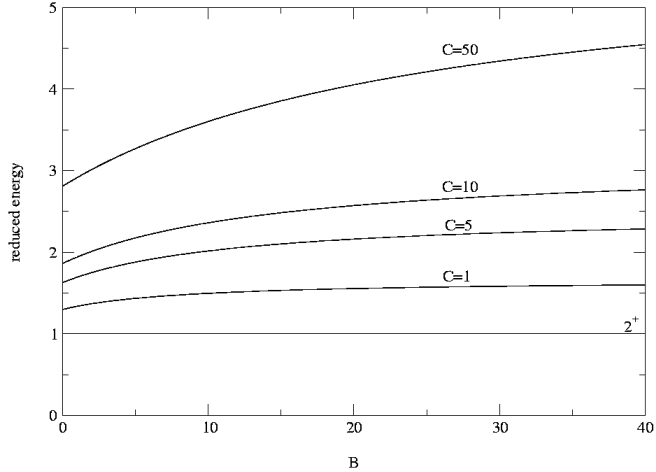,width=0.48\textwidth}
\caption{Reduced energies of the ($J^\pi=2^+,n_\gamma=1,n_\beta=0$) state
(solid line) as a function of $B$, for various values of the strength
of the harmonic potential in $\gamma$, $C$.
The first $2^+$ of the ground state band (dashed line) is reported for
reference.}
\label{pi6f2}
\end{figure}

We display in fig. \ref{pi6f1} and \ref{pi6f2} (see captions, 
adapted from \cite{FL}) the behaviour of some energy levels with respect to 
the parameters of the potential. 
The Davydov (rigid) model is recovered when $B\rightarrow\infty$
as can be seen from fig. \ref{pi6f1}.

\subsection{Bonatsos' {\it at al.} solution or Z(5)}
\begin{figure}[!t] 
\epsfig{file=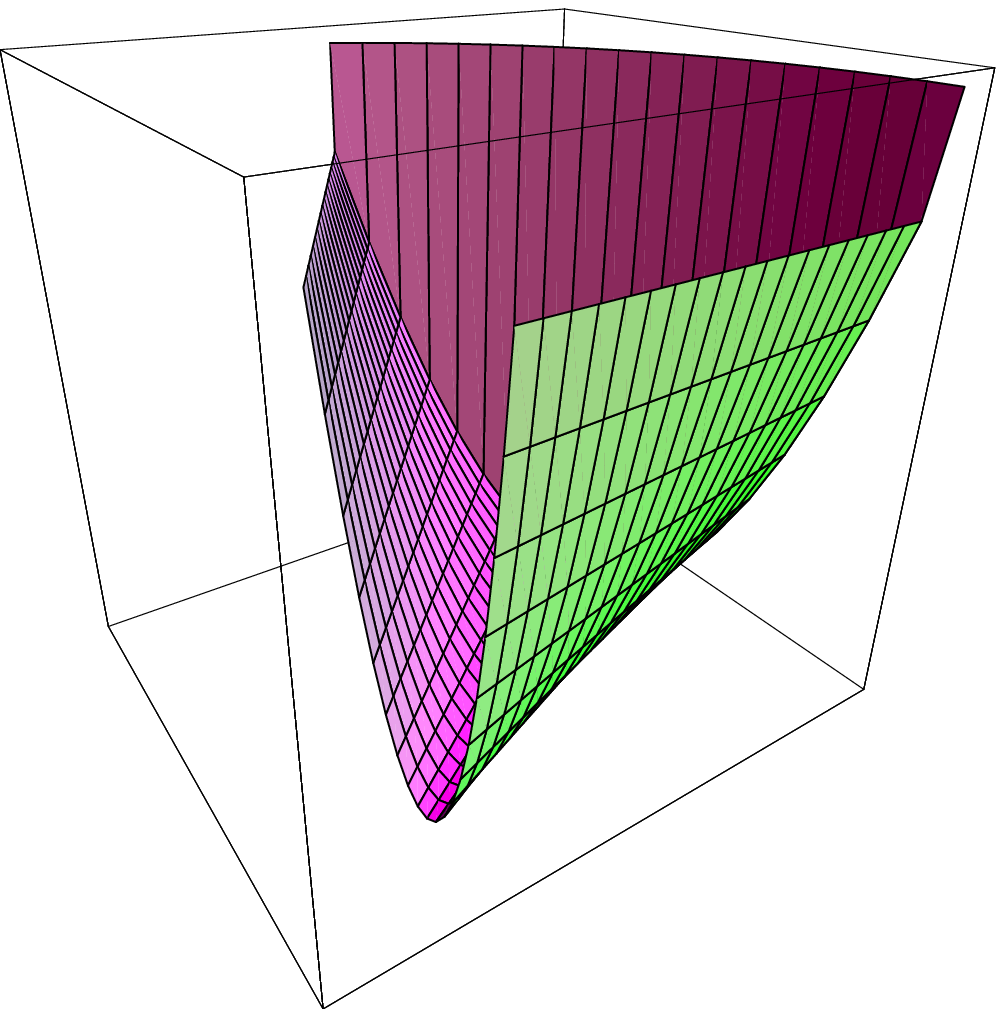,scale=0.65} 
\caption{Plot $V(\beta,\gamma)$ of the potential used the so-called 
Z(5) solution discussed by Bonatsos and collaborators (\cite{BonZ5}): 
a harmonic oscillator in $\gamma$ with a minimum in $\gamma_0=\pi/6$ 
combined with a square well potential in $\beta$ (the lowest corner of 
the frame box corresponds to the origin of polar coordinates).} 
\label{z5fig} 
\end{figure} 
A critical point symmetry for a shape phase transition from prolate 
($\gamma=0^o$) to oblate ($\gamma=60^o$) shapes was introduced in 
\cite{BonZ5} and called Z(5). 
The authors argue that in such a transition the triaxial region is crossed
and the middle lies at $\gamma=30^o$.
Unless the previous solution the present one does not use an exact 
separation of variables, but rather an approximate one with 
$u(\beta,\gamma)=u(\beta)+v(\gamma)$, where the $\beta$ potential is taken 
to be a square well and the $\gamma$ potential is taken as a harmonic
oscillator with the minimum in $\gamma=30^o$.
The infinite potential well in $\beta$ may be thought as the
critical point between a triaxial vibrator (with the minimum close to 
$\beta=0$) and a triaxial rotator with a minimum for a non null value of 
$\beta$. In this aspect the model closely resembles the X(5) model.

The same trick adopted in the previous section (formula (\ref{sum})) is 
used here to treat the rotational degrees of freedom. The Schr\"odinger 
equation is approximatively separated into the set
\be
\Bigl[ -{1\over\beta^4} {\partial \over \partial \beta}\beta^4{\partial 
\over \partial \beta}+ {1\over 4\beta^4}(4L(L+1)-3R^2)+u(\beta)\Bigr] 
\xi_{L,R}(\beta)
=\epsilon_{\beta}\xi_{L,R}(\beta)
\label{bon-beta}
\ee
and
\be
\Bigl[ -{1\over\langle \beta^2\rangle\sin{3\gamma}} {\partial \over \partial 
\gamma}\sin{3\gamma}{\partial \over \partial \gamma}+u(\gamma)\Bigr] 
\eta(\gamma)=\epsilon_{\gamma}\eta(\gamma)\,,
\ee
where the angular momentum quantum number, $L$, and its projection on the
intrinsic y-axis, $R$, are explicitely contained in the first equation.
In the second equation the average of $\beta^2$ over $\xi(\beta)$ appears:
this approximation is strictly valid only when the potential $u(\beta)$ is
deep and consequently the mean square value of $\beta$ does not oscillate much
and remains constant also for a few lower excited states.
The energy is $\epsilon\simeq \epsilon_\beta +\epsilon_\gamma$.
With the transformation $\tilde{\xi}(\beta)=\beta^{3/2} \xi(\beta)$ and the
substitutions $\epsilon_\beta=k_\beta^2$ and $z=\beta k_\beta$, eq. 
(\ref{bon-beta}) becomes a Bessel equation with eigenfunctions:
\be
\xi_{s,\nu}(\beta)=c_{s,\nu}\beta^{-3/2}J_\nu(k_{s,\nu}\beta)\,,
\ee
where the non-integer index is
\be
\nu={\sqrt{4L(L+1)-3R^2+9}\over 2}\,.
\ee
The boundary condition at the wall of the potential well, $\xi(\beta_{w})=0$,
 determines the spectrum as a function of $x_{s,\nu}$, the s{\it th} 
zero of the Bessel function $J_{\nu}(z)$ 
\be
\epsilon_{\beta;s,\nu}=\Bigl({x_{s,\nu}\over \beta} \Bigr)\,.
\ee
The $\gamma-$part is obtained, limiting ourselves to small oscillations
around the minimum, from the solution of the equation
\be
\Bigl[-{\partial^2\over \partial x^2}+{1\over 2} c\langle\beta^2\rangle x^2 
\Bigr]\eta(x)=\epsilon_\gamma\langle\beta^2\rangle \eta(x)\,,
\ee 
where $\gamma=\pi/6+x$. The above equation is an harmonic oscillator equation 
with eigenvalues
\be
\epsilon_\gamma=\sqrt{{2c\over\langle\beta^2\rangle}}(n_\gamma+1/2)\,,
\ee
with $n_\gamma=0,1,2,...~~$. The eigenfunctions are expressed in terms of
Hermite polynomials
\be
\eta_{n_\gamma}(x)=\sqrt{b\over \sqrt{\pi}2^{n_\gamma}n_\gamma!}
H_{n_\gamma}(bx)e^{-b^2x^2/2}
\ee
with $b=(c\langle\beta^2\rangle/2)^{1/4}$. The index of the Hermite 
polynomials is now the number of $\gamma-$phonons.

\subsection{Jolos' solution}
\label{Jol-sec}
One of the main objectives of ref. \cite{Jolos} is to summarize the various 
researches on shape phase transitions in nuclei. Using the interacting boson
model and the coherent state formalism the author analyzes the phase 
diagram of a cold nucleus discussing the order of phase transitions. 
Within the same perspective the author summarizes also some of the recent
achievements in the solution of the Bohr hamiltonian. In particular he 
proposes an approximate solution for the critical point of the phase 
transition between spherical and triaxially deformed shapes.
A potential of the form
\be
V(\beta,\gamma) =u(\beta)+{1\over 2} D\beta^6\cos{(3\gamma)}^2
\ee
is used. Assuming $D$ large enough, one can consider only small oscillations
around $\gamma=\pi/6$, and defining $x=\gamma-\pi/6$, the equation becomes
\be
\Biggl\{-{\hbar^2\over 2B_m}\Biggl[{1\over\beta^4}
{\partial \over \partial \beta}\beta^4{\partial \over \partial \beta}+
{1\over \beta^2}\Bigl({\partial^2\over \partial x^2}-
{9DB_m\over \hbar^2}\beta^8x^2-L(L+1)
+{3\over 4}R^2 \Bigr)+u(\beta)-E
\Biggr\} \Psi(\beta,\gamma,\theta_i)=0\,,
\ee
where $R$ is once again the projection of $\hat{\vec L}$ on the intrinsic 
y-axis. If the wave function is factorized in the following way
\be
\Psi(\beta,\gamma,\theta_i)=
f(\beta)e^{-{B_m\omega\over 2\hbar}\beta^4x^2}
{1\over \sqrt{2^nn!\sqrt{\pi}}}H_n\Bigl(x\beta^2 \sqrt{B_m\omega\over \hbar}
\Bigr){\cal D}^L_{M,R}(\theta_i)\,,
\ee
where $\omega=2(D/B_m)^{1/2}$ and $H_n$ are Hermite polynomials, 
(having used the original notation), one can exploit the action of the 
rotational and differential operators in the variable $x$. 
The equation for the $\beta-$part of the problem is therefore
\be
\Biggl\{-{\hbar^2\over 2B_m}\Biggl[{1\over\beta^4}
{\partial \over \partial \beta}\beta^4{\partial \over \partial \beta}-
{1\over \beta^2}\Bigl( L(L+1)+{3\over 4}R^2 \Bigr)\Biggr]
 +u(\beta)+
\hbar\omega(n+{1\over 2})-E \Biggr\} f(\beta)=0\,,
\ee
where one can use the infinite square well model for the potential $u(\beta)$
and take the assumption $\beta^2\rightarrow \langle \beta^2\rangle$.
Hence the wavefunction for the $\beta$ part may be written as
\be
f(\beta)=\beta^{-3/2} c^i_{L,R} J_\nu (k^i_{L,R}\beta)\,,
\ee
where (notice the slight change of notation with respect to the original)
\be
\nu=\sqrt{L(L+1)-{3\over 4}R^2+{9\over 4}}
\ee
and
\be
k^i_{L,R}=x^i_{L,R}/\beta_w\,,
\ee
with $\beta_w$ is the position of the infinite wall and $x^i_{L,R}$ is 
the $i-$th zero, $i=1,2,3,...$ of $J_\nu$. 
Here $c^i_{L,R}=\sqrt{2}/\beta_w J'_\nu(x^i_{L,R})$. The energy is 
\be
E^i_{L,R}={\hbar^2\over 2B_m\beta_w^2} (x^i_{L,R})^2+\hbar\omega n 
\langle iLRn \mid \beta^2\mid iLRn \rangle.
\ee

\section{Algebraic methods}
We have preferred to dedicate a separate section to the solutions of the Bohr
hamiltonian obtained through group theoretical techniques, even if this
breaks the chronological order and necessarily implies a redoubling of some
topics, because of the insight that they provide. We focus here especially
on the results obtained with the su(1,1) algebra. More informations on this
argument may be found in refs. \cite{Wyb}, ch. 18 and \cite{cizpa,CWood}, 
while a thorough group theoretical analysis of the collective model is 
found in ref. \cite{Kem,MO1}: the works of Chac\'on, Moshinsky and Sharp 
and of Kemmer, Pursey and Williams are very often considered as 'classics' 
on the group theory of the collective model.
Recently Rowe, Turner and Repka \cite{RTR} gave an useful algorithm for the 
computation of SO(5) spherical harmonics.

\subsection{Rowe and Bahri's work}
\label{Rowe-alg}
An alternative solution of the harmonic oscillator and of the 
Davidson potential has been given in ref. \cite{Rowe1}. It is 
easy to recognize that the generalized coordinates of the
nuclear collective model, ${q_\nu}$ and their conjugate momenta 
${\pi_\nu}$, may be used to form operators that are closed under 
commutation. Having defined the scalars products $\beta^2\=\sum_\nu 
\mid q_\nu\mid^2$ and $\pi^2\=\sum_\nu \mid \pi_\nu\mid^2$ we can 
introduce the operators
\be
{\hat Z}_1=\pi^2 \qquad {\hat Z}_2\=\beta^2 \qquad 
{\hat Z}_3={1\over 2}\sum_\nu (q_\nu \cdot\pi_\nu +\pi_\nu \cdot q_\nu )
\ee
that span an $sp(1,{\mathbb R}) \sim su(1,1)$ algebra. They are in fact
closed under commutation:
\be
[\hat Z_1,\hat Z_2]=-4i \hat Z_3 \quad
[\hat Z_3,\hat Z_2]=-2i \hat Z_2 \quad
[\hat Z_3,\hat Z_1]=2i \hat Z_1\,.
\label{pro}
\ee
With the linear transformation
\be 
\hat X_1\={1\over 4} \bigl( \hat Z_1-\hat Z_2 \bigr) \qquad ~
\hat X_2\={1\over 2} \hat Z_3 \qquad ~
\hat X_3\={1\over 4} \bigl( \hat Z_1+\hat Z_2 \bigr)
\label{tra1}
\ee
one may recognize the standard $su(1,1)\sim so(2,1)$ commutation relations
\be
[\hat X_1,\hat X_2]=-i \hat X_3 \qquad ~
[\hat X_2,\hat X_3]=i \hat X_1 \qquad ~
[\hat X_3,\hat X_1]=i \hat X_2\,.
\ee 
It is also very useful to the define raising, lowering and weight operators
for this algebra (the so-called eigenoperator decomposition)
\be
\hat X_{\pm} = \hat X_1 \pm i \hat X_2 \qquad
\hat X_0 = \hat X_3
\label{tra2}
\ee
that obey the following commutation relations
\be
[\hat X_+,\hat X_-]=-2\hat X_0 \qquad [\hat X_0,\hat X_\pm]= \pm \hat 
X_\pm\,.
\ee
The action of the above operators on orthonormal bases states for the irreps 
of $su(1,1)$ ($\mid n,\lambda \rangle$ with n=0,1,2,..) 
is given by the equations:
\begin{eqnarray}
&\hat X_+&\mid n\lambda\rangle =\sqrt{(\lambda+n)(n+1)}\mid n+1,\lambda\rangle
\nonumber \\
&\hat X_-&\mid n+1,\lambda\rangle =\sqrt{(\lambda+n)(n+1)}\mid n\lambda\rangle
\\
&\hat X_0& \mid n\lambda\rangle ={1\over 2}(\lambda+2n)\mid n\lambda\rangle 
~~~~~~~~~~~~~~~
\nonumber
\label{set}
\end{eqnarray}
The Casimir operator is
$$
{\hat \mathbb C} \= {\hat X}_3^2-{\hat X}_1^2-{\hat X}_2^2\={\hat X}_0 
({\hat X}_0-1 ) -{\hat X}_+ {\hat X}_- \= $$
\be
{1\over 8}\Bigl( \hat Z_1\hat Z_2+  
\hat Z_2\hat Z_1\Bigr)-{\hat Z_3^2 \over 4}
\label{Cas1}
\ee
and takes the values
\be
{\hat \mathbb C}\mid n\lambda\rangle ={1\over 4}\lambda (\lambda-2)
\mid n\lambda\rangle\,.
\label{Cas2}
\ee
From what we have summarized here it follows that the harmonic oscillator
hamiltonian may be written as (where we are omitting some $\hbar \omega$ 
factor) 
\be
{\hat Z}_1+{\hat Z}_2\=4{\hat X}_3
\ee
which is diagonal in the basis given above. Its spectrum is 
\be
E_{n\lambda}\=(2n+\lambda)
\ee
where $\lambda=v+5/2$.
With the remarkable nonlinear transformation \cite{Rowe1,Rowe2}
\be
{\hat Z}_1\rightarrow {\hat Z}_1+ {\hbar^2\epsilon\over{\hat Z}_2 } \qquad
{\hat Z}_2\rightarrow {\hat Z}_2 \qquad {\hat Z}_3\rightarrow {\hat Z}_3
\ee
the new operators satisfy again the $sp(1,{\mathbb R}) \sim su(1,1)$ 
commutation relations. 
The spectrum
\be
E_{nv}\=(2n+1+\sqrt{(v+3/2)^2+\epsilon})
\ee
is found introducing the definition of $\lambda$ that comes from 
the comparison between the eigenvalues of $\mathbb C$ obtained from
(\ref{Cas2}) and (\ref{Cas1}).

\subsection{Algebraic approach to Coulomb-like and Kratzer-like
potentials}
The preceeding treatment of the Davidson potential, together with
ch. 18 in ref. \cite{Wyb}, has served as a
source of inspiration for the treatment given in \cite{FV2}.
The spectrum of the Coulomb-like and Kratzer-like potentials may be 
derived in a similar fashion by noticing that the following operators
\be
\hat Z_1\=4\beta\Bigl(\pi^2+{B\over\beta^2}\Bigr) \qquad ~
\hat Z_2\=\beta \qquad ~
\hat Z_3\=2\bigl(\hat{\vec q}\cdot \hat{\vec \pi} -i\bigr)
\ee
are closed under commutation with the same relations of (\ref{pro}). 
With the Kratzer-like potential (that contains also the 
Coulomb-like case, when $B=0$) the operator $\beta{\cal H}$ is in 
fact expressible as a linear combination of the elements of the 
algebra of $su(1,1)$, namely in the form 
\be
\beta {\cal H} \=\hat Z_1/4-A.
\label{bH}
\ee
By defining again raising, lowering and weight operators we can write
the eigenvalue equation for the Bohr hamiltonian as
\be
\Bigl[ (1+4\epsilon)\hat X_1 -2A + (1-4\epsilon)\hat X_3 \Bigr] \Psi \=0
\ee
and following the procedure in \cite{Ald} we can perform a (1,3) 
hyperbolic rotation of an angle $\theta$
to diagonalize the eigenvalue equation. By choosing $tgh(\theta)= - (1+4
\epsilon)/(1-4\epsilon)$ (valid for $\epsilon <0$) we obtain a diagonal 
relation:
\be
\hat X_3 \tilde\Psi \= {A\over \sqrt{-4\epsilon}} \tilde \Psi\,,
\ee
where $\tilde \Psi$ is the rotated wavefunction.
The Casimir operator of the so(2,1) algebra is evaluated to be:
\be
\hat \mathbb{C}_2\= \hat \Lambda^2 + \hat X_- - \hat X_+ + B + 2 \,,
\ee
with eigenvalue $\tau(\tau+3)+B+2$.
The two last equations must be  compared with the two following eigenvalue
equations (for unitary representations $D^+$ \cite{Baru}):
\bea
\hat X_3 \mid \phi,\xi\rangle &\=& (\xi-\phi)\mid \phi,\xi\rangle\nonumber \\
\hat \mathbb{C}_2 \mid \phi,\xi\rangle &\=& \phi(\phi+1)\mid \phi,\xi\rangle\,.
\eea
This comparison yields a spectrum of the form:
\be
\epsilon_{\tau,\xi} \=-{A^2/4 \over (\sqrt{(\tau+3/2)^2+B}+1/2+\xi)^2}
\ee
that coincides with the one found from the direct solution of the differential 
equation with a Kratzer-like potential.

The algebra associated with the SO(5) group plays the role of a degeneracy 
algebra \cite{Cord}, while the group SO(2,1) is associated with the spectrum 
generating algebra.
For what concern either the problem considered here and the one in the previous
subsection the chain of subalgebras that gives the labels of the set of 
orthonormal states $\{\mid \xi \tau \alpha L M \rangle \}$ is given as 
\cite{Rowe1,Rowe2,FV1}:
\be
\begin{array}{ccccccccc}
SU(1,1)&\times &SO(5)&\supset& U(1)&\times & SO(3)&\supset & SO(2)\\
\lambda&~& \tau &\alpha &\xi&~&L&~&M \\
\end{array}
\ee
where $\lambda$ is an SU(1,1) lowest weight and $\alpha$ indexes the SO(3)
multiplicity. These basis diagonalize the above problems.  
We can thus state that the problem studied so far displays a 
SO(2,1)$\times$SO(5) dynamical algebra.

\subsection{Quasidynamical SO(2) symmetry for triaxial nuclei}

\begin{figure}[!t]
\epsfig{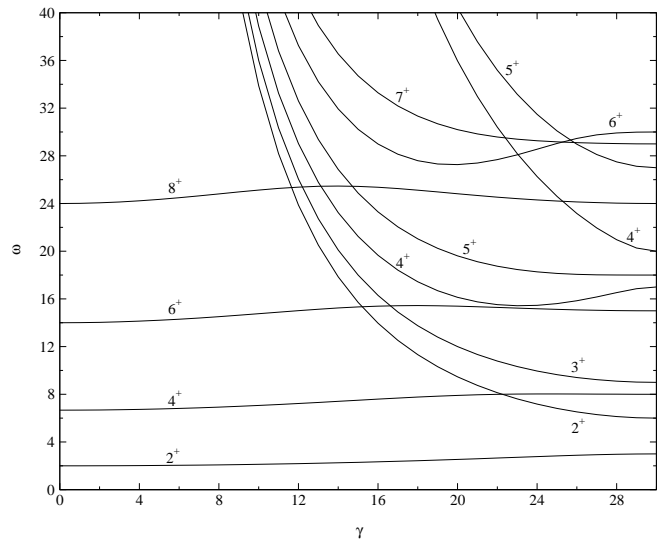}
\caption{Rotational part of the spectrum of the rigid 
triaxial rotor (Davydov's model).}
\label{rig}
\end{figure}
In fig. \ref{rig} we plot the eigenvalue of the rigid triaxial rotor
hamiltonian (see sect. V) as a function of $\gamma_0$.  
For $\gamma=0$ and $\gamma=\pi/3$ the 
projection of the third component of the angular momentum on the 
intrinsic axis 3 gives a good quantum number ($K$), while for $\gamma=\pi/6$
the eigenvalue of the projection on the intrinsic axis 1 is a good 
quantum number ($R$). In the intermediate regions, none of them may be taken 
as a good quantum number. 
Moving from $\gamma=30^o$ towards $\gamma=0^o$,
different groups of states may be classified into bands: a first band ($0^+,
2^+, 4^+,...$) tend to the finite axial rotor values;
a second band ($2^+,3^+,4^+,...$) is identified by its behaviour
when $\gamma\rightarrow 0$ (in fig. \ref{rig} this group of states somewhat
cluster around $\gamma\sim 10^o-12^o$); the beginning of a third band 
($4^+,5^+,....$) is seen to escape to infinity at a quicker pace (exiting
fig. \ref{rig} at around $\gamma\sim 20^o$).
The experimental observation that a classification in $\beta$ and $\gamma$
bands seems an almost universal feature of nuclear spectra reinforces
this choice. The labeling with the $K$ quantum number is often encountered 
in the literature, although for what we have said here it is not adequate.
One may describe this situation in terms of a quasidynamical symmetry 
\cite{RoL,RoWi} of a somewhat strange character: at $\gamma=0^o$ the group
SO(2) is a symmetry of the system, associated with $K$, while at $\gamma=30^o$ 
another SO(2) group is a symmetry of the system, associated with $R$, being
the chain U(5)$\supset$SO(3) common to the whole sector $0\le\gamma\le \pi/6$.
In the intermediate region $0<\gamma< \pi/6$ the SO(2) symmetry is broken
(badly broken in a 'classical' sense),
but it must be noticed (see fig. \ref{rig}) that the structure of the 
rotational spectrum present at $\gamma=0^o$ {\it persists} in the whole 
sector without being altered in a dramatic way. Only a smooth and slight 
change may be seen. On the other side the structure of the 'maximally' 
triaxial rotor at $\gamma=30^o$ {\it persists} also in the region around 
$\gamma\sim 20^o-30^o$. In the intermediate region these groups of states
escape to infinity, as already said. It must be further noticed that the 
regions where the various states that comes from the axial rotor side are 
more affected is exactly the region where the states coming from the 
$\gamma=30^o$ triaxial rotor diverge. The strange character of this 
quasidynamical symmetry mentioned above is that (at variance with the 
case discussed by Rowe and collaborators \cite{RoL,RoWi},
where a true phase transition was present between two exactly solvable limits 
associated with different symmetries and different group structures) here 
we are dealing with a smooth transition between two limits which formally 
have the same underlying group structure, SO(2), and there is no critical 
point in between.
Therefore we conclude that the use of a label that mimics the $K$ 
quantum number, that 
retain the formal division in $\beta$ and $\gamma$ bands typical of 
an axial rotor, is not only justified by the empirical observation that 
non-axial nuclei display the same classification in bands, but it is also 
supported in view of arguments based on a group theoretical approach. 
It is not clear at present if a quantization procedure around a tilted 
axis may help to further shed light on this aspect.

It is understood, however, that $K$ has been introduced as a convenient 
label (for basis states and eigenvalues) that comes into play in the
branching rules from SO(5) to SO(3) \cite{Kem,RTR}. 

\section{Recent developments of the collective model}
We cannot completely neglect, although we will just touch the argument 
briefly, that the collective model has recently experienced new important 
developments that are connected with the various solutions that we have 
summarized in the present review or that furnish new strategies to get 
numerical solutions. These developments have in common the 
{\it Streben} to simplification and tractability.

\subsection{Caprio's simplified approach to the CM}
The more general hamiltonian of the geometric collective model contains 
a  series expansion in terms in the surface deformation coordinates and
their conjugate momenta. The Bohr hamiltonian corresponds to the truncation
of all the kinetic terms with order higher than the second. 
Caprio studied a simplified version \cite{Mark2} of the hamiltonian in which
only the leading order kinetic term and the three lowest leading order
potential terms are retained, namely:
\be
H={1\over B_2} [\pi\times\pi]^{(0)}+
{C_2 \over \sqrt{5}} \beta^2-\sqrt{2 \over 35} C_3 \beta^3
\cos{3\gamma}+{C_4\over 5}\beta^4\,.
\ee
This hamiltonian has a structure rich 
enough to encompass rotational, vibrational and $\gamma-$unstable cases, 
but it's still very complex and contains four parameters. 
Caprio exploited analytic scaling relations to reduce the number of 
parameters from four to two 
and summarized the predictions of the model on two-dimensional contour 
plots. He used two 'simple' observations to make the reduction. Firstly,
an overall multiplication of the hamiltonian by a constant does not affect 
eigenvalues and wave functions. Secondly, with the transformation 
\be
V(\beta,\gamma)\rightarrow V'(\beta,\gamma)=a^2V(a\beta,\gamma)
\ee
all the eigenvalues are multiplied by $a^2$ and the wave functions 
are radially dilated by $1/a$. Other simplifications and graphs
are given in the text to which we refer the reader \cite{Mark2} for
a more detailed discussion.
The author mentions also the difficulty in numerical diagonalization 
of the problem in the basis of the five-dimensional harmonic oscillator,
emphasizing that a large number of basis functions must be used in order to
have correct results. A possible solution to this problem is outlined in the
following section.

\subsection{Rowe's tractable version of the CM}
The Bohr-Mottelson collective model may in principle be solved
numerically for any potential that is an analytic functions of the two
variables, but the expansion of rotational wave functions in a spherical
vibrational basis is very slowly convergent. The diagonalization is 
therefore very difficult. Rowe proposed a method in which the basis
is constructed with analytic wave functions whose $\beta-$part are 
centered around a non-zero value \cite{Rowe-new}. The convergence 
with this method is much faster for a wide variety of model hamiltonians
and this method represents an important improvement over older models.
This approach, that we will only briefly sketch, is formulated in an 
algebraic way using the group
SU(1,1)$\times$SO(5) as a dynamical group and recognizing that SO(5)-
invariant hamiltonians are diagonalizable using SU(1,1) as a spectrum 
generating algebra (see \cite{Rowe-new} for details). The dynamical subgroup
chain employed is SU(1,1)$\times$SO(5)$\subset$ SO(5)$\subset$ SO(3).
The Davidson potential \cite{Dav,Elli,Elli2,Rowe1}, that is an exactly 
solvable case, is employed for what concern the dependence on $\beta$. 
Its exact wave functions are (approximately) centered around the minimum of 
the potential well and give a good starting point for diagonalization. 
This simple version may be applied
directly to all the cases which lie between the spherical vibrator and the
Wilets and Jean ($\gamma-$flat) limit.

A more general extension of use of the basis obtained from the solution 
of the Bohr hamiltonian with the Davidson potential is also treated.
A periodic potential in gamma $V(\gamma)=-\chi \cos{3\gamma}$ is added to the 
hamiltonian. The reader is referred to the original paper for the details
and the discussion. This model is very flexible because of the rich variety
of spectra that may generate and it is more physical because of the relaxation
of the hypothesis of rigidity. Furthermore it allows for a tractable 
diagonalization procedure.

In principle this model, with suitable choice of potentials and parameters,
should be able to cover the whole complexity of nuclear collective 
features associated with the quadrupole degree of freedom.

\subsection{Rigid triaxial model with independent inertia moments}
In the rigid triaxial rotor model irrotational flow moments of inertia
have been usually assumed. In Ref. \cite{Wood} the rigid triaxial rotor
model is studied including the three components of the moments of inertia
as parameters that may be fitted to experimental data. The rigid rotor 
hamiltonian is
\be
H\= \sum_{i=1}^3 A_i\hat Q_i^2\,,
\label{rh1}
\ee
where the three parameters $A_i$ are defined as
\be
A_i={1\over 2{\cal I}_i}, \qquad i=1,2,3
\ee
and ${\cal I}_1\ge{\cal I}_2\ge{\cal I}_3$. The operators $\hat Q_i$
are the components of the angular momentum in the body-fixed frame of 
reference.
The hamiltonian (\ref{rh1}) may be rewritten in terms of diagonal and
non-diagonal operators (using the definitions $\hat Q_\pm=\hat Q_1 \pm 
i \hat Q_2$) as
\be
H\= A\hat Q^2+F \hat Q_3+G (\hat Q_+^2+\hat Q_-^2)
\ee
where $A=(A_1+A_2)/2$, $F=A_3-A$ and $G=(A_1-A_2)/4$.
A few low-lying eigenstates may obtained analytically from the diagonalization
of the hamiltonian in the rigid rotational basis $\mid LMK\rangle$
and they are summarized in Table \ref{ta5}. 
\begin{table}[!t]
\begin{tabular}{|c|c|}
\hline
$L$&$E(L)$\\\hline
0&0\\
2&$6A+2F \pm 2\sqrt{F^2+12G^2}$\\
3&$ 12 A+4F$\\ \hline
\end{tabular}
\caption{Low-lying eigenstates of the model described in \cite{Wood}.
Their total angular momentum is also indicated. A few other cases might 
be solved.}
\label{ta5}
\end{table}
Some of the eigenvectors and electric quadrupole properties are also discussed
in the paper and from an analysis it emerges that the components of the  
moment of inertia are not irrotational. For further discussion we refer the
reader to the cited work.

\section{Acknowledgments and apology} 
I wish to thank M.Caprio (Yale), K.Heyde (Gent) and A.Vitturi (Padova) 
for their help, encouragement and collaboration and to D.J.Rowe (Toronto) 
for his kind and keen interest in my work. I also thank G.L\'evai (Debrecen)
and R.F.Casten (Yale) for the comments on a preliminary draft of this work. 
Discussions with D.Bonatsos (Tessaloniki), F.Iachello (Yale) and
N.Pietralla (SUNY) are also acknowledged. Financial support 
and warm hospitality are acknowledged from the institutions wherein 
various parts or stages of this work has been carried on: 
FWO-Vlaanderen and Universiteit Gent (Belgium), 
Universit\`a di Padova and INFN (Italy) 
and Universidad de Sevilla (Spain).  

We apologize once again with all the researchers whose works have been 
omitted, mistreated or simply forgotten in the present paper, noticing
again that the main purpose was a discussion of solutions of the Bohr 
hamiltonian. 
To help the inquisitive reader and to render justice as far as we can
we included at the end of the bibliography an additional disordered list 
of interesting papers whose work for various reasons 
has not been summarized here 
\cite{f1,g,g1,g2,g3,g4,g6,g7,g8}.

\section*{APPENDIX A}
\subsection*{Number of repetitions} 
\renewcommand{\theequation}{A . \arabic{equation}}          
\setcounter{equation}{0} 
 
Within a given set of quantum numbers (a given $\tau$ in the general  
$\gamma-$unstable case, or $N$ in the harmonic oscillator case) the same  
$L$ may appear more than once. We define the repetitions of $L$ as the  
number of counts of that $L$ minus one.  
Despite the fact that this appendix could seem a purely academic  
{\it divertissement}, its novel number theoretical content is not 
completely detached from its physical significance: if a model 
predicts two states that are degenerate in energy to have the same $L$  
quantum number it may be very difficult to experimentally disentangle  
the two. The procedure of assignation of quantum numbers to measured 
spectra is not free from its own uncertainties and  
the outcome is a crucial test for theories. It seems thus very important 
to have a detailed knowledge of what we should expect from our  
theoretical calculation to look through the data. It will be clear that, 
in the following cases, the occurrence of repetitions is rather rare and it 
appears frequently only for high lying states so that   
likely is not prejudicial to any analysis that was carried on so far. 
 
\subsection*{General $\gamma-$unstable case} 
We will begin with the general $\gamma-$unstable case: the problem is  
to count how many repeated  
multiplicity of SO(3) appears within a given irrep of SO(5), or to count  
the repetitions of $L$'s for a given $\tau$. 
Looking at table I in the text we can see that up to $\tau=5$ every $L$  
appears just one time, so the number of repetitions is 0. For $\tau=6$, the  
quantum number $L=6$ appears two times, so the number of repetitions is 1. 
Continuing along this line one easily finds (by hand in the first few cases) 
that the number of repetitions, starting from $\tau=6$ is: 
\be 
1,1,2,4,5,7,10,12,15,19,22,26,31,35,40,... 
\label{1seq}
\ee  
This sequence has been calculated with a simple computer code  
and it corresponds to the sequence obtained from 
\be 
R(\tau)=\left\{ 
\begin{array}{lr}
1+\Bigl[ {n(n-3)\over 6} \Bigr] &\tau \geq 6\\
0 &\tau  < 6
\end{array}
\right.
\ee  
with $n=\tau-3=3,4,5,...$, where square brackets denote the integer part,  
or floor function. Within the same range of $n$ this is also connected to  
the number of solutions to the equation $x+y+z=0 ~(~mod~ n~)$ with  
$0\leq x<y<z<n$. For instance if $n=3$ the only possible solution is $x=0$,  
$y=1$ and $z=2$. This sequence arises also in connection with the so-called 
'orchard problem' (See \cite{Slo}).

\subsection*{Harmonic oscillator case} 
For the case of harmonic oscillator a further analysis may be done. 
For a given $N$ a certain number of pairs 
of $(n_\beta,\tau)$ satisfy $N=2n_\beta+\tau$. For each pair a value of 
$\tau$ is associated with a set of quantum numbers $L$ (the same of the  
preceding section). Hence the set of $L$'s corresponding to each $N$ is  
larger that in the former case (the degree of degeneration is higher) and 
the sequence of numbers of repetitions, starting from $N=4$, may be written  
as follows 
\be  
2,3,7,11,18,26,36,48,63,79,99,121,146,...
\label{2seq} 
\ee 
while in 
all the cases with $N\le 3$ there are no repetitions as can be seen from 
Table II. 
Unlike the previous case it was not possible to make a connection between 
the present sequence and any previously known sequence \cite{Fors}, nor it  
was possible to find a simple formula to express the $N$th term of the  
sequence. Note that the number of 
repetitions in this case is not simply given by
\be
{\bar R}(N) =\sum_{n_\beta=0}^{[N/2]} R(\tau=N-2n_\beta)
\label{Rbarra}
\ee
because it happens that many repetitions come from different 
$(n_\beta,\tau)$ pairs (otherwise they would have started from $N=6$).
It is furthermore intriguing to notice that when the terms of the sequence
(\ref{2seq}) are subtracted with the corresponding terms of the sequence 
that may be built from formula (\ref{Rbarra}), the remaining sequence 
(starting from $N=4$) is
\be
2,3,6,10,15,21,28,36,45,55,66,....
\ee
and apart from the first two elements the differences between two of any other
couple of adjacent terms increases arithmetically (!). 

Two very simple computer codes, IRREP and IRHAR, written in Fortran77, are  
available from the author or from WWW \cite{sito}. They count the number of 
times that a given quantum number occurs, and give thus the number of 
repetitions as in (\ref{1seq}) and (\ref{2seq}) .

\section*{APPENDIX B}
\subsection*{Meaning of the terms "softness" and "rigidity"} 
\renewcommand{\theequation}{B . \arabic{equation}}          
\setcounter{equation}{0} 
The purpose of the present Appendix is to clarify the meaning that should 
be attributed to the words ``softness'' and ``rigidity'' in the context of 
the collective model. 

In general the solution of the Schr\"odinger equation depends on the 
potential that is chosen. In the case of the Bohr hamiltonian (restricted 
to the quadrupole degree of freedom) the potential is a function of 
 $0\le \beta$ and $0\le \gamma \le \pi/3$. Every couple of values 
${\beta, \gamma}$ univocally corresponds to a given quadrupole surface. 
The state of the quantum system is represented by wavefunctions that have a
dependence on $\beta$, $\gamma$ or both. When the potential is very deep 
and narrow its wavefunctions (especially the more low-lying ones) are very 
localized in the region of the minimum and they may be associated with just 
one value of $\beta$.
The object (that lives in the configuration space) that is described by 
such state has therefore a rigid shape in the sense that its surface 
doesn't exhibit fluctuations.
On the contrary a more shallow potential gives rise to extended wavefunctions.
This means that the real object may be thought as a superposition of 
different quadrupole surfaces each with its own weight. The term "soft" 
was used to describe such an object. This system exhibits 
non-null fluctuations around the mean square deformation.

Sometimes it is argued by analogy with the work of Wilets and 
Jean \cite{Wilet} that the independence of the potential on one variable is 
associated with the term softness. The dependence or independence on a 
given variable should not be used to establish the softness or rigidity of 
some state. The unstable case of Wilets and Jean (for what concern the 
$\gamma$ d.o.f., that is a sort of angular variable and not a radial variable)
 is, among the many possible soft cases, the most extreme one.

For example we may decide to set the potential in $\beta$ to a constant on 
its whole semi-infinite domain.

In this $\beta-$independent case the 
eigenstates lie in the continuum and the eigenfunctions
have a periodic behaviour at infinity. It is thus not possible to associate
a finite value to the deformation of such a state!
The situation in the paper of Wilets and Jean is different: two cases were 
discussed, the infinite square well and the displaced harmonic oscillator. 
They are both examples of a soft behaviour,
because the two quantum mechanical potentials give rise to solutions endowed 
with extended wavefunctions.
One can see that while the displaced oscillator has a true dependence on 
$\beta$, the infinite square well has a dependence hidden in the boundary 
conditions (that are essential to fix the spectrum).
The analogy with the $\gamma-$unstable case is thus rather feeble, because
a $\beta-$unstable counterpart has not a definite physical meaning. The 
word "unstable" refers to the fact that the potential has no minima (and 
hence there's no "stability" in the
classical sense) and should be read as a synonymous of "independent".

In summary the word "soft" applies to the usual quantum mechanical case, 
while the word "rigid" must be associated with very peculiar cases for 
which either the value of $\beta$
(and/or $\gamma$) is fixed or the potential is of a very special kind. 
At the beginning of the
study of the Bohr-Mottelson model many "rigid" cases were discussed, mainly for
the need of simplicity and for the lack of more refined ("soft") solutions.

\section*{APPENDIX C}
\subsection*{Electromagnetic transitions} 
\renewcommand{\theequation}{C . \arabic{equation}}          
\setcounter{equation}{0} 
In some of the spectra displayed in the paper, electromagnetic 
transition rates have appeared. Although it is not the subject of the
present monograph, it is useful to discuss to some extent how they may 
be derived, especially because BE(2) values, being very sensitive to the
details of the wave functions, are often considered as crucial quantities
for the application of the aforementioned models to nuclei.  

The electromagnetic operator that is often encountered in this 
context reads:
\be
T(E2)=t\beta\Bigl[ D_{\mu,0}^{(2)}\cos{\gamma}+{1\over \sqrt{2}}
(D_{\mu,2}^{(2)}+D_{\mu,-2}^{(2)})\sin{\gamma}\Bigr]
\ee
where $t$ is a scale factor. This is the first order approximation of
the more general E2 operator \cite{GG,EG}. In the geometric collective 
model this operator is written in the laboratory frame coordinates as an
expansion:
\be
T(E2;\mu)=a\alpha_{2,\mu}+b[\alpha\times \alpha]^{(2)}_\mu+
{\cal{O}}(\alpha^3)\,,
\ee
where each term is coupled to a total angular momentum equal to 2.
It has been often argued \cite{Pep,Capslo} that a second order approximation
\be
T(E2)=a\beta\Bigl[ D_{\mu,0}^{(2)}\cos{\gamma}+{1\over \sqrt{2}}
(D_{\mu,2}^{(2)}+D_{\mu,-2}^{(2)})\sin{\gamma}\Bigr]-b\sqrt{2\over 7}
\beta^2\Bigl[ D_{\mu,0}^{(2)}\cos{2\gamma}-{1\over \sqrt{2}}
(D_{\mu,2}^{(2)}+D_{\mu,-2}^{(2)})\sin{2\gamma}\Bigr]
\ee
leads to non-trivial modifications in the calculations of the matrix 
elements.
Transition rates are then calculated in the following way:
\be
B(E\lambda;J_i\rightarrow J_f)={\langle J_i \mid T(E\lambda) \mid J_f
\rangle \over 2J_i+1}
\ee
where $\lambda$ is the angular momentum that characterizes the transition.
Usually B(E2) values have been normalized taking the value of the B(E2)
for the transition $2_1^+\rightarrow 0_1^+$ as 100.

Finally we mention that, for E0 transition, the electromagnetic operator 
is 
\be
T(E0)\propto \beta^2\,.
\ee

\end{document}